\documentclass[final,1p,times]{elsarticle}

\usepackage{amssymb}
\usepackage{amsmath}
\usepackage{braket}
\usepackage{bbm}
\usepackage{upgreek}
\usepackage{xcolor}
\usepackage{xspace}

\usepackage{multirow}

\biboptions{sort&compress}

\usepackage{hyperref}

\newcommand*\patchAmsMathEnvironmentForLineno[1]{%
\expandafter\let\csname old#1\expandafter\endcsname\csname #1\endcsname
\expandafter\let\csname oldend#1\expandafter\endcsname\csname
end#1\endcsname
 \renewenvironment{#1}%
   {\linenomath\csname old#1\endcsname}%
   {\csname oldend#1\endcsname\endlinenomath}%
}
\newcommand*\patchBothAmsMathEnvironmentsForLineno[1]{%
  \patchAmsMathEnvironmentForLineno{#1}%
  \patchAmsMathEnvironmentForLineno{#1*}%
}
\AtBeginDocument{%
\patchBothAmsMathEnvironmentsForLineno{equation}%
\patchBothAmsMathEnvironmentsForLineno{align}%
\patchBothAmsMathEnvironmentsForLineno{flalign}%
\patchBothAmsMathEnvironmentsForLineno{alignat}%
\patchBothAmsMathEnvironmentsForLineno{gather}%
\patchBothAmsMathEnvironmentsForLineno{multline}%
}

\usepackage{lineno}

\newcommand{\decay}[2]{\ensuremath{#1\!\to #2}\xspace}         
\newcommand{\comment}[1]{}

\def\CP      {\ensuremath{CP}\xspace}                 
\def\PB      {\ensuremath{B}\xspace}                 
\def\B       {\ensuremath{\PB}\xspace}
\def\Bd      {\ensuremath{\B^0}\xspace}
\def\Bu      {\ensuremath{\B^+}\xspace}
\def\Bbar    {\kern 0.18em\overline{\kern -0.18em \PB}{}\xspace}
\def\Bdb     {\ensuremath{\Bbar^0}\xspace}
\def\Bc      {\ensuremath{\B_c^+}\xspace}
\def\Lb      {\ensuremath{\Lambda_b^0}\xspace}
\def\PK      {\ensuremath{K}\xspace}                 
\def\kaon  {\ensuremath{\PK}\xspace}
\def\Kbar  {\kern 0.2em\overline{\kern -0.2em \PK}{}\xspace}
\def\Kp    {\ensuremath{\kaon^+}\xspace}
\def\Km    {\ensuremath{\kaon^-}\xspace}
\def\Kstar  {\ensuremath{\kaon^{*}}\xspace}
\def\Kstarz  {\ensuremath{\kaon^{*0}}\xspace}
\def\Kstarzb {\ensuremath{\Kbar^{*0}}\xspace}
\def\Kstarp  {\ensuremath{\kaon^{*+}}\xspace}
\def\KS    {\ensuremath{\kaon^0_S}\xspace}
\def\Pmu         {\ensuremath{\mu}\xspace}                 
\def\mumu       {\ensuremath{\Pmu^+\Pmu^-}\xspace}
\def\Ppi         {\ensuremath{\pi}\xspace}                 
\def\pion  {\ensuremath{\Ppi}\xspace}
\def\pip   {\ensuremath{\pion^+}\xspace}
\def\pim   {\ensuremath{\pion^-}\xspace}
\def\piz   {\ensuremath{\pion^0}\xspace}
\def\Ps      {\ensuremath{s}\xspace}                 
\def\squark    {\ensuremath{\Ps}\xspace}
\def\Bs      {\ensuremath{\B^0_\squark}\xspace}
\def\ellell       {\ensuremath{\ell^+\ell^-}\xspace}
\def\ee       {\ensuremath{e^+e^-}\xspace}
\def\emu       {\ensuremath{e^\pm\mu^\mp}\xspace}

\def\mup       {\ensuremath{\mu^+}\xspace}

\def\mun       {\ensuremath{\mu^-}\xspace}
\def\rhoz  {\ensuremath{\rho^{0}}\xspace}
\def\rhop  {\ensuremath{\rho^{+}}\xspace}

\def\Et       {\ensuremath{E_{\rm T}}\xspace}
\def\pt       {\ensuremath{p_{\rm T}}\xspace}
\def\qsq       {\ensuremath{q^2}\xspace}

\def\jpsi       {\ensuremath{J/\psi}\xspace}
\def\psitwos       {\ensuremath{\psi(2S)}\xspace}

\def\invfb   {\ensuremath{\mbox{\,fb}^{-1}}\xspace}
\def\invab   {\ensuremath{\mbox{\,ab}^{-1}}\xspace}
\def\mub{\ensuremath{\mathrm{\,\upmu b}^{-1}}\xspace}
\newcommand{\tev}{\ensuremath{\mathrm{\,Te\kern -0.1em V}}\xspace}
\newcommand{\gev}{\ensuremath{\mathrm{\,Ge\kern -0.1em V}}\xspace}
\newcommand{\GeV}{\ensuremath{\mathrm{\,Ge\kern -0.1em V}}\xspace}
\newcommand{\mev}{\ensuremath{\mathrm{\,Me\kern -0.1em V}}\xspace}
\newcommand{\MeV}{\ensuremath{\mathrm{\,Me\kern -0.1em V}}\xspace}
\newcommand{\kev}{\ensuremath{\mathrm{\,ke\kern -0.1em V}}\xspace}
\newcommand{\ev}{\ensuremath{\mathrm{\,e\kern -0.1em V}}\xspace}
\newcommand{\gevc}{\ensuremath{{\mathrm{\,Ge\kern -0.1em V\!/}c}}\xspace}
\newcommand{\mevc}{\ensuremath{{\mathrm{\,Me\kern -0.1em V\!/}c}}\xspace}
\newcommand{\gevcc}{\ensuremath{{\mathrm{\,Ge\kern -0.1em V\!/}c^2}}\xspace}
\newcommand{\gevgevcccc}{\ensuremath{{\mathrm{\,Ge\kern -0.1em V^2\!/}c^4}}\xspace}
\newcommand{\mevcc}{\ensuremath{{\mathrm{\,Me\kern -0.1em V\!/}c^2}}\xspace}

\newcommand{\btosll}{\ensuremath{b\to s\ell^+\ell^-}\xspace}
\newcommand{\btosmm}{\ensuremath{b\to s\mu^+\mu^-}\xspace}
\newcommand{\btosee}{\ensuremath{b\to s e^+e^-}\xspace}
\newcommand{\btosemu}{\ensuremath{b\to s e^+\mu^-}\xspace}

\newcommand{\btodll}{\ensuremath{b\to d\ell^+\ell^-}\xspace}

\newcommand{\eg}{\ensuremath{e.\,g.}\xspace}
\newcommand{\ie}{\ensuremath{i.\,e.}\xspace}
\newcommand{\vs}{\ensuremath{vs.}\xspace}

\def\ctl       {\ensuremath{\cos{\theta_l}}\xspace}
\def\thetal       {\ensuremath{\theta_l}\xspace}
\def\ctk       {\ensuremath{\cos{\theta_K}}\xspace}
\def\thetak       {\ensuremath{\theta_K}\xspace}

\def\CP                {\ensuremath{C\!P}\xspace}
\def\EOS               {\texttt{EOS}\xspace}

\renewcommand{\Re}{\ensuremath{{\cal R}e}\xspace}
\renewcommand{\Im}{\ensuremath{{\cal I}m}\xspace}

\catcode`\@=11 
\newcommand{\parenbar}{\mathpalette\p@renb@r}
\def\p@renb@r#1#2{\vbox{%
  \ifx#1\scriptscriptstyle \dimen@.7em\dimen@ii.2em\else
  \ifx#1\scriptstyle \dimen@.8em\dimen@ii.25em\else
  \dimen@1em\dimen@ii.4em\fi\fi \offinterlineskip
  \ialign{\hfill##\hfill\cr
    \vbox{\hrule width\dimen@ii}\cr
    \noalign{\vskip-.3ex}%
    \hbox to\dimen@{$\mathchar300\hfil\mathchar301$}\cr
    \noalign{\vskip-.3ex}%
    $#1#2$\cr}}}
\catcode`\@=12 

\newcommand{\op}[1]{\ensuremath{\mathcal{O}_{#1}}\xspace}

\journal{Progress in Particle and Nuclear Physics}

\begin{document}

\begin{frontmatter}



\title{Flavor Anomalies in Heavy Quark Decays}


\author[tudo]{J.\ Albrecht}
\ead{johannes.albrecht@tu-dortmund.de}
\address[tudo]{TU Dortmund, Experimentelle Physik 5, Otto-Hahn-Stra\ss{}e 4, 44227 Dortmund, Germany}

\author[tum]{D.\ van Dyk}
\ead{danny.van.dyk@gmail.com}
\address[tum]{TU Munich, Theoretische Physik T31, James-Franck-Str. 1/I, 85748 Garching, Germany}

\author[rwth]{C.\ Langenbruch\corref{cor}}
\ead{christoph.langenbruch@cern.ch}
\address[rwth]{RWTH Aachen, I. Physikalisches Institut B, Sommerfeldstr. 14, 52056 Aachen, Germany}

\cortext[cor]{Corresponding author}

\begin{abstract}
Recent measurements of $b$-hadron decays show a pattern of 
consistent tensions with
the respective Standard Model~(SM) predictions. These tensions appear 
both in the sector of rare flavour-changing neutral currents and in tree-level semileptonic $b$-hadron decays. 
Flavour-changing neutral-current decays are loop-suppressed in the SM and are thus 
very susceptible to contributions from new heavy particles and/or new interactions beyond the SM. 
  
In rare semileptonic decays tensions are observed in measurements of branching fractions and angular observables,
as well as in lepton flavour universality tests. 
Lepton flavour universality is also tested by comparing tree-level $\decay{b}{c\ell^-\bar{\nu}_\tau}$ processes involving third generation leptons $\ell=\tau$ to semileptonic decays with light leptons $\ell=e,\mu$ in the final state. These tests also
show tensions between measurements and the SM prediction.
    
Taken together, these tensions constitute the so-called \textit{flavour anomalies} in $b$-hadron decays,
and could be first signs of New Physics (NP) beyond the SM, if established beyond any reasonable doubt. 
This article reviews both the current experimental status of the \textit{flavour anomalies} and developments
for the relevant theoretical predictions. The review concludes with a discussion of future prospects for the field.
\end{abstract}




\end{frontmatter}

\tableofcontents

\clearpage
\section{Introduction}
\label{sec:introduction}
Since the conception of the Standard Model of particle physics (SM),
semileptonic and radiative $b$-hadron decays have been of great phenomenological
interest.

On the one hand, rare neutral-current induced $b$-hadron decays have been
considered among the most powerful probes for effects of beyond-the-SM (BSM) physics
since the early 1990s~\cite{Ali:1994bf}. In the SM,  they are only allowed at the loop-level and hence  
hypothetical BSM contributions at tree level can easily overcome the loop-suppressed SM contribution.
Thus, rare flavour-changing neutral current decays were originally used to search for such large, 
tree-level induced BSM effects. 
With the enormous data samples obtained by the Tevatron and the
$B$-factories and, later on, the LHC experiments, 
these types of decays 
are now used as precision tests of the SM.

On the other hand, exclusive semileptonic charged-current induced $b$-hadron decays have been an
indispensible ingredient in the determination of the Standard Model
quark mixing parameters, encoded in the unitary Cabbibo-Kobayashi-Maskawa (CKM)
matrix.
Together with other experimental constraints they allow to overconstrain
the small set of independent CKM matrix parameters~\cite{Ciuchini:1999xh,Hocker:2001xe}. These analyses, referred to as CKM metrology,
are therefore powerful probes of the CKM unitarity paradigm of the SM.
Another paradigm of the SM is lepton-flavour universality of the gauge interactions. 
It can readily be tested in charged-current induced decays to high precision, due to the large
experimental data sets of, and very accurate theoretical predictions for, these processes.

With the advent of the first signs of tensions in both types of decays,
the so-called \textit{flavour anomalies}, 
we see a strong and enduring renewal of interest 
in and scrutiny of analyses of $b$-hadron decays --- both on the experimental and the theoretical
side. In this review we collect and summarize past efforts from both sides
to understand the \textit{flavour anomalies} in exclusive $b$-hadron decays and furthermore give an outlook on future developments.
Inclusive $b$-hadron decays are of great relevance to the field, both to corroborate the anomalies in exclusive decays
with independent systematic effects and to complement the sensitivity to BSM couplings. 
Measurements of rare inclusive radiative and semileptonic decays have mostly been performed by the $B$-factory experiments and 
currently exhibit significant experimental uncertainties.
Updates by the Belle~II experiment are eagerly awaited. 
For this reason we limit this review to cover exclusive $b$-hadron decays only.

Long standing tensions are also present between the exclusive and inclusive determinations of the CKM matrix elements $V_{\rm cb}$ and $V_{\rm ub}$, and for \CP\ asymmetries in charmless two-body $B\to K\pi$ decays. In the kaon sector the situation on $\epsilon^\prime/\epsilon$ is unclear at present and could give rise to a further flavour anomaly. As this review concentrates on recent results in the $b$-hadron sector we do not discuss these measurements here.  

The experiments primarily contributing to studies of $b$-hadron decays that give rise to the \textit{flavour anomalies} are
the $B$-factory experiments BaBar~\cite{Aubert:2001tu,TheBABAR:2013jta} and Belle~\cite{Abashian:2000cg},
the Tevatron experiments CDF~\cite{Acosta:2004yw} and D\O~\cite{Abazov:2005pn},
and the Large Hadron Collider (LHC) experiments \mbox{ATLAS}~\cite{Aad:2008zzm}, CMS~\cite{Chatrchyan:2008aa}, and LHCb~\cite{Alves:2008zz,Aaij:2014jba}. 
The $B$-factory experiments operated at $e^+e^-$ colliders (BaBar at the PEP-II collider located at the Stanford Linear Accelerator Center and Belle at the KEKB collider at the KEK laboratory) running on the $\Upsilon(4S)$ resonance to produce $B^+B^-$ and $\Bd\Bdb$ pairs with a $b\bar{b}$ production cross-section of $1.05\,{\rm nb}$ on-resonance.  
Until the end of data taking in 2008 and 2010 the BaBar and Belle experiments recorded data samples of $426\invfb$ and $711\invfb$ on the $\Upsilon(4S)$ resonance. These datasets correspond to $471\,{\rm M}$ and $772\,{\rm M}$ $B\bar{B}$ pairs, respectively. 

The experiments at the LHC profit from the very large $b\bar{b}$ production cross-section in high energy $pp$ collisions. 
During the LHC Run~1 in the years 2011 and 2012 the LHC operated at centre-of-mass energies of $\sqrt{s}=7\tev$ (in 2011) and $8\tev$ (in 2012). 
In the following LHC Run~2, during the years 2015--2018, the LHC energy was increased to $\sqrt{s}=13\tev$. 
These energies correspond to $b\bar{b}$ production cross-sections of around $300\mub$ at $\sqrt{s}=7\tev$ and $560\mub$ at $\sqrt{s}=13\tev$~\cite{Aaij:2016avz,Aaij:2017qml}. 

The LHCb experiment is optimised for the analysis of $b$ and $c$-hadron decays. 
As a single arm forward spectrometer it covers the forward pseudorapidity range of $2<\eta<5$, 
where the production cross-section for $b$ and $c$-hadrons is largest. 
During the LHC Run~1 and Run~2 the LHCb collaboration collected a data sample corresponding to integrated luminosities of $3\invfb$ and $5.7\invfb$, respectively\footnote{Note that the LHCb experiment operates at a lower, constant, instantaneous luminosity than ATLAS and CMS to reduce pile-up (luminosity levelling).}. 
The general purpose experiments ATLAS and CMS cover a more central pseudorapidity region and are optimised for the search for directly produced new particles at high transverse momentum. 
Even though trigger thresholds are higher than at the LHCb experiment, 
ATLAS and CMS provide valuable measurements, especially for decays involving dimuons in the final state. 
The ATLAS and CMS experiments collected data samples corresponding to integrated luminosities of around $25\invfb$ during the LHC Run~1 and around $150\invfb$ during Run~2. 

Following the Long Shutdown~(LS)~2, during which the LHC experiments are upgraded, the LHC Run~3 is expected to start in 2022. 
The upgraded LHCb experiment~\cite{Bediaga:2012uyd} plans to take a data sample corresponding to an integrated luminosity of $50\invfb$ during the LHC Runs~3 and~4. 
With the planned LHCb Upgrade~II~\cite{Aaij:2244311,Bediaga:2018lhg}, the collaboration foresees to collect a data sample corresponding to $300\invfb$ in the LHC Runs after LS~4. 
The ATLAS and CMS experiments plan to collect samples corresponding to $300\invfb$ during LHC Run~3 and ultimately $3000\invfb$ during the High Luminosity LHC Runs following LS~3. 

The Belle~II experiment~\cite{Abe:2010gxa} located at the SuperKEKB $e^+e^-$ collider started taking data with the  instrumented detector in March 2019. 
The target instantaneous luminosity at SuperKEKB is a factor 40 larger than what was achieved at KEKB. 
Over the next ten years the Belle~II experiment expects to collect a data sample corresponding to $50\invab$, around a factor 50 larger than the sample available from the $B$-factories. 
With this large data sample and systematic uncertainties that are largely orthogonal to the experiments at hadron colliders, 
Belle~II will play an important role in the clarification of the \textit{flavour anomalies}.\\

This review is structured as follows. 
In Sec.~\ref{sec:effectivetheory} we introduce the theory framework for the description of the processes relevant for the \textit{flavour anomalies}. 
Section~\ref{sec:btosgamma} discusses rare radiative processes. While no clear anomalies are currently present in the sector of radiative decays, they give strong constraints on effective couplings relevant also for rare semileptonic decays. 
Rare (semi)leptonic decays are covered in Sec.~\ref{sec:btosll}, which includes the very rare purely leptonic decays (Sec.~\ref{sec:leptonic}), the more abundant semileptonic decays (Sec.~\ref{sec:semileptonic}), and lepton flavour universality tests in rare decays (Sec.~\ref{sec:lfu}). 
In Sec.~\ref{sec:treelevel} we present tree-level $\decay{b}{c\ell^-\bar{\nu}_\ell}$ decays with particular focus on the anomalies in $\decay{b}{c\tau^-\bar{\nu}_\tau}$ transitions. 
Finally, we summarise and give an outlook in Sec.~\ref{sec:conclusions}.

\section{Theory Framework}
\label{sec:effectivetheory}

Our theoretical understanding of $b$-hadron decays is complicated by the
multitude of relevant energy scales. These include the electroweak
scale $\mu_W \sim M_H$, the natural scale of the $b$-quark $\mu_b \sim m_b$,
and the hadronic scale 
$\Lambda_\text{had} \sim \text{a few hundred MeV}$.
Effective Field Theories (EFTs) are indispensable tools to disentangle
the different scales and make reliable and precise predictions for $b$-hadron
decays possible.\\

The Weak Effective Field Theory (WET) is the one such such EFT needed
to understand the decays of free $b$ quarks within the Standard Model (SM).
To this end one removes such SM quantum fields as dynamic degrees of freedom
that cannot be on-shell in $b$ decays, 
due the limited release of energy
$E \leq m_b$~\cite{Aebischer:2017gaw,Jenkins:2017jig}. The effects of these fields
--- the Higgs field, the top quark, and the $W$
and $Z$ bosons --- are encoded in the Wilson coefficients~\cite{Wilson:1972ee} of the
effective field theory.
The remainder of the SM fields are present in a set of local field operators starting
at mass dimension six. Consequently, the WET is a nonrenormalisable field theory.
Here, as in the literature, we will not discuss operators of mass dimension eight or larger,
which are suppressed by at least the maximal energy release $\mu_b^2 / \mu_W^2 \sim 0.2\%$.

The Wilson coefficients can be realiably calculated at the scale $\mu_W$ in perturbation theory,
in a process known as ``matchting'' of the SM amplitudes onto free-quark
matrix elements of the local operators.
The description of the $b$-hadron decays, however, requires hadronic matrix elements of
the operators that naturally live at the scale $\mu_b$ or below. The gap between both scales is bridged
by means of Renormalisation Group Evolution (RGE) of the WET operators and their respective Wilson coefficients
The RGE makes it possible to simultaneously calculate the Wilson coefficients at the scale $\mu_W$,
evolve them to the scale $\mu_b$, and use the hadronic matrix elements of the operators at
the latter scale. For an encompassing review of the SM effective theory and its RGE we refer to Ref.~\cite{Buchalla:1995vs}.
For the purpose of this review, we wish to highlight that the SM WET coefficients are lepton-flavour
universal, and that Wilson coefficients for operators with non-diagonal lepton-flavour indices are suppressed
by $m_\nu^2/M_W^2$ in the SM.\\

However, the WET also provides a framework to describe effects Beyond the Standard Model (BSM)
if one assumes the absence of non-standard forces and matter fields below the electroweak scale.
The lack of evidence for such BSM effects below the electroweak scale in direct searches has
made the WET the standard interface for the interpretation of $b$-quark decay observables.
While there are competing standards for a complete basis of WET operators at mass dimension
six~\cite{Aebischer:2017gaw,Jenkins:2017jig}, we instead use a basis that is more tailored toward the phenomenology
of the two types of $b$ decays discussed here: charged-current and neutral-current semileptonic
decays, and neutral-current radiative decays. Nevertheless, the insights gained in one basis
of WET operators can be translated onto other bases. To facilitate this process, the Wilson Coefficients
eXchange Format (WCxF)~\cite{Aebischer:2017ugx} has been agreed upon, which is supported by a majority of the public
codes that predict $b$ decay observables.\\

\begin{table}[t]
    \centering
    \begin{tabular}{c c c c c c c}
        \hline
        $\boldsymbol{i}$ &
            1U &
            2U &
             7 &
             8 &
             9 &
            10 \\\hline\hline
        $\boldsymbol{C_i}$ &
            -0.290 &
            +1.010 &
            -0.337 &
            -0.183 &
            +4.27  &
            -4.17  \\
        \hline
    \end{tabular}
    \caption{The values of the Wilson coefficients $C_i(\mu = 4.2\,\GeV)$ with $i=1U,2U,7,8,9,10$ in the SM
    to NNLL accuracy. The Wilson coefficients of the operators $i=3\dots6,7',8'$ are negligibly small
    in the SM. The remaining Wilson coefficients are zero in the matching to the SM at the present accuracy.
    The values have been obtained from the \EOS software, and evolved from the matching scale $\mu_0 = 120\,\GeV$.
    Parametric uncertainties appear due to, \eg, the value of the $t$ mass.
    }
    \label{tab:wet-btoD-WCs}
\end{table}
\paragraph{Neutral-current decays} For the description of neutral-current induced
$b\to D$, $D=s,d$ transitions we use the effective
Lagrangian
\begin{equation}
\label{eq:wet-btoD-Leff}
\begin{aligned}
    {\cal L}^{\rm eff}_{b\to D} = \frac{4 G_{\text{F}}}{\sqrt{2}}
        & \lambda_t^D \left[
            \sum_{i=7,\dots,8'} C_i \op{i} + \sum_{i=9,\dots,T5} C_i \op{i} + \sum_{i=1c,2c,3,\dots,6} C_i \op{i}
          \right]\\
    +   & \lambda_u^D \bigg[
            C_{1u} \op{1u} + C_{2u} \op{2u} - C_{1c} \op{1c} - C_{2c} \op{2c}
          \bigg]
    + \text{h.c.}
\end{aligned}
\end{equation}
Here $G_\text{F}$ is the Fermi constant, 
$\lambda^D_{U} \equiv V_{Ub}^* V_{UD}^{\phantom{*}}$, $U=u,c$, and the sets of local operators at mass dimension six 
are organized as radiative operators ($i=7,\dots,8'$), semileptonic operators ($i=9,\dots,T5$), and
four-quark operators ($i=1c,2c,1u,2u,3,\dots,6$). The radiative operators are defined as:
\begin{equation}
    \label{eq:wet-btoD-radiative}
    \begin{aligned}
        \op{7}  = & \frac{e  }{16\pi^2} \overline{m}_b(\mu) [\bar{s} \sigma^{\mu\nu}     P_R b] \, F_{\mu\nu}\,, &
        \op{7'} = & \frac{e  }{16\pi^2} \overline{m}_b(\mu) [\bar{s} \sigma^{\mu\nu}     P_L b] \, F_{\mu\nu}\,, \\
        \op{8}  = & \frac{g_s}{16\pi^2} \overline{m}_b(\mu) [\bar{s} \sigma^{\mu\nu} t^A P_R b] \, G^A_{\mu\nu}\,, &
        \op{8'} = & \frac{g_s}{16\pi^2} \overline{m}_b(\mu) [\bar{s} \sigma^{\mu\nu} t^A P_L b] \, G^A_{\mu\nu}\,. \\
    \end{aligned}
\end{equation}
Here $t^A$ represents an $\text{SU}(3)_C$ generator, $P_{R(L)} \equiv (\mathbbm{1} \pm \gamma_5)/2$ are chiral
projectors, $F$ and $G^A$ are the photon and gluon field strengths,
and $e$ and $g_s$ are the electromagnetic and strong coupling constants,
respectively. All four operators are normalised to the running $b$-quark mass in the $\overline{\text{MS}}$ renormalisation
scheme, which is also used for all Wilson coefficients from here on.\\
The four-quark operators that arise in the SM are defined in the Chetyrkin/Misiak/M\"unz basis~\cite{Chetyrkin:1997gb}
\begin{equation}
    \label{eq:wet-btoD-4quark}
    \begin{aligned}
        \op{1U} = & [\bar{U} \gamma^\mu t^A P_L b] \, [\bar{s} \gamma_\mu t^A P_L U]\,, &
        \op{2U} = & [\bar{U} \gamma^\mu     P_L b] \, [\bar{s} \gamma_\mu     P_L U]\,, \\
        \op{3}  = & [\bar{s} \gamma^\mu     P_L b] \, \sum_q [\bar{q} \gamma_\mu     q]\,, &
        \op{4}  = & [\bar{s} \gamma^\mu t^A P_L b] \, \sum_q [\bar{q} \gamma_\mu t^A q]\,, \\
        \op{5}  = & [\bar{s} \gamma^\mu \gamma^\nu \gamma^\rho     P_L b] \, \sum_q [\bar{q} \gamma_\mu \gamma_\nu \gamma_\rho     q]\,, &
        \op{6}  = & [\bar{s} \gamma^\mu \gamma^\nu \gamma^\rho t^A P_L b] \, \sum_q [\bar{q} \gamma_\mu \gamma_\nu \gamma_\rho t^A q]\,.
    \end{aligned}
\end{equation}
where $U=u,c$, and the sums iterate over the quark flavours $q=u,d,s,c,b$.
The semileptonic neutral-current operators are commonly defined as:
\begin{equation}
    \label{eq:wet-btoD-sl}
    \begin{aligned}
        \op{9\ell}   = & \frac{e^2}{16\pi^2} [\bar{s} \gamma^\mu           P_L b] \, [\bar{\ell} \gamma_\mu          \ell]\,, &
        \op{10\ell}  = & \frac{e^2}{16\pi^2} [\bar{s} \gamma^\mu           P_L b] \, [\bar{\ell} \gamma_\mu \gamma_5 \ell]\,, \\
        \op{S\ell}   = & \frac{e^2}{16\pi^2} [\bar{s}                      P_R b] \, [\bar{\ell}                     \ell]\,, &
        \op{P\ell}   = & \frac{e^2}{16\pi^2} [\bar{s}                      P_R b] \, [\bar{\ell}            \gamma_5 \ell]\,, \\
        \op{T\ell}   = & \frac{e^2}{16\pi^2} [\bar{s} \sigma^{\mu\nu}          b] \, [\bar{\ell} \sigma_{\mu\nu}     \ell]\,, & 
        \op{T5\ell}  = & \frac{e^2}{16\pi^2} [\bar{s} \sigma^{\mu\nu} \gamma_5 b] \, [\bar{\ell} \sigma_{\mu\nu}     \ell]\,.
    \end{aligned}
\end{equation}
The four chirality flipped operators $\op{9'}$ through $\op{P'}$ are obtained from $\op{9}$ through $\op{P}$
by replacing $P_{R(L)}$ with $P_{L(R)}$.
The Wilson coefficients of operators \op{i} with $i=9^\prime,10^\prime,S^{(\prime)},P^{(\prime)},T,T5$ vanish
in the SM. The approximate values of the Wilson coefficients of the remaining lepton-flavour universal SM-like operators
are shown in Tab.~\ref{tab:wet-btoD-WCs}, at the renormalisation scale $\mu_b$. Although these coefficients
have values of order $1$, we remind the reader that the definition of the corresponding operators already accounts
for suppressive effect, such as CKM factors $\lambda_U^D$, loop-level suppression, and
normalisation to the Fermi constant $G_F$. All these render flavour-changing neutral $b\to D$ transitions
very rare processes in the SM. The WET can be extended to include semileptonic operators
$b\to D \ell_1^+\ell_2^-$, that is, lepton-flavour violating operators with $\ell_1 \neq \ell_2$. The Wilson
coefficients of these operators are power-suppressed by $m_\nu^2/M_W^2$ terms in the SM.\\

The RGE evolution of the remaining SM operators involves
large numerical effects due to the four-quark operators, and is known to next-to-next-to-leading
logarithmic (NNLL) accuracy~\cite{Adel:1992hh,Buchalla:1995vs,Gambino:2003zm,Bobeth:2003at,Gorbahn:2004my}.
However, the RGE of the basis at dimension six, that is including the BSM-only operator set,
is presently only known to leading logarithmic (LL) accuracy~\cite{Aebischer:2017gaw,Jenkins:2017jig,Jenkins:2017dyc,Dekens:2019ept}.\\
Beyond local radiative and semileptonic operators, Eq.~\eqref{eq:wet-btoD-Leff} also contains a set of hadronic
operators. While these do not \textit{directly} contribute to the decay processes, they nevertheless
play an important role. Non-local operators, defined via the time-ordered products of the local WET operators
with the electromagnetic current,
\begin{equation}
    \label{eq:wet-btoD-nonlocal}
    T_i(x, 0) \equiv T\left\lbrace j_\text{e.m.}^\mu(x), \op{i}(0) \right\rbrace
\end{equation}
enter the decay amplitudes at the same level of the coupling strength $\alpha_e$ as the
radiative and semileptonic operators. They require process-specific treatment, and are
further discussed in dedicated sections on the theory of the processes at the centre 
of this review.\\

\paragraph{Charged-current decays} For the description of charged-current induced
$b\to c \ell\nu$ transitions we use the effective Lagrangian
\begin{equation}
    {\cal L}^{\rm eff}_{b\to c\ell\bar\nu} =
        \frac{4 G_F}{\sqrt{2}} V_{cb} \sum_i C_i(\mu) \op{i} + \text{h.c.}\,.
\end{equation}
A commonly used basis of dimension-six operators with left-handed neutrinos reads
\begin{equation}
    \label{eq:wet-btoc-sl}
    \begin{aligned}
        \op{VL,\ell} = & [\bar{c} \gamma^\mu  P_L b]\, [\bar{\ell} \gamma_\mu P_L \nu]\,, &
        \op{VR,\ell} = & [\bar{c} \gamma^\mu  P_L b]\, [\bar{\ell} \gamma_\mu P_L \nu]\,, \\
        \op{SL,\ell} = & [\bar{c}             P_L b]\, [\bar{\ell} P_L \nu]\,, &
        \op{SR,\ell} = & [\bar{c}             P_R b]\, [\bar{\ell} P_L \nu]\,, \\
        \op{T,\ell}  = & [\bar{c} \sigma^{\mu\nu} b]\, [\bar{\ell} \sigma_{\mu\nu} P_L \nu]\,.
    \end{aligned}
\end{equation}
In the SM, all Wilson coefficients vanish with the exception of
$i=VL$, with $C_{VL} = 1 + {\cal O}(\alpha_e)$ universally for all leptons.
Universal electromagnetic corrections to the matching of $C_{VL}$ --- the so-called Sirlin factor~\cite{Sirlin:1980nh}
--- have been computed.
This basis of dimension-six operators does not suffer from numerically large anomalous mass dimensions
due to the absence of strong interaction effects. 
Consequently, the RGE evolution to LL accuracy suffices for
their description\cite{Aebischer:2017gaw,Jenkins:2017jig,Jenkins:2017dyc,Dekens:2019ept}.
Contrary to the $b\to D$ transitions, non-local operators do not play a role in $b\to c\ell\nu$ transitions
at leading-order in the electromagnetic interaction.
\section{Rare radiative $B$ decays}
\label{sec:btosgamma}
\subsection{Theory}
\label{sec:btosgammatheory}
The theoretical description of exclusive rare radiative $b\to D\gamma$ decays involves
computation of the decay amplitudes $A(\bar{B}\to V \gamma_{L(R)})$ within
the framework of the WET at mass dimension six.
Here $\bar{B}$ is a heavy-light meson containing a single valence $b$ quark;
$V$ is a vector meson with appropriate valence quark content, and $\gamma_L$
and $\gamma_R$ are left- and right-handed on-shell photon states.
The WET operators $\op{7}$ and $\op{7'}$ produce a left-handed and a right-handed
photon, respectively. Their hadronic matrix elements can be parametrized in
terms of the two form factors $T_1$ and $T_2$; see Eq.~\eqref{eq:btosll:V-FFs}.
An algebraic relation connects both form factors for on-shell photons: $T_1(q^2 = 0) = T_2(q^2 = 0)$~\cite{Ball:2004rg}.
In the absence of further hadronic contributions, the form factors would therefore
cancel out in ratios of the amplitudes. Hence, the ratio of the two photon-polarisations
\begin{equation}
    \frac{A(\bar{B}\to V \gamma_{R})}{A(\bar{B}\to V \gamma_{L})} \sim \frac{C_{7'}}{C_7} \equiv r_\gamma
\end{equation}
probes the ratio $C_{7'}/C_{7}$. In the SM, $C_{7'}$ is chirally supressed,
$C_{7'} = m_s / m_b C_7$, and hence $r_\gamma \ll 1$.
Observables that can probe the photon polarisation are therefore an excellent
quasi-null test of the SM.\\

The hadronic form factors for $B\to K^*$, $B_s\to \phi$ and $B\to \rho$ transitions are
directly available at $q^2 = 0$ from light-cone sum rules (LCSRs)~\cite{Straub:2015ica,Gubernari:2018wyi,Gubernari:2020eft}.
The uncertainties for all transitions are $\sim10\%$ in LCSR calculation with light-meson DAs~\cite{Straub:2015ica},
and $\sim 30\%$ in LCSR calculations with $B$-meson DAs~\cite{Gubernari:2018wyi,Gubernari:2020eft}.
The form factors can be computed from first principles through numerical simulations of
QCD on space time lattices (LQCD). At the moment, this approach is restricted to the phase
space region $q^2 \leq 15\,\GeV^2$ for both $B\to K^*$ and $B_s\to \phi$ modes~\cite{Horgan:2013hoa}.
In this region, the form factors have relative uncertainties of $\sim 10\%$, which can be systematically improved
in future analyses. The extrapolation to $q^2 = 0$ incurs additional uncertainties, increasing the
total uncertainty to $\sim 20\%$ for both modes. No predictions for $B\to \rho$ transitions are available at the moment.
Exclusive $B\to V\gamma$ modes are presently predicted within the narrow-width approximation, i.e., the
$V$ state is treated as an asymptotically stable state rather than a hadronic resonance. This incurs additonal
uncertainties. A pilot study of this effect on the LCSR calculations has been carried out in Ref.~\cite{Descotes-Genon:2019bud},
finding a $20\%$ upward shift of the central values of the form factors obtained from LCSRs with $B$-meson DAs.
The hadronic form factors for $B\to A\gamma$ decays with $A$ an axial meson have been computed in LCSRs
in the case of the $K_1$ state~\cite{Aliev:2010ki}. However, the three-body decay $K_1\to K\pi\pi$ and nearby
resonances make it presently impossible to provide a reliable theoretical prediction of the $B\to K_1(\to K\pi\pi)\gamma$
observables.\\

The $b\to D\gamma$ amplitudes receive further contributions from the non-local operators
$T_i$ defined in Eq.~\eqref{eq:wet-btoD-nonlocal}. These non-local operators
arise from insertions of four-quark operators if $i=1,\dots,6$ and the chromomagnetic operators if $i=8,8'$.
Hadronic matrix elements of the non-local operators can partially be obtained in the framework
of QCD factorisation (QCDF) for spectator-scattering topologies~\cite{Beneke:2001at,Beneke:2004dp}.
Within the QCDF framework and to leading power in $1/m_b$, transition amplitudes factorise into hard kernel
that are perturbatively calculable, and universal hadronic inputs in form of light-cone distribution
amplitudes (LCDAs). The hard kernels are functions of the momentum distributions of the partons within
the external states, and the kernels can exhibit a singular behaviour in the kinematic endpoints
of the partonic phase space. The factorisation formula holds if in the convolutions of the kernels
with the LCDAs this singular behaviour cancels. Cases in which the singular behaviour is not cancelled are
labelled ``endpoint divergent'', and the corresponding contributions to the amplitude cannot be reliably
predicted within the QCDF framework.\\
The transition amplitudes for the radiative $b\to s$ decays $B^{0(+)}\to K^{*0(+)}\gamma$ and $B_s^0\to \phi\gamma$
factorise to leading power in $1/m_b$ for the all four-quark operators $i=1,\dots,6$, but \textit{not}
for the chromomagnetic operators $i=8,8'$~\cite{Beneke:2001at}. The transition amplitudes for the radiative $b\to d$
decays $B^{0(+)}\to \rho^{0(+)}\gamma$ suffer additional endpoint divergence due to the
effect of the four-quark operators $i=1,2$ at leading power in $1/m_b$, due to insertions
in annihilation topologies that are not possible for the $b\to s$ transition~\cite{Beneke:2004dp}.\\
The endpoint divergent contributions due to the four-quark operators and due to the chromomagnetic
operators have been calculated in Ref.~\cite{Ball:2006eu} and Ref.~\cite{Dimou:2012un}, respectively. Both
calculation have been carried out in the framework of light-cone sum rules with light-meson
distribution amplitudes.\\
Within the QCDF framework, effects of soft-gluons attached to light-quark loops induced by the
four-quark operators cannot be computed. They have been estimated using LCSRs~\cite{Muheim:2008vu,Ball:2006eu,Khodjamirian:2010vf}.
For the $B\to V$ modes, Ref.~\cite{Khodjamirian:2010vf} predicts a sizable effect. However,
it has recently been shown that the predictions of ref.~\cite{Khodjamirian:2010vf} suffer from
an incomplete treatment of the three-particle contributions of the $B$-meson DAs~\cite{Gubernari:2020eft}.
The computation using the complete three-particle DAs indicates that the soft-gluon effects
are negligible for $q^2\leq 0$~\cite{Gubernari:2020eft}; see the more detailed discussion in Sec.~\ref{sec:btosll:theory}.\\

\paragraph{Isospin asymmetries}
The contributions from weak-annihilation topologies in particular drive~\cite{Ball:2006eu}
the SM predictions of the isospin asymmetries~\cite{Kagan:2001zk} in these decays. The isospin
asymmetries are defined as
\begin{align}
  {\cal A}_{\rm I} = & \frac{c_V^2\,\Gamma(\decay{B^{0}}{V^{(*)0}\gamma})-\Gamma(\decay{B^{+}}{V^{(*)+}\gamma})}{c_V^2\,\Gamma(\decay{B^{0}}{V^{(*)0}\gamma})+\Gamma(\decay{B^{+}}{V^{(*)+}\gamma})}\label{eq:radiativeisospin}\\
  =& \frac{c_V^2\,{\cal B}(\decay{B^{0}}{K^{(*)0}\gamma})-(\tau_0/\tau_+){\cal B}(\decay{B^{+}}{K^{(*)+}\gamma})}{c_V^2\,{\cal B}(\decay{B^{0}}{K^{(*)0}\gamma})+(\tau_0/\tau_+){\cal B}(\decay{B^{+}}{K^{(*)+}\gamma})},\nonumber
\end{align}
Here $c_V$ accounts for the valence quark content in the neutral vector mesons, with $c_V = 1$ for $V=K^*$ and $c_V = \sqrt{2}$ for
$V=\rho$. The most recent predictions for the isospin asymmetries read~\cite{Lyon:2013gba}
\begin{align}
  {\cal A}_{\rm I}^{K^*\gamma}  = & (4.9 \pm 2.6)\%\,, &
  {\cal A}_{\rm I}^{\rho\gamma} = & (5.2 \pm 2.8)\%\,.
\end{align}

\paragraph{$\CP$ asymmetries}
The amplitudes in exclusive rare $b\to D \gamma$ decays produce direct \CP\ asymmetries, since the
hadronic contributions can exhibit both~\cite{Grossman:2017thq} a strong-phase difference
(due to differences in intermediate off-shell $\bar{u}u$ and $\bar{c}c$ quark states)
and a weak-phase difference (due to differences in the phases of the CKM factors $\lambda_u^D$ and $\lambda_t^D$)
in the four-quark operator contributions. The SM predictions of the direct \CP\ asymmetries
\begin{equation}
    {\cal A}_{\CP}^{V\gamma} =  \frac{{\cal B}(\bar{B}\to \bar{V} \gamma) - {\cal B}(B\to V \gamma)}{{\cal B}(\bar{B}\to \bar{V} \gamma) + {\cal B}(B\to V \gamma)}\,,
\end{equation}
are rather small~\cite{Paul:2016urs} at
\begin{align}
    {\cal A}_{\CP}^{K^*\gamma}  = & 0.003 \pm 0.001\,.
\end{align}

The neutral decay modes $\decay{\parenbar{B}^0}{\Kstarz(\to K_S \pi^0)\gamma}$ and $\parenbar{B}_s^0\to \phi(\to K^+K^-)\gamma$
provide access to a time-dependent \CP\ asymmetries, since the final states can be 
reached either through direct decay of the initial $\parenbar{B}$ meson or through decay after the initial $\parenbar{B}$ meson oscillates~\cite{Atwood:1997zr}. 
These time-dependent \CP\ asymmetries can be cast in the form:
\begin{align}
    {\cal A}_{\CP}^{f}(t) = & \frac{{\cal S}^{f} \sin(\Delta M t) - {\cal C}^{f} \cos(\Delta M t)}
    {\cosh(\Delta \Gamma t/2) - {\cal A}_{\Delta \Gamma}^{f} \sinh(\Delta\Gamma t/2)}\,.\label{eq:timedependentcpasym}
\end{align}
In the above, $\Delta M$ and $\Delta \Gamma$ are the mass difference and decay width difference of the
respective $B_q$-$\bar{B}_q$ meson system, and ${\cal S}^f$, ${\cal C}^f$ and ${\cal A}_{\Delta \Gamma}^f$ are
final-state $f$ specific coefficients. They can be expressed as
\begin{align*}
    {\cal S}^{f}                =& \frac{2\Im(\lambda_f)}{1+|\lambda_f|^2} &
    {\cal C}^{f}                =& \frac{1-|\lambda_f|^2}{1+|\lambda_f|^2} &
    {\cal A}_{\Delta\Gamma}^{f} =& \frac{2\Re(\lambda_f)}{1+|\lambda_f|^2}
\end{align*}
for a final \CP\ eigenstate $f$ and where we abbreviate
\begin{equation}
    \lambda_f = \frac{q}{p}\frac{\bar{A}_f}{A_f}\,.\\
\end{equation}
Here $\parenbar{A}_f$ denotes the decay amplitude or its \CP\ conjugate for the common final state $f$,
and $q$ and $p$ are $B$-meson mixing parameters. In the limit $|q/p|\to 1$ the coefficient
${\cal C}^{V\gamma}$ can be identified with the direct \CP\ asymmetry $a_{\rm CP}^{V\gamma}$.
Out of the three coefficients, only two are independent quantities and therefore physical observables,
due to the algebraic identity
\begin{equation}
    1 = |{\cal S}^f|^2 + |{\cal C}^f|^2 + |{\cal A}_{\Delta\Gamma}^f|^2\,.
\end{equation}
The two coefficients ${\cal S}^f$ and ${\cal A}_{\Delta\Gamma}^f$ of the time-dependent \CP\ asymmetry are sensitive to the photon polarization $r_\gamma$, 
\begin{align}
    {\cal S}^V & \sim \frac{2 r_\gamma}{1 + r_\gamma^2}\,, &
    {\cal A}_{\Delta\Gamma}^V & \sim \frac{2 r_\gamma}{1 + r_\gamma^2}\,,
\end{align}
with two different proportionality factors.

\paragraph{The dielectron mode $B\to Ve^+e^-$}
While not strictly a radiative decay, the decay $B\to \Kstar e^+e^-$ can probe the radiative decay via the decay chain
$B\to K^*\gamma^*(\to e^+e^-)$ at very small dielectron mass, $q^2 \ll 1\,\GeV^2$. In this region of the dielectron
phase space, longitudinally polarized dielectron states are negligible compared to the transverse polarisations.
The dielectron final state therefore provides excellent access to $r_\gamma$ through the rich set of
angular observables accessible in the four-body final state~\cite{Paul:2016urs}.

\subsection{Experimental results}
\label{sec:radiative_isospin}
\subsubsection{Decay rates and asymmetries}
\paragraph{Isospin asymmetries}
Recently, the Belle collaboration published the first evidence for isospin violation between the decays $\decay{\Bu}{\Kstarp\gamma}$ and $\decay{\Bd}{\Kstarz\gamma}$ at $3.1\,\sigma$~\cite{Horiguchi:2017ntw}. The experimental result of
\begin{align}
    {\cal A}_I^{K^*\gamma} =& (6.2\pm 1.5_{\text{stat}}\pm 0.6_{\text{syst}}\pm 1.2_{f_{+-}/f_{00}})\%
\end{align}
is in good agreement with the SM prediction~\cite{Lyon:2013gba} and the measurement by the BaBar collaboration~\cite{Aubert:2009ak}. 
The $B$-factory experiments have also measured the isospin asymmetry for the $b\to d\gamma$ decays $\decay{\Bu}{\rhop\gamma}$ and $\decay{\Bd}{\rhoz\gamma}$. The results~\cite{Aubert:2008al,Taniguchi:2008ty} 
\begin{align}
{\cal A}_I^{\rho\gamma} =& -0.43^{+0.25}_{-0.22}\pm 0.10,\\ 
{\cal A}_I^{\rho\gamma} =& -0.48^{+0.21}_{-0.19}{}^{+0.08}_{-0.09}\nonumber  
\end{align}
are in good agreement with the SM prediction~\cite{Lyon:2013gba}. 

In addition, Belle, BaBar and LHCb provide precise measurements of exclusive $b\to s\gamma$ and $b\to d\gamma$ branching fractions, the most precisely determined modes are given in Tab.~\ref{tab:bfbtosgamma}. 
The measurements are found to be in good agreement with the SM predictions~\cite{Paul:2016urs,Straub:2018kue,Ball:2006eu}, that are affected by significant uncertainties from the hadronic form factors, as discussed in Sec.~\ref{sec:btosgammatheory}. 
\begin{table}
    \centering
{ \renewcommand*{\arraystretch}{1.25}
    \begin{tabular}{lrrr}\hline
    Decay & \multicolumn{1}{c}{${\cal B}$} & \multicolumn{1}{c}{Ref.} &  ${\cal B}$~SM~\cite{Paul:2016urs,Straub:2018kue,Ball:2006eu}\\\hline\hline
    $\decay{\Bd}{\Kstarz\gamma}$     & $(4.47 \pm 0.10 \pm 0.16 )\times 10^{-5}$ & BaBar~\cite{Aubert:2009ak} & \multirow{2}{*}{$(4.18\pm 0.84)\times 10^{-5}$}\\
     $\decay{\Bd}{\Kstarz\gamma}$    & $(3.96\pm 0.07 \pm 0.14)\times 10^{-5}$ & Belle~\cite{Horiguchi:2017ntw} & \\
    $\decay{\Bu}{\Kstarp\gamma}$     & $(4.22 \pm 0.14 \pm 0.16)\times 10^{-5}$ & BaBar~\cite{Aubert:2009ak} & \multirow{2}{*}{$(4.25\pm 0.88)\times 10^{-5}$}\\
    $\decay{\Bu}{\Kstarp\gamma}$     &  $(3.76\pm 0.10\pm 0.12)\times 10^{-5}$ & Belle~\cite{Horiguchi:2017ntw} & \\
\hline
    $\decay{\Bs}{\phi\gamma}$     & $(3.6\pm 0.5\pm 0.3\pm 0.6)\times 10^{-5}$ & Belle~\cite{Dutta:2014sxo} & \multirow{2}{*}{$(4.02\pm 0.52)\times 10^{-5}$}\\
    $\decay{\Bs}{\phi\gamma}$     & $(3.38\pm 0.34 \pm 0.20)\times 10^{-5} $ & LHCb~\cite{Aaij:2012ita} & \\
\hline
    $\decay{\Bu}{\rhop\gamma}$     & $(1.20^{+0.42}_{-0.37}\pm 0.20)\times 10^{-6}$ & BaBar~\cite{Aubert:2008al} & \multirow{2}{*}{$(1.16\pm 0.26)\times 10^{-6}$}\\
    $\decay{\Bu}{\rhop\gamma}$     & $(0.87^{+0.29}_{-0.27}{}^{+0.09}_{-0.11})\times 10^{-6}$ & Belle~\cite{Taniguchi:2008ty} & \\
    $\decay{\Bd}{\rhoz\gamma}$     & $(0.97^{+0.24}_{-0.22}\pm 0.06)\times 10^{-6}$ & BaBar~\cite{Aubert:2008al} & \multirow{2}{*}{$(0.55\pm 0.13)\times 10^{-6}$}\\
    $\decay{\Bd}{\rhoz\gamma}$     & $(0.78^{+0.17}_{-0.16}{}^{+0.09}_{-0.10})\times 10^{-6}$ & Belle~\cite{Taniguchi:2008ty} & \\ 
    \hline\end{tabular}}
    \caption{Branching fractions of exclusive $b\to s\gamma$ and $b\to d\gamma$ decays by the BaBar~\cite{Aubert:2009ak,Aubert:2008al}, Belle~\cite{Horiguchi:2017ntw,Taniguchi:2008ty,Dutta:2014sxo}, and LHCb~\cite{Aaij:2012ita} collaborations. The measurement of ${\cal B}(\decay{\Bs}{\phi\gamma})$ by the LHCb collaboration~\cite{Aaij:2012ita} uses the world average of ${\cal B}(\decay{\Bd}{\Kstarz\gamma})$. The SM predictions~\cite{Paul:2016urs,Straub:2018kue} are affected by significant hadronic uncertainties.\label{tab:bfbtosgamma}}
\end{table}

\paragraph{Direct \CP\ asymmetries}
As for the isospin asymmetries, the uncertainties from the hadronic form factors cancel in the direct \CP\ asymmetries ${\cal A}_{\CP}$ at leading order. 
The Babar~\cite{Aubert:2009ak}, Belle~\cite{Horiguchi:2017ntw}, and LHCb~\cite{Aaij:2012ita} collaboration find for the direct \CP\ asymmetries
\begin{align}
{\cal A}_{\CP}(\decay{B}{K^*\gamma}) =& (-0.4\pm 1.4\pm 0.3)\%, ~~~\text{\cite{Horiguchi:2017ntw}}\\
{\cal A}_{\CP}(\decay{B}{K^*\gamma}) =& (-0.3\pm 1.7\pm 0.7)\%,~~~\text{\cite{Aubert:2009ak}~and}\nonumber\\
{\cal A}_{\CP}(\decay{\Bd}{\Kstarz\gamma}) =& (+0.8\pm 1.7\pm 0.9)\%,~~~\text{\cite{Aaij:2012ita}}\nonumber
\end{align}
in good agreement with the SM~\cite{Paul:2016urs}. 

\subsubsection{Time-dependent \CP\ asymmetries}
As discussed in Sec.~\ref{sec:btosgammatheory}, the measurement of time-dependent asymmetries in rare $b\to s\gamma$ decays allow to probe photon polarisation. 
In the \Bd\ system, the decay width difference $\Delta\Gamma_d$ is negligible, so Eq.~\ref{eq:timedependentcpasym} simplifies to
\begin{align}
{\cal A}_{\CP}^{V\gamma}(t)    =& \frac{\Gamma(\decay{\Bdb}{V\gamma})-\Gamma(\decay{\Bd}{V\gamma})}{\Gamma(\decay{\Bdb}{V\gamma})+\Gamma(\decay{\Bd}{V\gamma})} = S^{V\gamma}\sin(\Delta M_d t) - C^{V\gamma} \cos(\Delta M_d t)
\end{align}
Experimental determinations of ${\cal S}^{V\gamma}$ and ${\cal C}^{V\gamma}$ require decay-time (or, in the case of the $B$-factories where $B\bar{B}$ meson pairs are produced coherently,  decay-time difference) dependent analyses that identify the \Bd\ production flavour (\textit{flavour tagging}). Flavour tagging relies on flavour specific decay signatures of the other $B$ (\eg\ the lepton charge in semileptonic $\decay{b}{c\ell^-\bar{\nu}_\ell}$ decays). 
The flavour tagging performance is given by the effective tagging power $\epsilon_{\rm eff}=\epsilon_{\rm tag}(1-2\omega_{\rm tag})^2$, 
which corresponds to the effective reduction of the sample size due to the efficiency to provide a flavour tag ($\epsilon_{\rm tag}$) and the mistag probability ($\omega_{\rm tag}$). 
Typical effective tagging powers at BaBar and Belle are around $30\%$~\cite{Bevan:2014iga}.

\begin{figure}
  \centering
\includegraphics[width=0.49\textwidth]{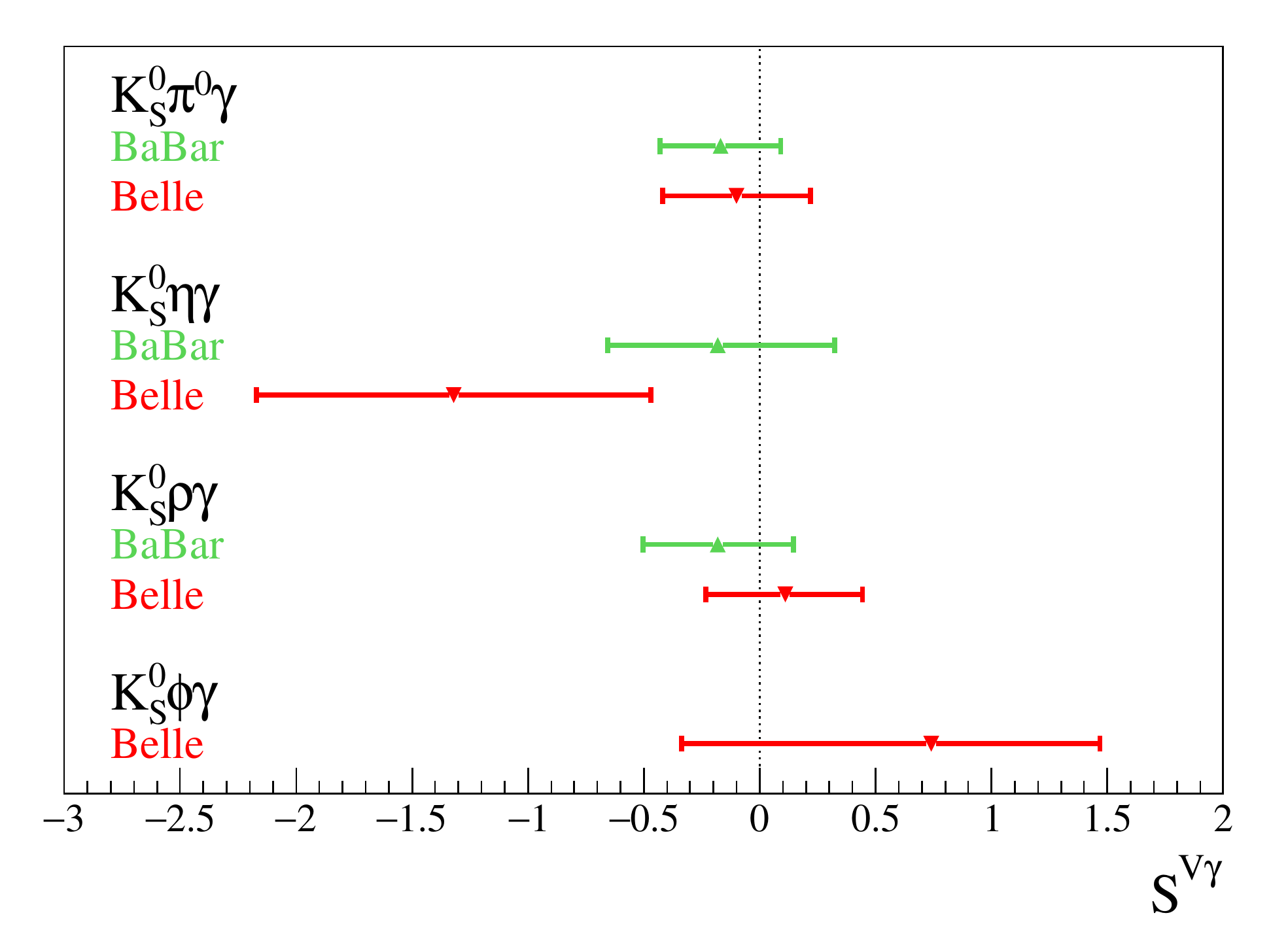}
\includegraphics[width=0.49\textwidth]{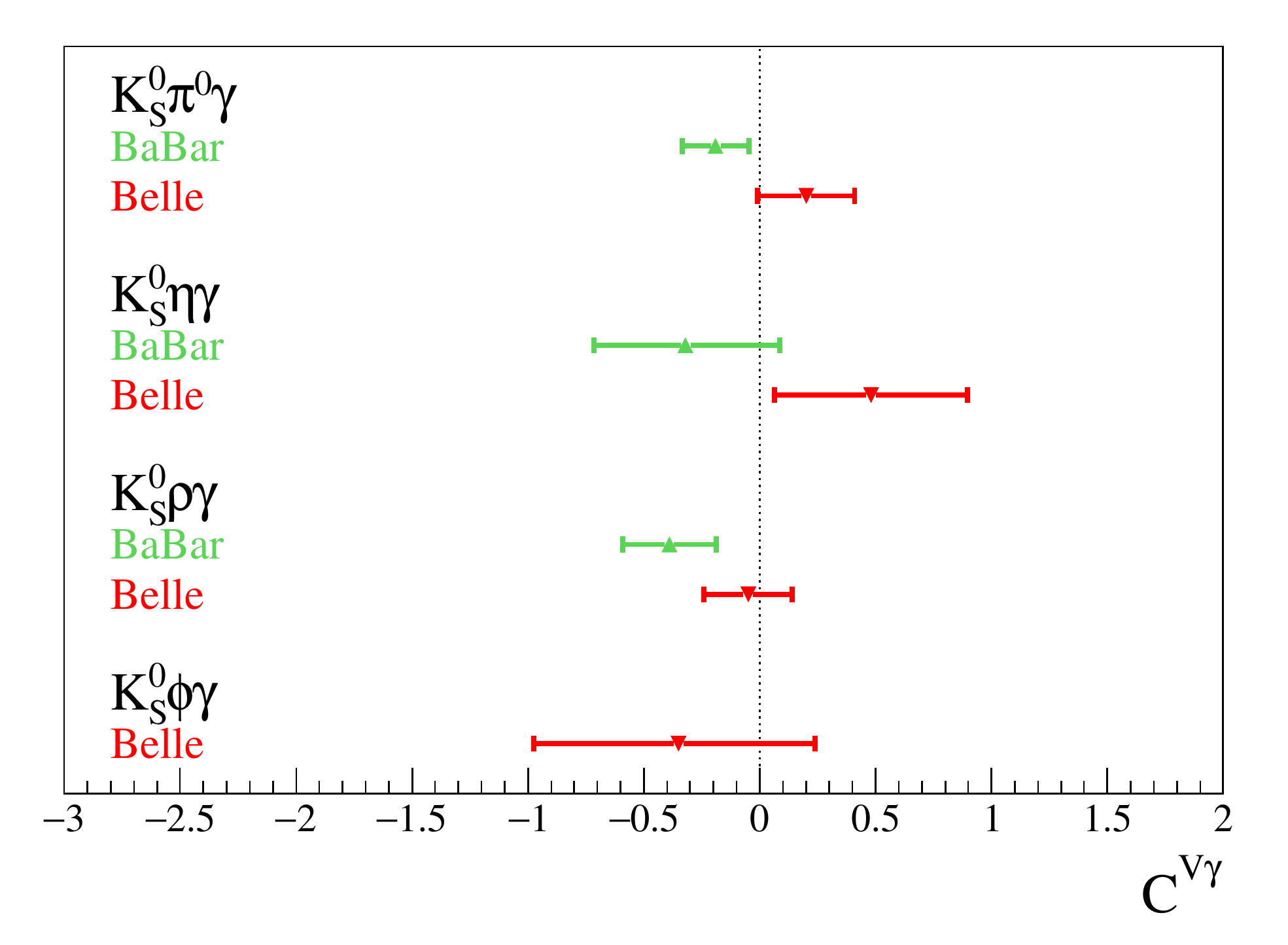}
  \caption{
  Measurements of the time-dependent \CP-asymmetries (left) ${\cal S}^{V\gamma}$ and (right) ${\cal C}^{V\gamma}$ by the BaBar and Belle collaborations in the 
  $\KS\piz\gamma$~\cite{Aubert:2008gy,Ushiroda:2006fi}, $\KS\eta\gamma$~\cite{Aubert:2008js,Nakano:2018lqo},  $\KS\rhoz\gamma$~\cite{Sanchez:2015pxu,Li:2008qma}, and
$\KS\phi\gamma$~\cite{Sahoo:2011zd} final states.\label{fig:radiativetimedependent}}
\end{figure}

The BaBar and Belle collaborations have determined ${\cal S}^{V\gamma}$ and ${\cal C}^{V\gamma}$ in the final states $\KS\piz\gamma$ (including the $\Kstarz\gamma$)~\cite{Aubert:2008gy,Ushiroda:2006fi}, $\KS\eta\gamma$~\cite{Aubert:2008js,Nakano:2018lqo}, and $\KS\rhoz(\to\pip\pim)\gamma$~\cite{Sanchez:2015pxu,Li:2008qma}. 
The Belle collaboration has also performed a measurement using the $\KS\phi\gamma$ final state~\cite{Sahoo:2011zd}. 
The results are given in Fig.~\ref{fig:radiativetimedependent}, they are in good agreement with the SM prediction of predominantly left-handed photon polarisation and thus small ${\cal S}^{V\gamma}$. 
No direct \CP\ violation is observed. 
So far, no measurements of these decays were performed at the LHC, due to the more challenging reconstruction of final states containing \KS\ (and \piz) mesons, and the less efficient 
flavour tagging. 
The Belle collaboration furthermore performed a time-dependent analysis of the $b\to d\gamma$ decay $\decay{\Bd}{\rhoz\gamma}$~\cite{Ushiroda:2007jf}. 
The results for this highly suppressed mode exhibit significant uncertainties and are in agreement with the SM expectation~\cite{Ball:2006eu}. 

In the \Bs\ system, the width difference $\Delta\Gamma_s$ is significant and therefore sensitivity to ${\cal A}_{\Delta\Gamma}$ is retained.  
It should be noted that ${\cal A}_{\Delta\Gamma}$ can be accessed without the need to perform identification of the \Bs\ production flavour. The LHCb collaboration performed a first untagged measurement using the decay $\decay{\Bs}{\phi\gamma}$, resulting in a ${\cal A}_{\Delta\Gamma}^{\phi\gamma}$ value consistent with the SM within $2\,\sigma$~\cite{Aaij:2016ofv}. 
More recently, LHCb performed a first tagged determination of the observables ${\cal S}^{\phi\gamma}$, ${\cal C}^{\phi\gamma}$, and ${\cal A}^{\phi\gamma}_{\Delta\Gamma}$ with $\decay{\Bs}{\phi\gamma}$ decays~\cite{Aaij:2019pnd}. 
The effective tagging power 
in this measurement is found to be $\epsilon_{\rm eff}=\epsilon_{\rm tag}(1-2\omega_{\rm tag})^2=(4.99\pm 0.14)\%$. 
The observables 
are determined to be 
\begin{align}
    {\cal S}^{\phi\gamma} =& 0.43\pm 0.30 \pm 0.11,\\
    {\cal C}^{\phi\gamma} =& 0.11\pm 0.29 \pm 0.11, ~~~\text{and}\nonumber\\
    {\cal A}^{\phi\gamma}_{\Delta\Gamma} =& -0.67^{+0.37}_{-0.41} \pm 0.17,\nonumber
\end{align}
consistent with the SM prediction~\cite{Muheim:2008vu} within $1.3$, $0.3$ and $1.7\,\sigma$, respectively. 

\subsubsection{Photon polarisation from $\decay{\Bu}{K_1^+(\to \Kp\pip\pim)\gamma}$}
\begin{figure}
  \centering
\raisebox{-0.5\height}{\includegraphics[width=0.49\textwidth]{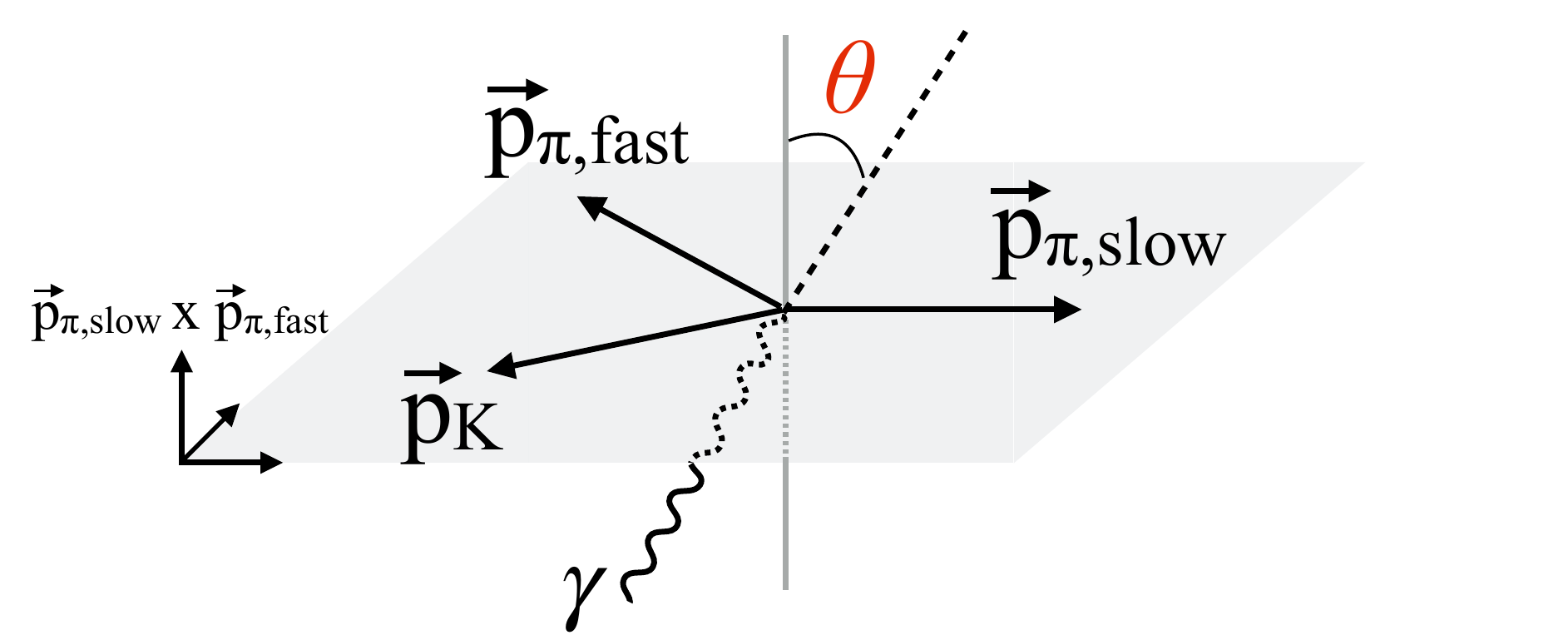}}
\raisebox{-0.5\height}{\includegraphics[width=0.49\textwidth]{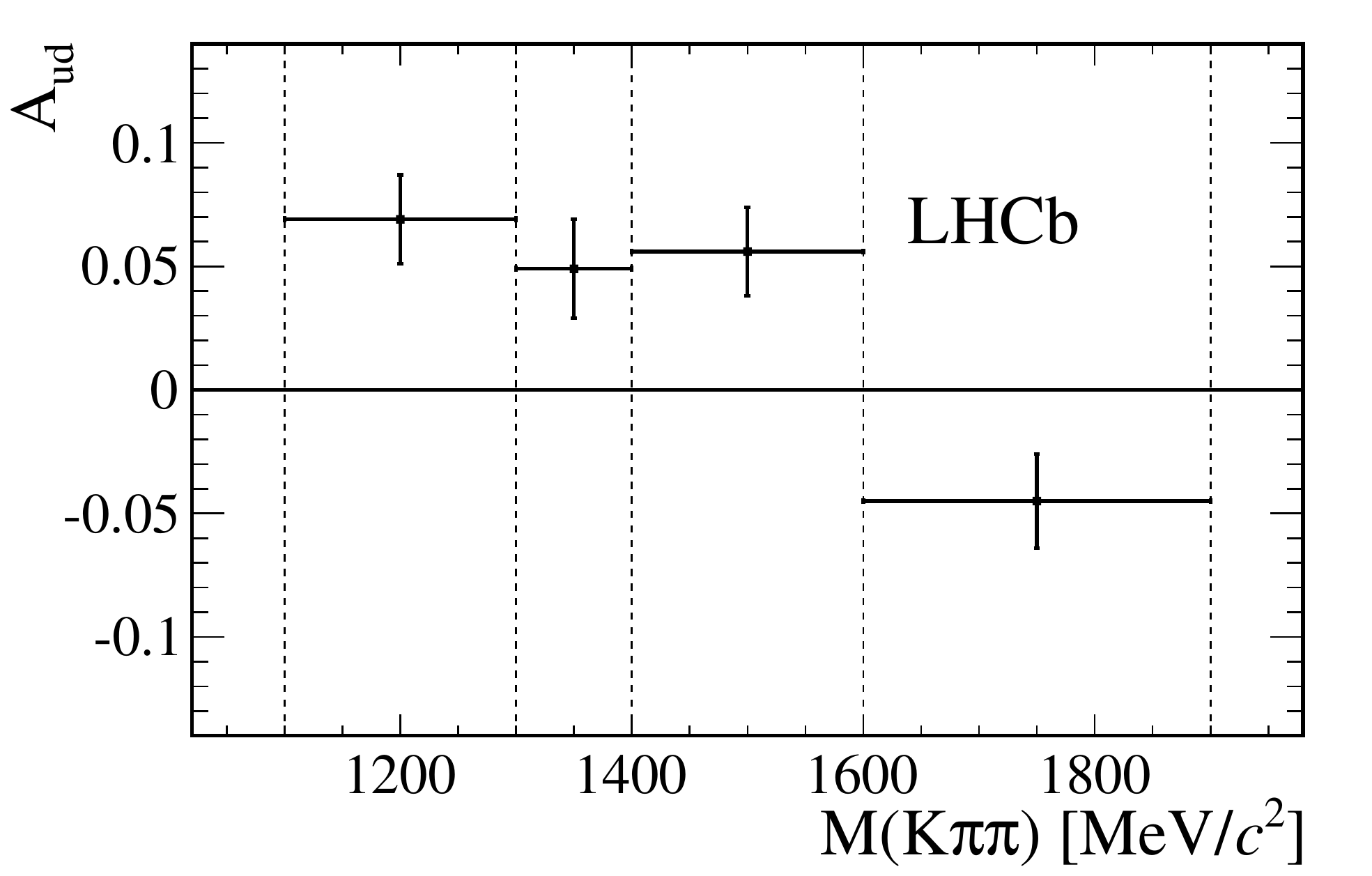}}
  \caption{
  (Left) The angle $\theta$ is defined as the angle between the direction opposite the photon $\vec{p}_\gamma$ and the normal $\vec{p}_{\pi,\rm slow}\times \vec{p}_{\pi,\rm fast}$ to the $\Kp\pip\pim$ plane in the $\Kp\pip\pim$ center-of-mass frame. (Right) The up-down asymmetry ${\cal A}_{\rm ud}$ in four bins of $m_{K\pi\pi}$. 
  Figures reproduced from Ref.~\cite{Aaij:2014wgo}.\label{fig:updownasymmetry}}
\end{figure}

The radiative decay $\decay{\Bu}{\Kp\pip\pim\gamma}$ allows access to information on the photon polarisation through the angle $\theta$, 
defined as the angle of the photon with respect to the plane spanned by the charged hadrons in the $\Kp\pip\pim$ center-of-mass frame (see Fig.~\ref{fig:updownasymmetry}). 
The up-down asymmetry ${\cal A}_{\rm ud}$, \ie\ the asymmetry between photons emitted in the upwards and downwards direction, is proportional to the photon polarisation asymmetry
\begin{align}
    {\cal A}_{\rm ud} = \frac{\int_{0}^{+1}\cos\theta\frac{{\rm d}\Gamma}{{\rm dcos}\theta}-\int_{-1}^{0}\cos\theta\frac{{\rm d}\Gamma}{{\rm dcos}\theta}}{\int_{-1}^{+1}\cos\theta\frac{{\rm d}\Gamma}{{\rm dcos}\theta}} \propto \frac{|{\cal C}_7|^2-|{\cal C}_7^\prime|^2}{|{\cal C}_7|^2+|{\cal C}_7^\prime|^2} \propto \frac{1 - |r_\gamma|^2}{1 +|r_\gamma|^2}
\end{align}
where however the proportionality factor depends on the resonances contributing to the $\Kp\pip\pim$ system~\cite{Gronau:2002rz}. 
Several different strange resonances, \eg\ $K_1^+(1270)$, $K_1^+(1400)$, $K^{*+}(1410)$ etc., contribute to the final state, a study of the hadronic system can be found in Ref.~\cite{Sanchez:2015pxu}. 
The LHCb collaboration performed a measurement of ${\cal A}_{\rm ud}$ in four bins of $m_{\Kp\pip\pim}$~\cite{Aaij:2014wgo}. 
The results given in Fig.~\ref{fig:updownasymmetry} 
excludes the null-hypothesis of zero photon polarisation at a significance of $5.2\,\sigma$.
The use of an amplitude analysis technique to separate contributions from the different strange resonances is being investigated~\cite{Bellee:2019qbt,Bediaga:2018lhg}. 
The absence of theoretical predictions for hadronic matrix elements with a $K\pi\pi$ system
prohibit presently the use of ${\cal A}_{\rm ud}$ in phenomenological studies.

\subsubsection{Baryonic $\decay{b}{s\gamma}$ decays}
Baryonic $\decay{b}{s\gamma}$ transitions offer unique opportunities to study photon polarisation due to the non-zero spins of the initial and final state hadrons~\cite{Mannel:1997xy,Hiller:2001zj}. 
The angular differential decay distributions for the radiative baryon decay $\decay{\Lb}{\Lambda\gamma}$ are given by~\cite{Legger:2006cq}
\begin{align}
\frac{{\rm d}\Gamma}{{\rm dcos}\theta_\gamma} \propto& 1 - \frac{1-|r_\gamma|^2}{1+|r_\gamma|^2}P_{\Lb}\cos\theta_\gamma, ~~~\text{and}\\
\frac{{\rm d}\Gamma}{{\rm dcos}\theta_p} \propto& 1-\frac{1-|r_\gamma|^2}{1+|r_\gamma|^2}\alpha_{p,\frac{1}{2}}\cos\theta_p,\nonumber
\end{align}
where $\theta_\gamma$ denotes the angle between the photon direction and the $\Lb$ spin, 
$\theta_p$ the angle between the direction of the proton and the $\Lambda$ (in the $\Lambda$ rest frame), 
$\alpha_{p,\frac{1}{2}}=0.732\pm 0.014$~\cite{Ablikim:2018zay,Ireland:2019uja,pdg2020}
the $\Lambda$ asymmetry\footnote{This value was recently updated from the previous world average of $\alpha_{p,\frac{1}{2}}=0.642\pm 0.013$.}, 
and $P_{\Lb}$ the $\Lb$ production polarisation. 

Recently, the LHCb collaboration performed the first observation of a baryonic $b$-hadron decay with the observation of the decay $\decay{\Lb}{\Lambda\gamma}$~\cite{Aaij:2019hhx}.  
The branching fraction of the decay is measured to be~\cite{Aaij:2019hhx}
\begin{align}
    {\cal B}(\decay{\Lb}{\Lambda\gamma}) =& (7.1\pm 1.5_{\rm syst}\pm 0.6_{\rm stat}\pm 0.7_{\rm ext.})\times 10^{-6},
\end{align}
in agreement with SM predictions~\cite{Wang:2008sm,Mannel:2011xg,Gutsche:2013pp}. 
An angular analysis of the decay should be possible with the full LHCb Run~2 data sample. 

\subsubsection{$\decay{\Bd}{\Kstarz\ee}$ at low $q^2$}
At low invariant di-lepton masses $m_{\ell\ell}^2=q^2\ll 1\gevgevcccc$ the $\decay{b}{s\ellell}$ transition $\decay{\Bd}{\Kstarz\ellell}$ is dominated by virtual photon contributions (\ie\ ${\cal C}_7$ in the SM). 
The angular distributions of the decay are thus sensitive to the photon polarisation and allow to probe potential contributions from right-handed currents~\cite{Paul:2016urs,Becirevic:2011bp}. 
The LHCb experiment has performed an angular analysis of the decay $\decay{\Bd}{\Kstarz\ee}$ 
in the $q^2$ region\footnote{Note that the lower boundary can only be reached with the di-electron mode due to the small electron mass.} $0.0008<q^2<0.257\gevgevcccc$.
The angular analysis uses a folding technique to simplify the angular distribution discussed in detail in Sec.~\ref{sec:angularanalyses},
specifically the decay angle $\phi$ is transformed such that $\tilde{\phi}=\phi+\pi$ for $\phi < 0$ and $\tilde{\phi}=\phi$ otherwise. 
Neglecting contributions from the S-wave and lepton mass effects (which is a good approximation for electrons), the angular distribution is then given by
\begin{align}
  \frac{1}{{\rm d}(\Gamma+\bar{\Gamma})/{\rm d}q^2}\frac{{\rm d}^4(\Gamma+\bar{\Gamma})}{{\rm d}q^2\,{\rm dcos}\thetal\,{\rm d}\tilde{\phi}\,{\rm dcos}\thetak} = \frac{9}{16\pi} \biggl[ & 
    \tfrac{3}{4}(1-F_{\rm L}) \sin^2\thetak + F_{\rm L} \cos^2\thetak\\
    + & \tfrac{1}{4}(1-F_{\rm L}) \sin^2\thetak\cos 2\thetal - F_{\rm L} \cos^2\thetak\cos 2\thetal \nonumber\\
    + & (1-F_{\rm L})A_{\rm T}^{\rm Re} \sin^2\thetak\cos\thetal\nonumber\\
    + & \tfrac{1}{2}(1-F_{\rm L})A_{\rm T}^{(2)} \sin^2\thetak\sin^2\thetal\cos 2\tilde{\phi}\nonumber\\
    + & \tfrac{1}{2}(1-F_{\rm L})A_{\rm T}^{\rm Im} \sin^2\thetak\sin^2\thetal\sin 2\tilde{\phi}\biggr].\nonumber
\end{align}
Here, the angular observables $A_{\rm T}^{(2)}$, $A_{\rm T}^{\rm Re}$, and $A_{\rm T}^{\rm Im}$ are related to the angular observables discussed in Sec.~\ref{sec:angularanalyses}
via $A_{\rm T}^{(2)}=P_1$, $A_{\rm T}^{\rm Re}=2P_2$, and $A_{\rm T}^{\rm Im}=-2P_3^{\rm CP}$. 
The results, given in Tab.~\ref{tab:kstareelowq2} are in excellent agreement with SM predictions~\cite{Straub:2018kue,Straub:2015ica}
and provide the world's most precise constraints on photon polarisation. 

\begin{table}
  \centering
{ \renewcommand*{\arraystretch}{1.25}
  \begin{tabular}{lrr}\hline
    Observable & Measurement~\cite{Aaij:2020umj} & SM prediction~\cite{Straub:2018kue,Straub:2015ica} \\\hline\hline
   $F_{\rm L}$  & $0.044\pm 0.026\pm 0.014$ & $0.051\pm 0.013$\\
   $A_{\rm T}^{(2)}$  & $+0.11 \pm 0.10 \pm 0.02$ & $-0.0001\pm0.0004$\\
   $A_{\rm T}^{\rm Re}$  & $-0.06\pm 0.08 \pm 0.02$ & $+0.033\pm 0.020$\\
   $A_{\rm T}^{\rm Im}$  & $+0.02 \pm 0.10 \pm 0.01$ & $-0.00012\pm 0.00034$\\
  \hline\end{tabular}}
  \caption{Angular observables from an angular analysis of the decay $\decay{\Bd}{\Kstarz\ee}$ by the LHCb collaboration~\cite{Aaij:2020umj}, together with the SM predictions from Refs.~\cite{Straub:2018kue,Straub:2015ica}.\label{tab:kstareelowq2}}
\end{table}

\subsubsection{Experimental prospects}
The Belle~II experiment will perform precise measurements of 
time-dependent \CP\ violation 
in both $b\to s\gamma$ and $b\to d\gamma$ decays. 
With the full Belle~II data set the collaboration expects sensitivities of $\sigma({\cal S}^{\KS\piz\gamma})=0.035$ and $\sigma({\cal S}^{\rho\gamma})=0.07$~\cite{Kou:2018nap}, respectively. 
Furthermore, Belle~II expects to measure
branching fractions for exclusive $b\to d\gamma$ modes to a precision of around $5\%$, and \CP- and isospin-asymmetries for exclusive $b\to s\gamma$ decays to a precision of $0.5\%$ or better with the full data sample~\cite{Kou:2018nap}. 

Angular analyses of baryonic radiative decays are uniquely possible at LHCb. 
With the full Upgrade~II data sample, the LHCb collaboration expects to measure the photon polarisation asymmetry in the decay $\decay{\Lb}{\Lambda\gamma}$ with $4\%$ precision~\cite{Bediaga:2018lhg}. 
For the angular analysis of the rare decay $\decay{\Bd}{\Kstarz\ee}$ at low $q^2$, the LHCb collaboration expects statistical sensitivities for $A_{\rm T}^{(2)}$ and $A_{\rm T}^{\rm Im}$ of $2\%$ with the full Upgrade~II data sample~\cite{Bediaga:2018lhg}. 
Furthermore, LHCb plans to perform a competitive measurement of ${\cal S}^{\KS\pi\pi\gamma}$, projecting around 200\,k signal events with the Upgrade~II data sample~\cite{Bediaga:2018lhg}. 

\section{Rare (semi)leptonic $B$ decays}
\label{sec:btosll}

\subsection{Leptonic $B$ decays}
\label{sec:leptonic}
\subsubsection{Theory}
\label{sec:leptonicTH}

As discussed in Sec.~\ref{sec:effectivetheory}, in the SM $b\to D\ellell$, $D=d,s$ transitions are highly suppressed,
since they only emerge at the one-loop level and involve small CKM factors.
For the purely leptonic decays, helicity suppression is an additional factor~\cite{DeBruyn:2012wk} in the SM, which can be lifted
by BSM effects through the scalar and pseudoscalar operators with $i=S,S',P,P'$ in Eq.~\eqref{eq:wet-btoD-sl}.
The matching calculations for the SM operator have now reached the level of NLO in the electroweak coupling and
NNLO in the strong coupling $\alpha_s$ in the matching calculations~\cite{Bobeth:2013uxa,Bobeth:2013tba,Hermann:2013kca}.\\

Amplitudes for the purely leptonic decays $B_q \to \ellell$ depend on only a single hadronic matrix element to leading
order in the electromagnetic coupling~\cite{DeBruyn:2012wk}. This matrix element is commonly parametrised in terms of the
$B_q$-meson decay constant $f_{B_q}$~\cite{Aoki:2019cca}
\begin{equation}
    \bra{0} \bar{q} \gamma^\mu \gamma_5 b \ket{\bar{B}_q(p)} = i f_{B_q} p^\mu\,.
\end{equation}
The decay constant has been determined in lattice QCD simulations. Several Lattice QCD analyses with
$N_f = 2 + 1 + 1$ light quark flavours are available~\cite{Dowdall:2013tga,Carrasco:2013naa,Bussone:2016iua,Bazavov:2017lyh,Hughes:2017spc}.
Their world averages
\begin{align}
    f_{B_d} = & 190.0 \pm 1.3\,\MeV\,, &
    f_{B_s} = & 230.3 \pm 1.3\,\MeV\,,
\end{align}
are dominated by a single analysis by the Fermilab/MILC collaboration~\cite{Bazavov:2017lyh}.\\

The full set of time-integrated observables in flavour-specific measurements of $\parenbar{B}_s \to \ellell$ decays
has been discussed in Ref.~\cite{DeBruyn:2012wk}. In addition to the time-integrated branching ratio
\begin{equation}
    \braket{\mathcal{B}(\parenbar{B}_s^0\to\ellell)}
        \equiv \int_0^\infty {\rm d}t \mathcal{B}(\parenbar{B}_s^0\to\ellell)(t)\,,
\end{equation}
it includes the effective lifetime. The latter is defined as the first moment of the
decay-time distribution~\cite{DeBruyn:2012wk}
\begin{equation}
    \tau_\text{eff}^{\ellell}
        \equiv \frac{\int_0^\infty {\rm d}t \, t\, \mathcal{B}(\parenbar{B}_s^0\to\ellell)(t)}{\int_0^\infty {\rm d}t\, \mathcal{B}(\parenbar{B}_s^0\to\ellell)(t)}\,.
\end{equation}
Since the $\parenbar{B_q^0}$ are pseudoscalar mesons, the lepton pair with definite helicities is either
$\ell^+_L\ell^-_L$ or $\ell^+_R\ell^-_R$. Neither is a \CP\ eigenstate, and therefore only a time-dependent
rate asymmetry but no time-dependent \CP\ asymmetry can be defined. The rate asymmetry takes the same form
as Eq.~\eqref{eq:timedependentcpasym}, with three coefficients ${\cal C}^{\ellell}$, ${\cal S}^{\ellell}$
and ${\cal A}_{\Delta\Gamma}^{\ellell}$. To leading order in $\alpha_s$ the SM predictions for these
coefficients read~\cite{DeBruyn:2012wk}
\begin{align}
    {\cal S}^{\ellell} = & 0 = {\cal C}^{\ellell}, &
    {\cal A}_{\Delta\Gamma}^{\ellell} = & 1\,.
\end{align}
Power-enhanced electromagnetic effects introduce deviations from these prediction~\cite{Beneke:2019slt}. For the $\parenbar{B}_s^0$
system, the results read~\cite{Beneke:2019slt}
\begin{align}
    {\cal S}^{\mu^+_L\mu_L^-} = -{\cal S}^{\mu^+_R\mu_R^-} = & +0.6\%\,, &
    {\cal C}^{\mu^+_L\mu_L^-} = -{\cal C}^{\mu^+_R\mu_R^-} = & +0.1\%\,, &
    {\cal A}_{\Delta\Gamma}^{\mu^+\mu^-} = & 1 - 2.0\times 10^{-5}\,.
\end{align}

The SM predictions for the time-integrated observables are sensitive to the aforementioned power-enhanced
electromagnetic effects. For a soft-photon energy of $60\MeV$ or smaller the SM predictions of the time-integrated
rates read~\cite{Beneke:2019slt}
\begin{align}
    10^9\times    \mathcal{B}(\parenbar{B}_s^0 \to \mu^+\mu^-) & = 3.660 \pm 0.137\,, &
    10^{10}\times  \mathcal{B}(\parenbar{B}_d^0 \to \mu^+\mu^-) & = 1.027 \pm 0.052\,. &
\end{align}

The SM prediction of the ratio of branching fractions is very precise, since the decay constant, CKM factor and Wilson coefficient
cancel and to leading-order in $\alpha_e$ no new hadronic uncertainties arise. The ratios read
\begin{align}
    r^q_{\ell/\mu}
        & = \frac{\mathcal{B}(\bar{B}_q \to \ell^+\ell^-)}{\mathcal{B}(\bar{B}_q \to \mu^+\mu^-)}
        = \frac{m_\ell^2}{m_\mu^2} \sqrt{\frac{M_{B_q}^2 - m_\ell^2}{M_{B_q}^2 - m_\mu^2}}\,,
\end{align}
and evaluate to
\begin{align}
    r^s_{\tau/\mu}
        & = 212\,, &
    10^7 r^s_{e/\mu}
        & = 234\,, \\
    r^d_{\tau/\mu}
        & = 209\,, &
    10^7 r^d_{e/\mu}
        & = 234\,,
\end{align}
with negligible uncertainties.

\subsubsection{Experimental results}

Amongst the purely leptonic \btosll\ decays, the decays $\decay{\Bs}{\mumu}$ and $\decay{\Bd}{\mumu}$ play a special role, as they are experimentally the most easily accessible. The LHC experiments have taken the lead in the analysis of these decays, profiting from the very large $B$-meson production rate in high energy proton collisions and the excellent muon reconstruction and identification at the experiments. The decade long program to search for these decays and later discover $\decay{\Bs}{\mumu}$ is described below. 

The decay of $B$ mesons to electrons suffers from Bremsstrahlung losses which make the decay much harder to reconstruct, especially at the LHC. 
Decays to $\tau^-$ leptons face the challenge of 
a variety of possible $\tau^-$ decay modes, all including one or more neutrinos. 
Nonetheless, all six leptonic decay modes of $B$ mesons have been searched for at the LHC, the measurements will be discussed in the following.

\paragraph{Observation and current status of $\decay{\Bs}{\mumu}$}

The decays $\decay{\Bs}{\mumu}$ and $\decay{\Bd}{\mumu}$ have been considered prime discovery modes for new scalar of pseudo-scalar particles, the quest to find these decays has been ongoing since the first CLEO and ARGUS searches in the 1980s. The LHCb experiment found the first evidence for the decay $\decay{\Bs}{\mumu}$ in 2012~\cite{PhysRevLett.110.021801}, the first observation was published in a joint analysis from the CMS and LHCb collaborations~\cite{CMS:2014xfa}. The most precise measurements, that are discussed here, are provided by the three LHC experiments ATLAS~\cite{Aaboud:2018mst}, CMS~\cite{Sirunyan:2019xdu} and LHCb~\cite{Aaij:2017vad}, they are combined in an LHC average~\cite{LHCb:2020zud}. All three collaborations use approximately one half of their complete Run 1 and 2 data sets, updates using the full data set are expected soon. 

\begin{figure}
  \centering
\includegraphics[width=0.49\textwidth]{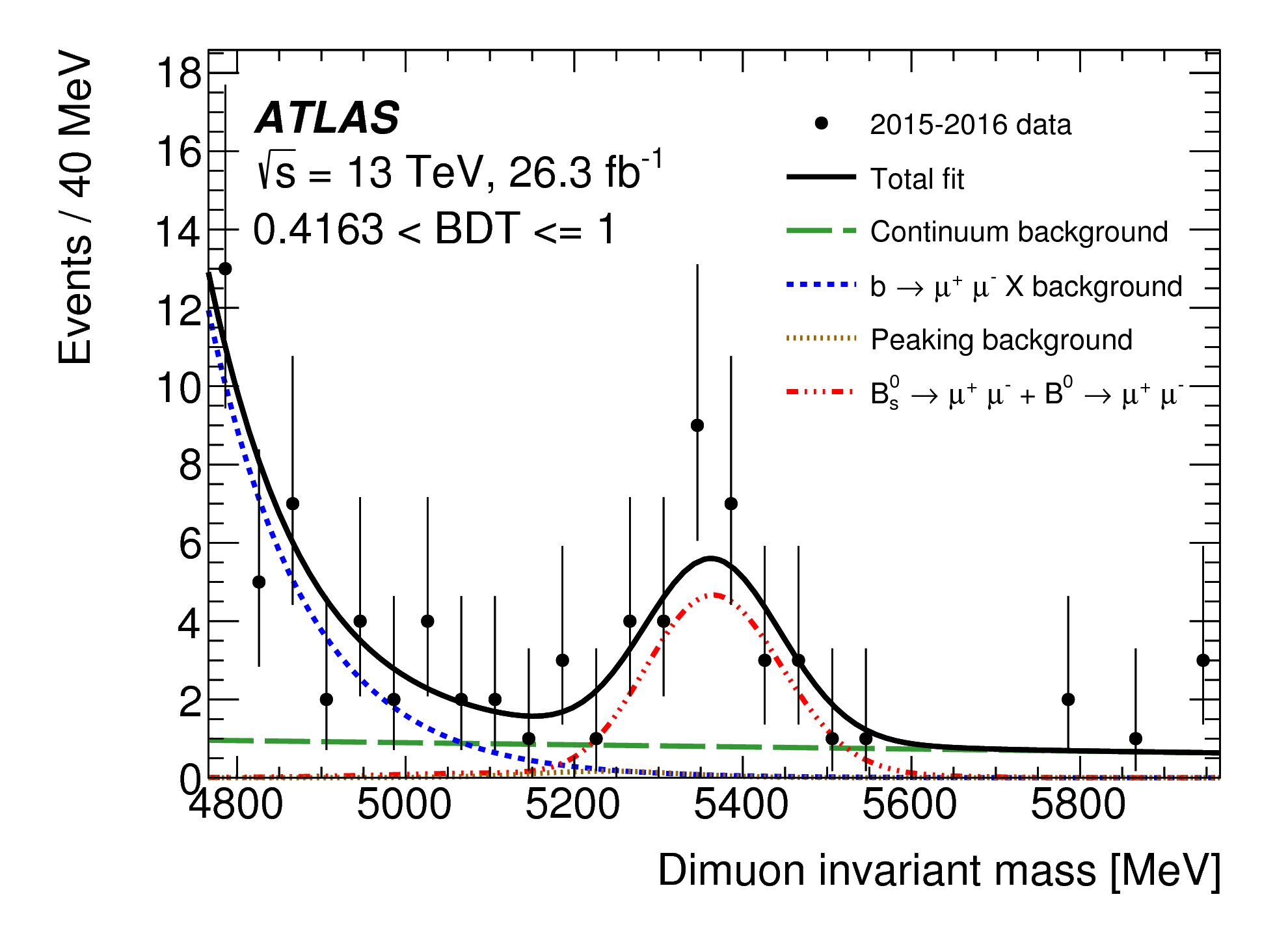}
\includegraphics[width=0.49\textwidth,height=0.35\textwidth]{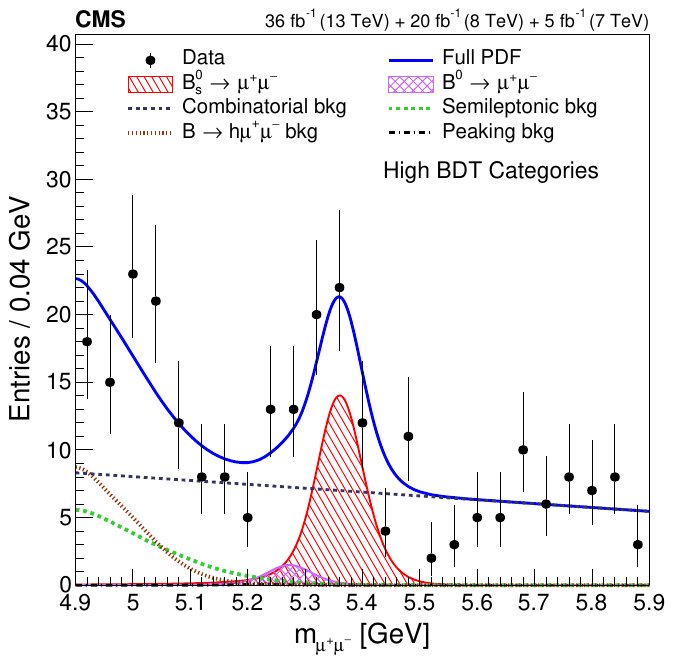}
\includegraphics[width=0.49\textwidth]{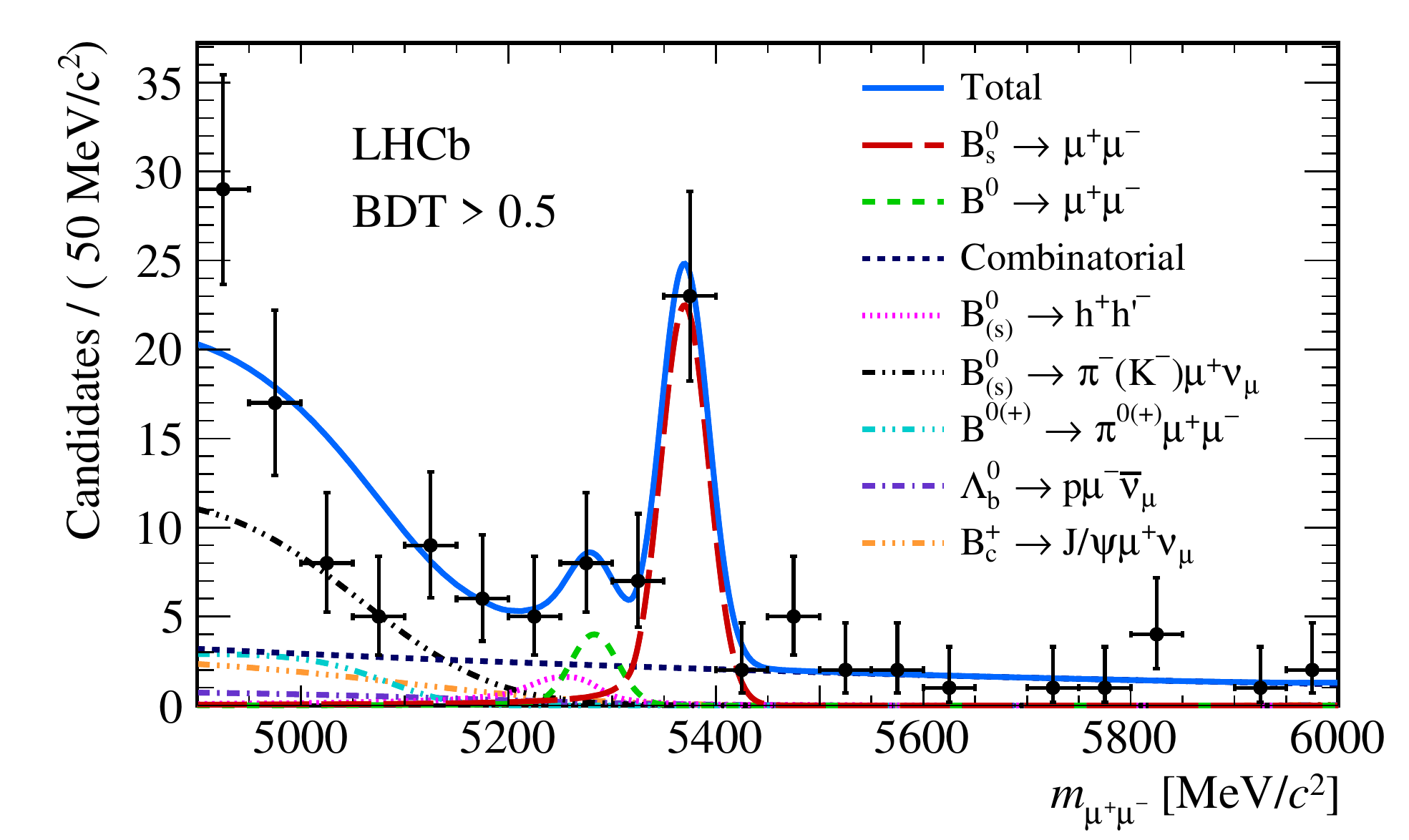}
   \caption{\label{fig:BsmmMass} Reconstructed invariant di-muon mass distributions from the ATLAS. CMS and LHCb experiments in a region of high $\decay{B}{\mumu}$ signal likelihood. Individual physical background components are indicated in the legend of each plot.  
   Figures reproduced from from~\cite{Aaboud:2018mst,Sirunyan:2019xdu,Aaij:2017vad}.}
\end{figure}

\begin{figure}
  \centering
\includegraphics[width=0.49\textwidth]{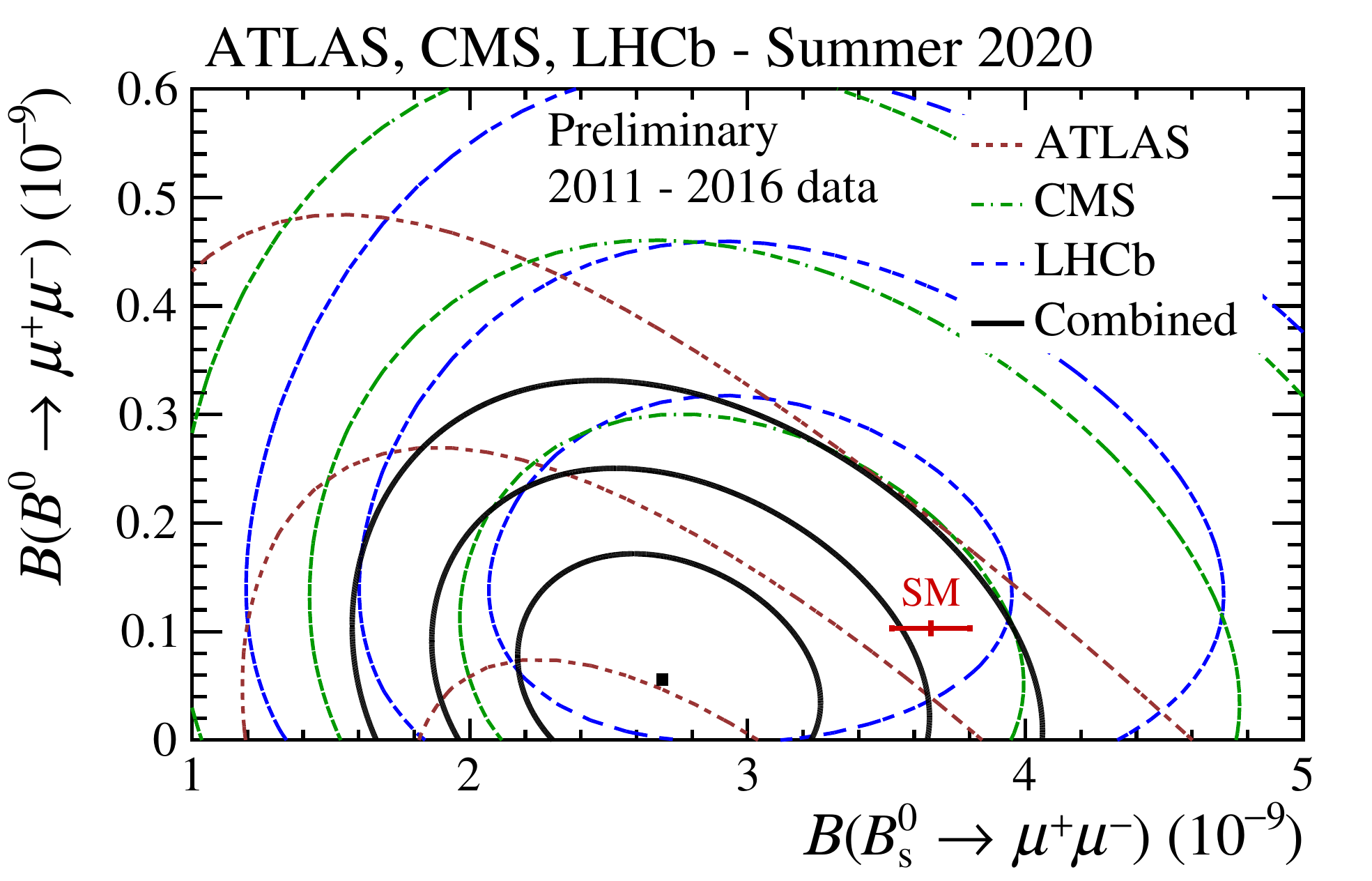}
\includegraphics[width=0.49\textwidth]{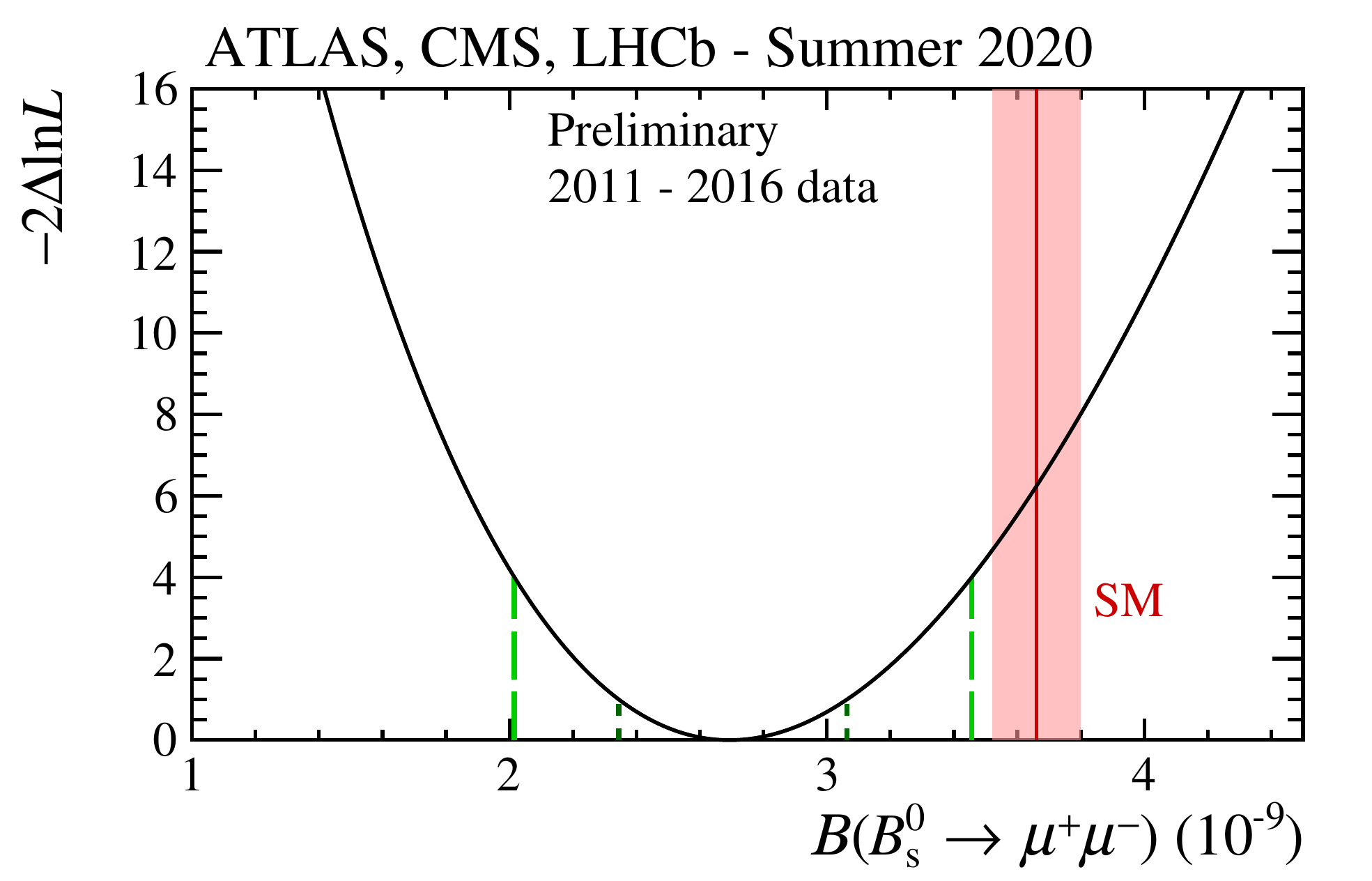}
  \caption{\label{fig:bsmmAv} (left) Two-dimensional likelihood contours of the results for the $\decay{\Bs}{\mumu}$ and $\decay{\Bd}{\mumu}$ decays for the ATLAS, CMS and LHCb experiments are shown with their combination. The red dashed line represents the ATLAS experiment, the green dot-dashed line the CMS experiment, the blue long-dashed line the LHCb experiment and the continuous line their combination.
For each experiment and for the combination, likelihood contours correspond to the values of {$-2\Delta ln \cal{L}$} = 2.3, 6.2 and 11.8 respectively. (right) Value of {$-2\Delta ln \cal{L}$} for $\decay{\Bs}{\mumu}$. The dark (light) green dashed lines
represent the 1$\sigma$ (2$\sigma$) interval, the red solid band shows the SM prediction with its uncertainty. Figures reproduced from from~\cite{LHCb:2020zud}. }
\end{figure}

The reconstructed di-muon invariant mass of all three experiments can be seen in Fig.~\ref{fig:BsmmMass}, which shows the region of high signal likelihood for all three experiments. The data show a clearly visible excess for all experiments, the individual statistical signal significance is determined to be 4.6\,$\sigma$ for ATLAS, 5.6\,$\sigma$ for CMS and 7.8\,$\sigma$ for LHCb. Dominant background sources are on the left mass sideband semileptonic $B$ decays that are partially reconstructed and hadronic two-body $B$ decays close to the $B^0$ mass. 
The measured branching fractions of $\decay{\Bs}{\mumu}$ and $\decay{\Bd}{\mumu}$ of the individual experiments are combined~\cite{LHCb:2020zud}, the 2D-likelihood contours can be found in Fig.~\ref{fig:bsmmAv}~(left), the one dimensional projection in Fig.~\ref{fig:bsmmAv}~(right). Consistent with the expectation, no evidence for the rarer decay $\decay{\Bd}{\mumu}$ is seen. A limit of  ${\cal{B}} (\decay{\Bd}{\mumu}) < 1.9  \times 10^{-10}$ is set at 95\% confidence level. 
The combined branching fraction of the decay $\decay{\Bs}{\mumu}$ is obtained to be ${\cal{B}} (\decay{\Bs}{\mumu}) = 2.69^{+0.37}_{-0.35}  \times 10^{-9}$. 
Including theoretical uncertainties, the one-dimensional compatibility with the SM prediction~\cite{Beneke:2019slt} is estimated to be $2.4\,\sigma$ for the $\Bs$ decay and $0.6\,\sigma$ for the $\Bd$ decay, while
the two-dimensional compatibility with the SM point is estimated to be of $2.1\,\sigma$. It will be interesting to see if this tension gets resolved or increased with the upcoming updates that utilize the full Run~1 and~2 data sets.  

\paragraph{Measurement of the effective $\decay{\Bs}{\mumu}$ lifetime}

\begin{figure}
  \centering
\includegraphics[width=0.49\textwidth]{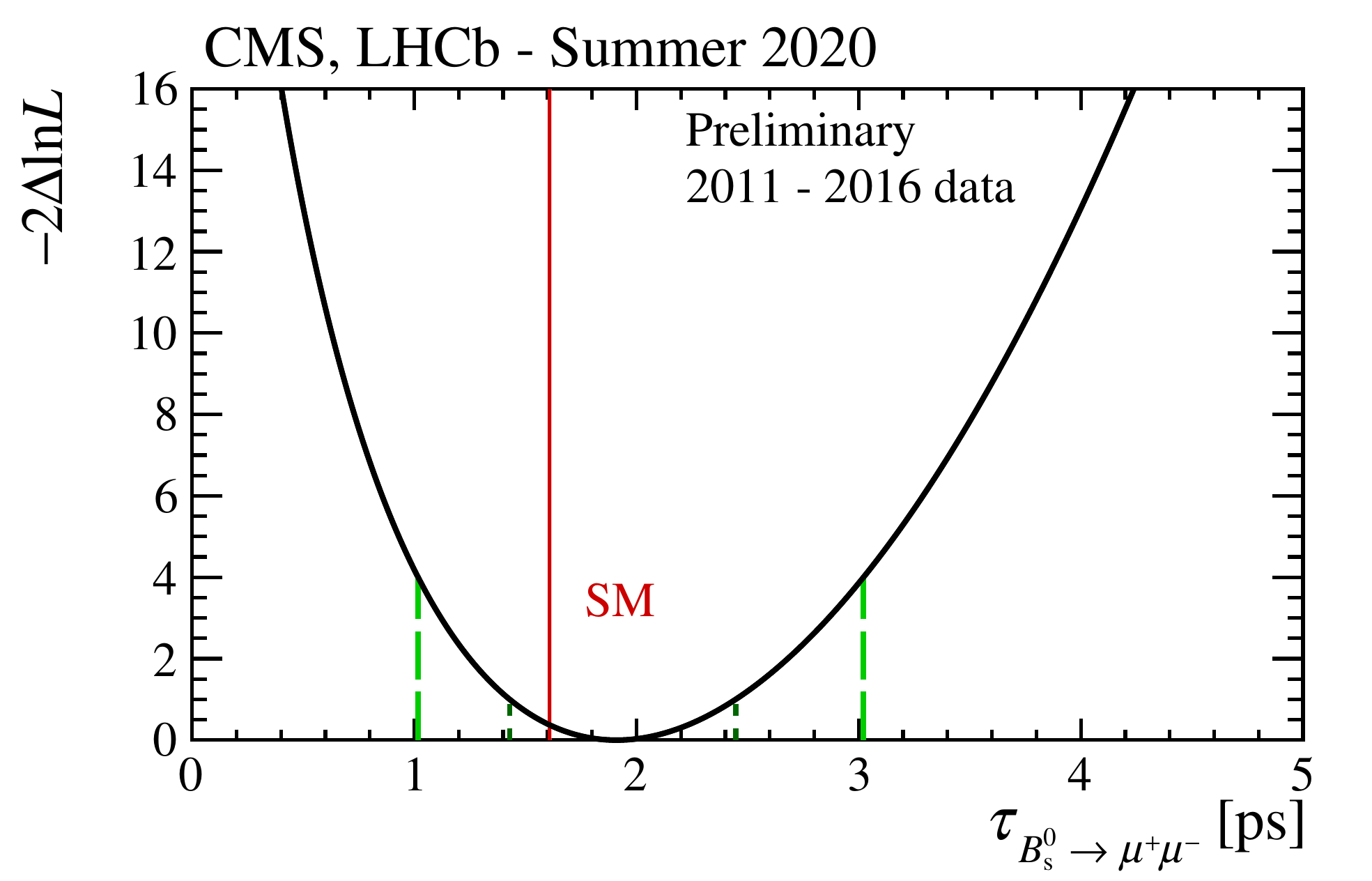}
\includegraphics[width=0.49\textwidth]{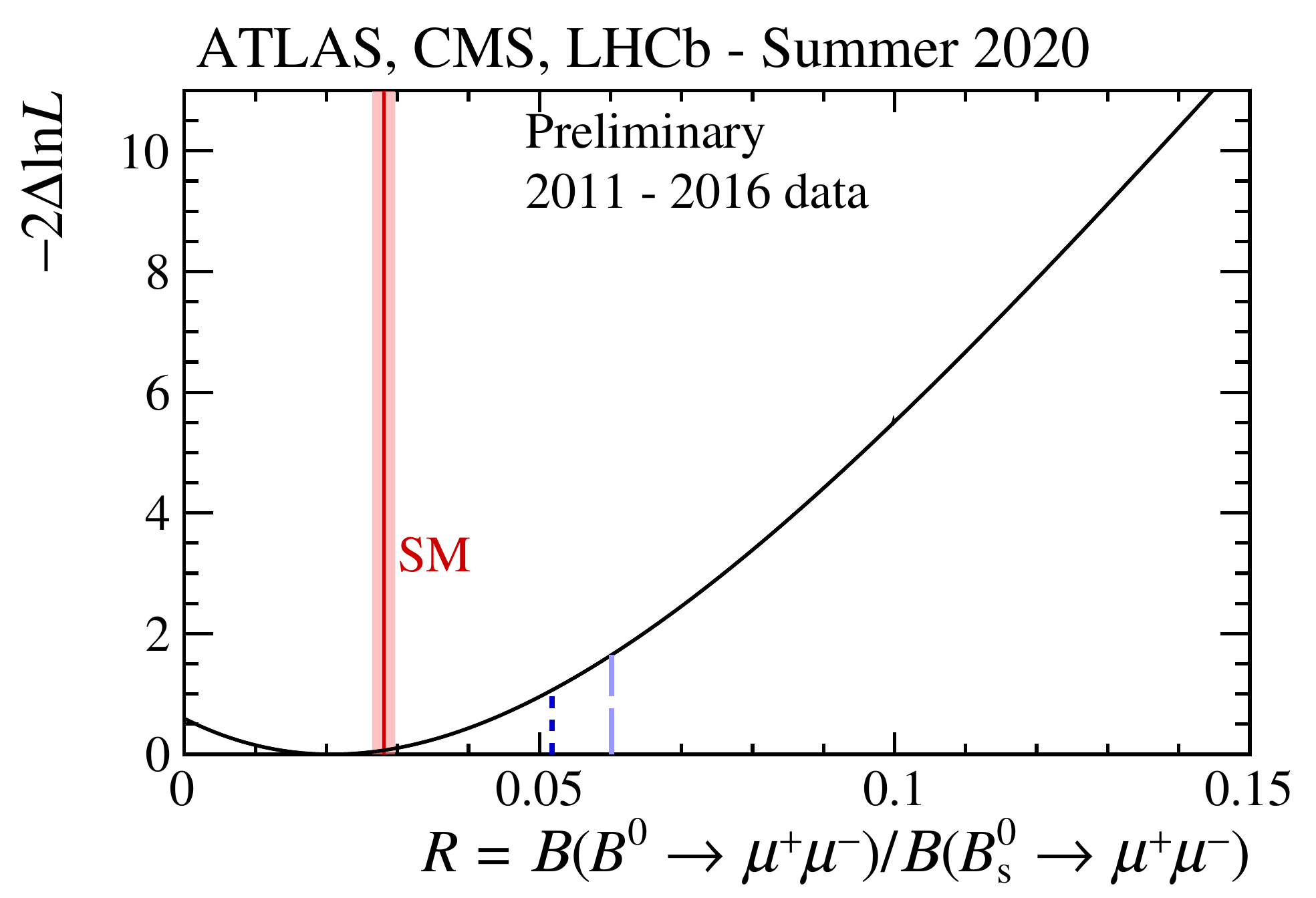}
  \caption{\label{fig:bsmmLifetime} 
  (left) Value of {$-2\Delta ln \cal{L}$} for the $\decay{\Bs}{\mumu}$ effective lifetime, the dark (light) green dashed lines represent the 1$\sigma$ (2$\sigma$) interval. 
    (right) Value of {$-2\Delta ln \cal{L}$} for the ratio of the $\decay{\Bd}{\mumu}$ and $\decay{\Bs}{\mumu}$ branching fractions, R, the light (dark) blue dashed line represents the 90\% (95\%) CL and the red solid band shows the SM prediction with its uncertainty.
Figures reproduced from from~\cite{LHCb:2020zud}. }
\end{figure}

As discussed in Sec.~\ref{sec:leptonicTH}, the effective $\decay{\Bs}{\mumu}$ lifetime offers an orthogonal approach to test the SM, first measurements have been performed by the CMS and LHCb collaborations. The LHCb measurement is performed from a fit to the background-subtracted decay-time distribution of signal candidates. The CMS measurement is determined with a two-dimensional likelihood fit to the proper decay  time and di-muon invariant mass; the model introduced in the likelihood fit adopts the per-event decay time resolution as a conditional parameter in the resolution model. For
both experiments, the measurement is fully dominated by its statistical uncertainty. The combination is found to to be $\tau_{\decay{\Bs}{\mumu}} = 1.91^{+0.37}_{-0.35}$\,ps, in excellent agreement with the SM. The likelihood contour is shown in Fig.~\ref{fig:bsmmLifetime}~(left). 

\paragraph{Prospects for $\decay{\Bd}{\mumu}$}

The decay $\decay{\Bd}{\mumu}$ has not yet been observed, and the ratio of the $\Bs$ and the $\Bd$ decay is an important test of the flavour structure of the underlying interaction. The ratio is determined by the combination of all three LHC experiments to be 
\begin{equation}
    R = \frac
        {{\cal B}(\decay{\Bd}{\mumu})}
        {{\cal B} (\decay{\Bs}{\mumu})} 
        < 0.060 
\end{equation}
at 95\% confidence level. The likelihood contour can be found in Fig.~\ref{fig:bsmmLifetime}~(right), so far in good agreement with the SM prediction.  
The next step, the discovery of the decay \decay{\Bd}{\mumu}, can hopefully be made by the combination of the LHC experiments in Run~3, chances for individual experiments are to observe the first evidence~\cite{Bediaga:2018lhg}. The following step is then a precision test of the ratio $R$, which will be a major task for the LHC experiments beyond Run~3.  

\paragraph{Search for $\decay{\B}{e^+e^-}$}

The decays $\decay{\Bs}{e^+e^-}$ and $\decay{\Bd}{e^+e^-}$ contain the same sensitivity to physics beyond the SM as their muonic siblings. In contrast to them, the SM rate is much stronger helicity suppressed, making an observation of the SM decay rate impossible in any current or planned collider experiment. This absence of \textit{Standard Model background} means these decays constitute a very clean probe for physics beyond the SM and constitute a test for the universality of leptonic couplings that is complementary to the ratios described later in this chapter. 

\begin{figure}
  \centering
\includegraphics[width=0.49\textwidth]{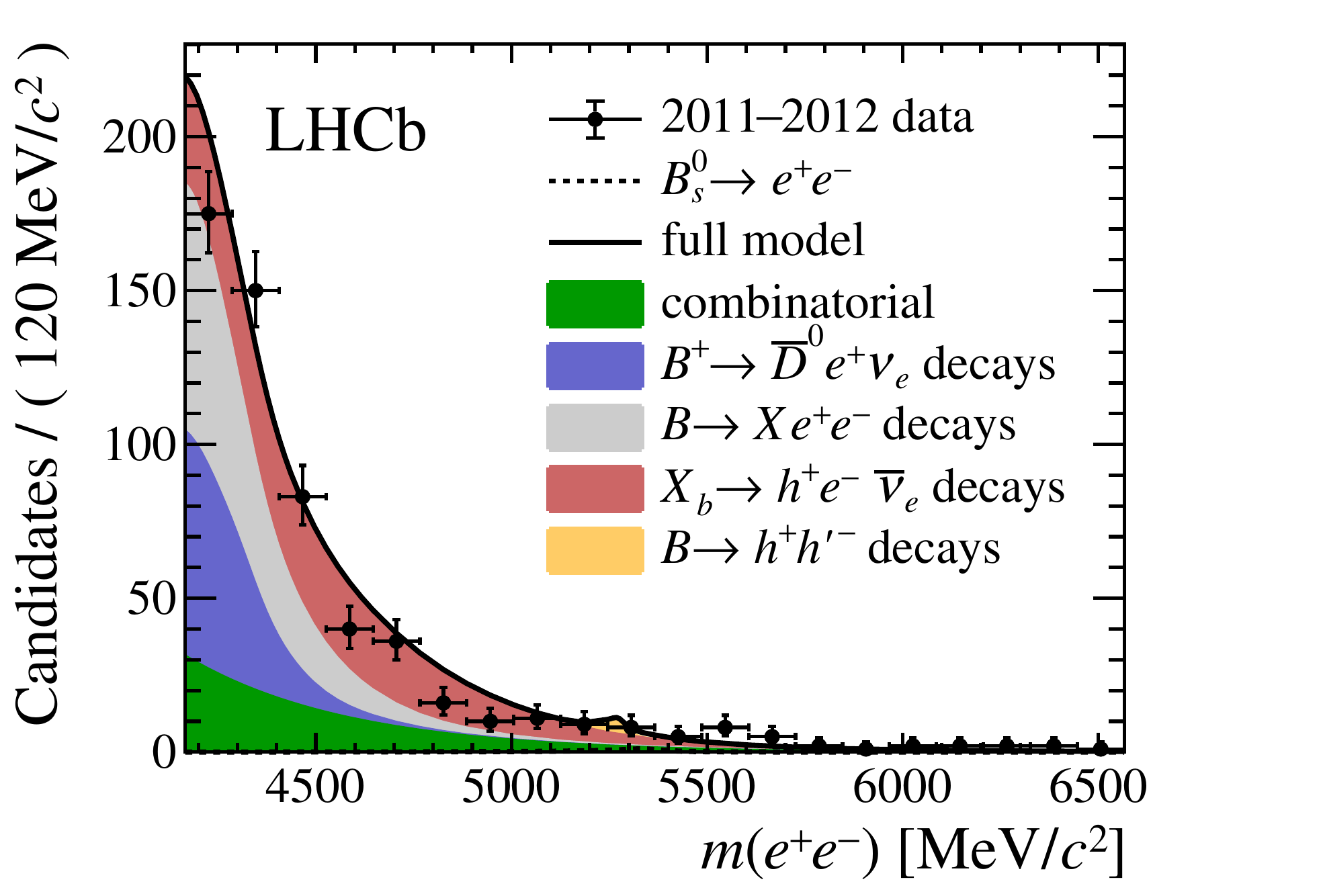}
  \caption{\label{fig:bs2ee} Reconstructed invariant di-electron mass distribution taken form Run~1 of the LHCb analysis in a region of high signal likelihood. No excess of signal candidates over the background expectation is seen. Individual physical background components are indicated in the legend. The Run 2 distributions looks comparable. Figure reproduced from from~\cite{PhysRevLett.124.211802}. }
\end{figure}

In the recent years, only the LHCb collaboration published a search for the decays $\decay{\Bs}{e^+e^-}$ and $\decay{\Bd}{e^+e^-}$, using about one half of the available data set~\cite{PhysRevLett.124.211802}. The reconstructed invariant di-electron mass distribution is shown exemplary for the Run~1 data set in Fig.~\ref{fig:bs2ee} for a region of high signal likelihood together with the fit projection. The data are well described by the background expectations, no excess of signal has been found. 
Upper limits of 
\begin{eqnarray}
\mathcal{B}  (\decay{\Bs}{\ee}) &<& 11.2 \times 10^{-9} \textrm{ and } \\
\mathcal{B} (\decay{\Bd}{\ee}) &<& 3.0 \times 10^{-9}
\end{eqnarray}
are found at 95\% CL. As the Standard Model expectation of these decays is far out of experimental reach, these modes remain a clean null test for  physics beyond the SM with new (pseudo)-scalar contributions. For example, flavour-universal NP scenarios can predict rates close to the current experimental bounds, see \eg\ Ref.~\cite{Fleischer_2017}.

\paragraph{Search for $\decay{\B}{\tau^+\tau^-}$}

In the case of $\decay{B}{\tau^+\tau^-}$, the helicity suppression is not very effective due to the large $\tau^-$ mass yielding branching fractions enhanced by roughly a factor 200 with respect to the muonic decay. 
As the decay involves only particles of the third generation, it could evolve into playing a significant role to test different new physics models, specially those with flavour violating Higgs couplings or tree level scalar of pseudo-scalar exchanges. 

\begin{figure}
  \centering
\includegraphics[width=0.49\textwidth]{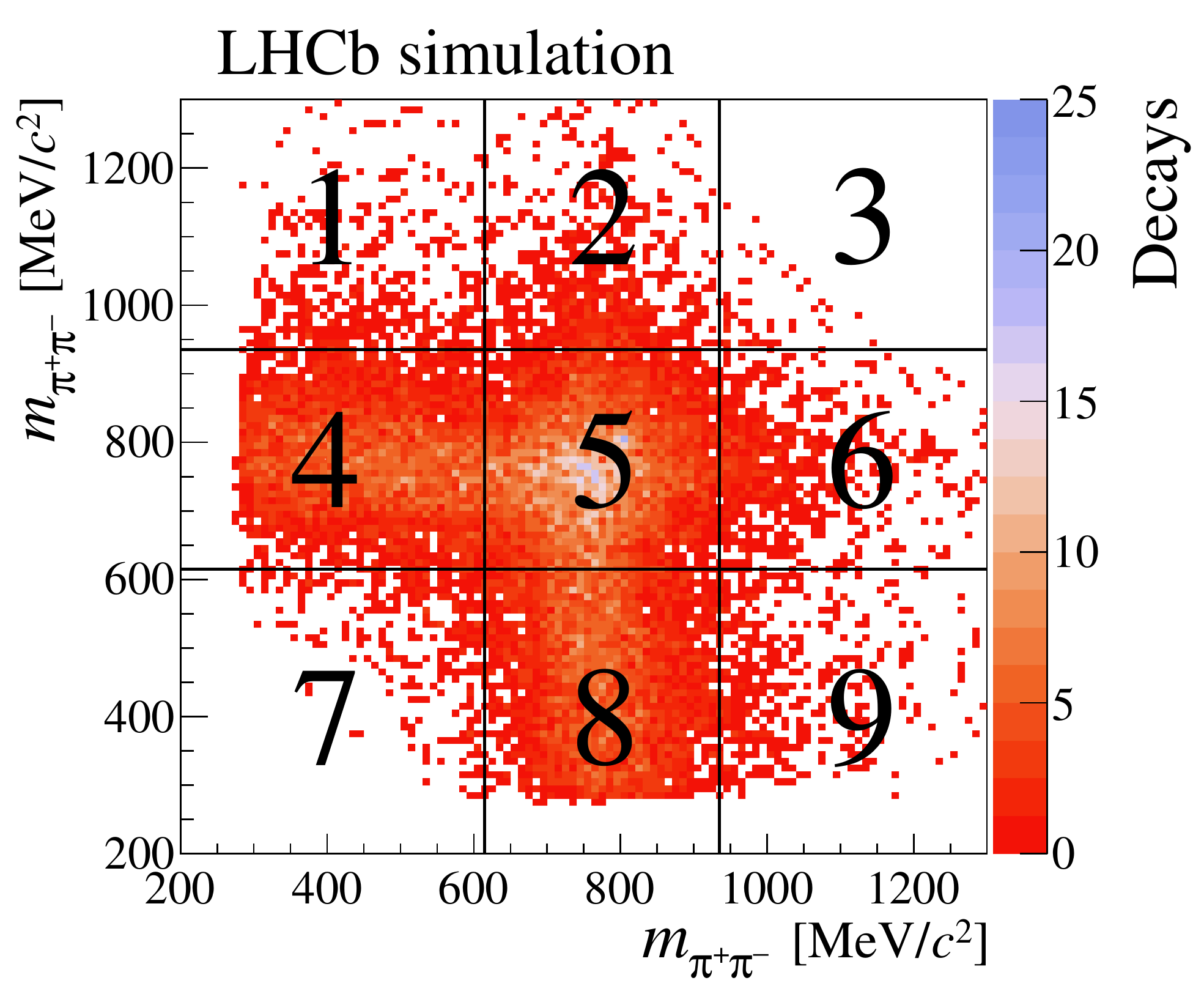}
\includegraphics[width=0.49\textwidth]{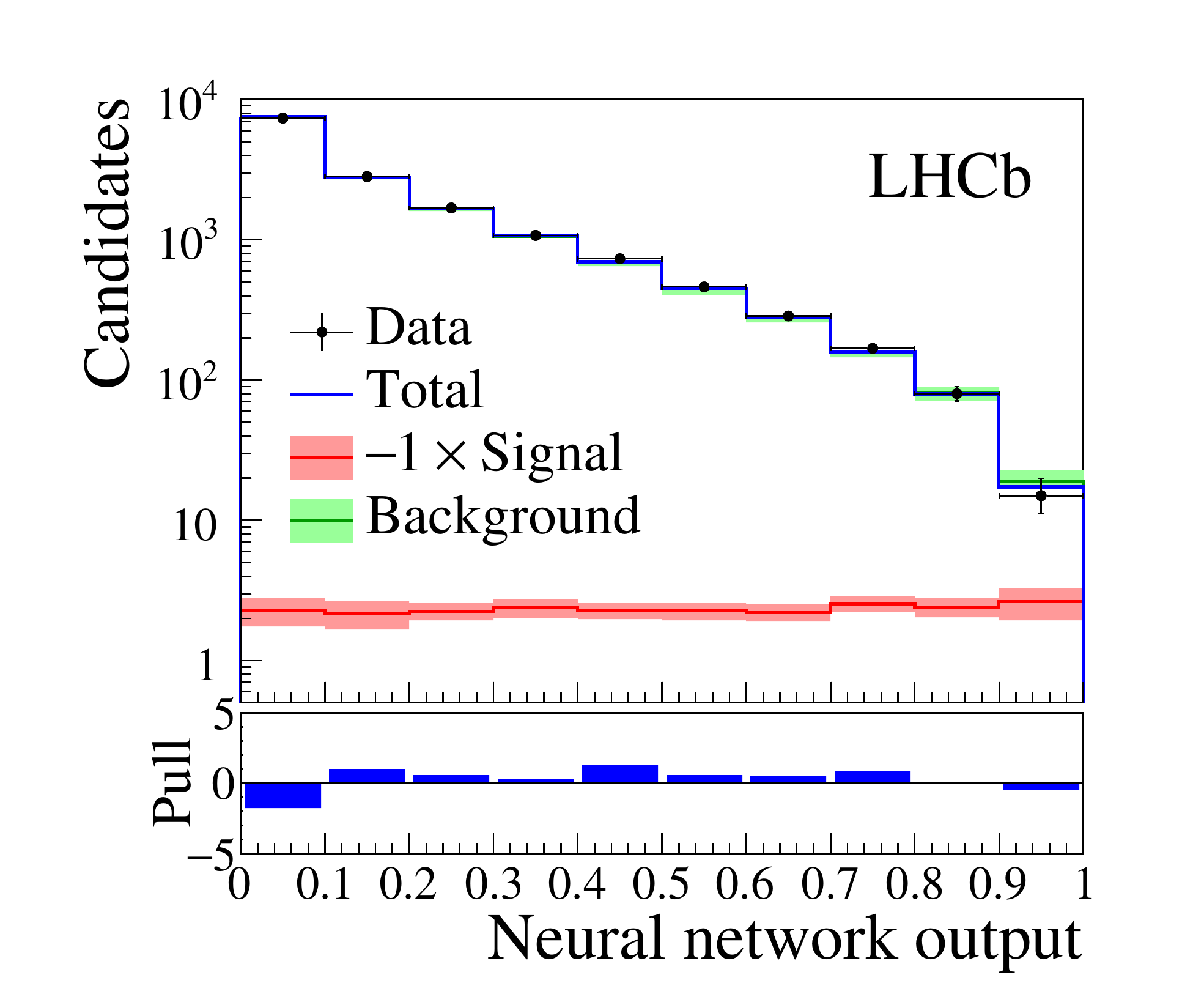}
  \caption{\label{fig:bstt} (left) Two-dimensional distribution of the invariant masses of the two oppositely charged two-pion combinations for
simulated $\decay{\Bs}{\tau^+\tau^-}$ candidates. The analysis is performed in regions in this plane, as indicated by vertical and horizontal lines. (right) Distribution of the NN output in the signal region (black points), with the total fit result (blue line) and the
background component (green line). The fitted $\decay{\Bs}{\tau^+\tau^-}$ signal component is negative and is therefore shown multiplied by -1
(red line). 
Figures reproduced from from~\cite{Aaij:2017xqt}.}
\end{figure}

However, the $\tau^-$ decay and its reconstruction poses interesting experimental challenges. 
The published LHCb analysis~\cite{Aaij:2017xqt} uses data from Run 1 only and reconstructs the $\tau^-$ candidates in the decay mode $\decay{\tau^-}{\pi^+ \pi^- \pi^- \nu_\tau}$. The three-prong decay allows to identify the $\tau^-$ decay vertices and hence more cleanly reconstruct the $B$ meson candidates.
For each $\tau^-$ candidate, the two-dimensional distribution of the invariant masses of the two oppositely charged two-pion combinations is divided into nine sectors, as illustrated in Fig.~\ref{fig:bstt}~(left), exploiting the information of intermediate resonances as the $\rho^0$. 
Because of the final-state neutrinos, the $\tau^+\tau^-$ mass provides only a weak discrimination between signal and background, and cannot be used to identify the signal. A central Neural Network (NN) is hence used to separate the signal component, as shown in Fig.~\ref{fig:bstt}~(right). 
The fitted $\decay{\Bs}{\tau^+\tau^-}$ signal component is negative and upper exclusion limits are determined to be  
\begin{eqnarray}
\mathcal{B}  (\decay{\Bs}{\tau^-\tau^+}) &<& 6.8 \times 10^{-3} \textrm{ and } \\
\mathcal{B} (\decay{\Bd}{\tau^-\tau^+}) &<& 2.1 \times 10^{-3}\nonumber
\end{eqnarray}
at 95\% CL. This precision is expected to improve by roughly a factor five with the data set of the upgraded LHCb experiment~\cite{Bediaga:2018lhg}. 
The Belle~2 experiment, that has recently started data-taking, might be able to explore
these modes in some depth, reaching limits on the branching fractions of order of $10^{-5}$ and $10^{-4}$ for $\Bd$ and $\Bs$ decays, respectively~\cite{Kou:2018nap}.

\paragraph{Outlook leptonic B-decays}

\begin{figure}
  \centering
\includegraphics[width=0.49\textwidth]{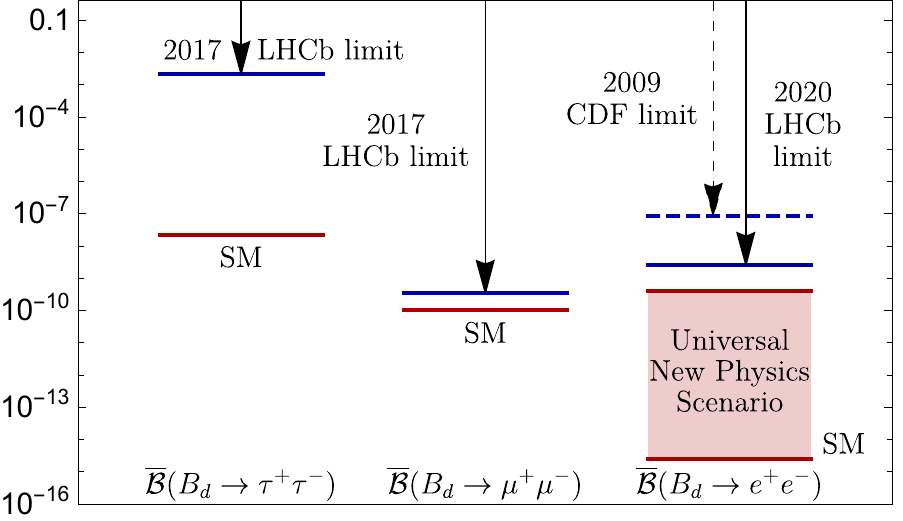}
\includegraphics[width=0.49\textwidth]{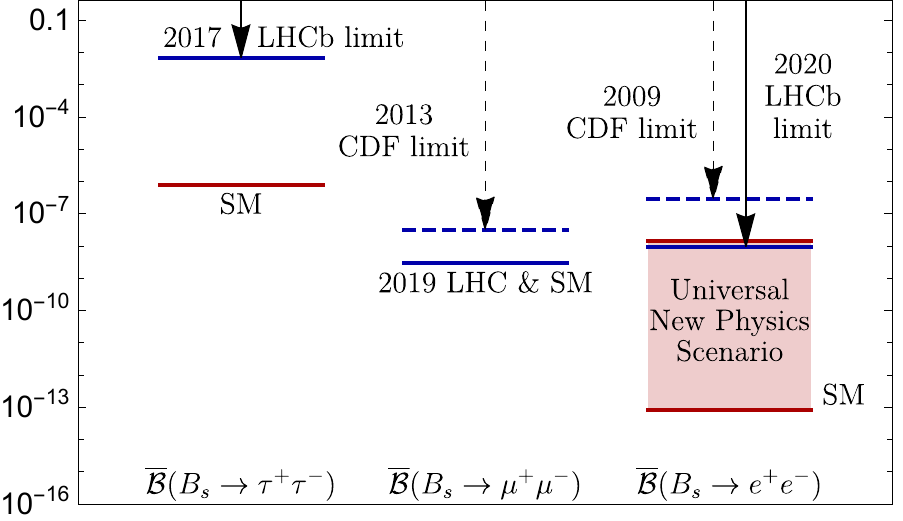}
  \caption{\label{fig:b2ll} Summary of the measurements and limits of all six purely leptonic $B$ meson decays. Only the decay $\decay{\Bs}{\mumu}$ is observed, the others are restricted by upper limits on their branching fractions. Regions of special interest will be tested soon for $\decay{\Bd}{\mumu}$, $\decay{\Bd}{\ee}$ and $\decay{\Bs}{\ee}$. 
  Figures updated from from~\cite{Fleischer_2017}. }
\end{figure}

The current experimental situation in purely leptonic $B$ decays, together with SM expectations, is shown in Fig.~\ref{fig:b2ll}. No purely leptonic $\Bd$ decay has been observed yet. The experimental sensitivity for the muonic decay $\decay{\Bd}{\mumu}$ is in reach of the SM prediction, the other two predictions are several orders of magnitude away from the experimental sensitivity. In the area of $\Bs$ decays, the pattern is similar: while the decay $\decay{\Bs}{\mumu}$ has been observed, the other two Standard Model expectations are out of experimental reach. Any observed signal in these decay modes would therefore be a clear indication of physics beyond the SM. This makes the area of purely leptonic $B$ decays a very fruitful field for future searches for traces of NP. 

\subsection{Semileptonic $B$ decays}  
\label{sec:semileptonic}
\subsubsection{Theory}
\label{sec:btosll:theory}

Exclusive semileptonic $b\to s\ellell$ transitions of the type $B\to M\ellell$, with $M=P,V$ either
a pseudoscalar ($P$) or a vector $(V)$ meson, are of great phenomenological interest. Unlike  the purely leptonic decays discussed in Sec.~\ref{sec:leptonicTH}, the semileptonic decays are
not helicity suppressed, which increases their branching fraction by orders of magnitude to
the $\mathcal{O}(10^{-7})$ level. This makes them better accessible experimentally due to their higher decay rates,
reducing statistical uncertainties and increasing phenomenological usefulness.
The hadronic final state also offers the opportunity for intricate
decay distributions that are sensitive to complementary combinations of the WET Wilson coefficients.
However, this phenomenological usefulness comes in lockstep with a more complicated theoretical
description and therefore larger theoretical uncertainties.\\

To leading order in the electromagnetic interaction, the matrix elements of semileptonic and radiative
operators factorise in hadronic and leptonic matrix elements. Schematically
\begin{multline}
    \label{eq:btosll-amp}
    A(\bar B\to M \ell^+\ell^-)
        \propto G_F\, \lambda_D^t
          \times \bigg[ (C_9 \,L^\mu_{V} + C_{10} \,L^\mu_{A})\  {\cal F}^{\mu} + (C_{9'} \,L^\mu_{V} + C_{10'} \,L^\mu_{A})\  {\cal F}^{\prime,\mu}\\
        -  \frac{L^\mu_{V}}{q^2} \Big\{  2 i m_b (C_7\,{\cal F}^{T,\mu} + C_{7'}\,{\cal F}^{T,\prime,\mu})  + 16\pi^2 {\cal H}^{\mu} \Big\}   \bigg]\ ,
\end{multline}
where we suppress small corrections due to $\lambda_D^u$ and retain only the SM-like operators and their chirality flipped counter parts. Scalar, pseudoscalar and tensor operators can be readily included, see \eg~Ref.~\cite{Bobeth:2012vn},
but have been dropped to increase legibility. The hadronic matrix elements are
\begin{align}
    {\cal F}_{B\to M}^{\mu}(k, q)        &\equiv  \bra{M(k)}\bar{s}\gamma_\mu P_L\, b \ket{\bar{B}(p)}\,, \\
    {\cal F}_{B\to M}^{\prime,\mu}(k, q) &\equiv  \bra{M(k)}\bar{s}\gamma_\mu P_R\, b \ket{\bar{B}(p)}\,, \\
    {\cal F}_{B\to M}^{T,\mu}(k, q)      &\equiv  \bra{M(k)}\bar{s}\sigma_{\mu\nu} q^\nu P_R\, b \ket{\bar{B}(p)}\,, \\
    {\cal F}_{B\to M}^{T\prime,\mu}(k, q)&\equiv  \bra{M(k)}\bar{s}\sigma_{\mu\nu} q^\nu P_L\, b \ket{\bar{B}(p)}\,, \\
    {\cal H}_{B\to M}^{\mu}(k, q)        &\equiv  i\!  \int d^4x\, e^{i q\cdot x}\,
                                          \bra{M(k)} T\big\{ j_\mu^{\rm em}(x), \sum_i C_i T_i(x) \big\} \ket{\bar B(p)}\,,
\end{align}
with the sum over operators $T_i$ with $i=1c,2c,3\dots6,8$. The hadronic matrix elements are classified either
as local matrix elements ${\cal F}^{(\prime)}$ and ${\cal F}^{T(\prime)}$ or as nonlocal matrix elements ${\cal H}$.
Both types of matrix elements are needed for reliable and accurate predictions of the amplitudes and therefore of
the observables in semileptonic decays. For phenomenological discussions, one often encounters the hadronic amplitudes of these
decays in the transversity basis. These scalar-valued amplitudes can be obtained by projecting the vector-valued
hadronic parts of eq.~\eqref{eq:btosll-amp} onto a basis of polarisation vectors. The number of independent
transversity amplitudes depends on the process under consideration. In the following, we will review the theory
status of the various hadronic matrix elements, separated in local and nonlocal matrix elements

\paragraph{Local hadronic matrix elements} The local matrix elements ${\cal F}_{B\to M}$ can be organized
schematically as
\begin{equation}
    \bra{M(k)} \bar{s} \Gamma^\mu b \ket{\bar{B}(p)} = \sum_i f^{B\to M}_i(q^2) S_{B\to M}^\mu(p, k)\,.
\end{equation}
Here $q\equiv p - k$, $\Gamma$ represents an arbitrary Dirac structure, $S_i$ represents a set of Lorentz structures,
and $f_i$ are the so-called hadronic form factors.
The form factors are scalar-valued functions of $q^2$ and are real-valued below the lowest pair production threshold
$q^2 \leq (M_B + M_K)$. The number of independent structures $S_i$ and independent and non-zero form factors
$f_i$ depends on the angular momentum and parity of the hadronic states involved.
As for the $B$-meson decay constants, the form factors are genuinely non-perturbative objects.
Studies of the form factors within the framework of QCD factorization (QCDF) found that --- for large
energies $E_M$ of the final hadron in the $B$ rest frame ---
ratios of the form factors are accessible to leading power in the heavy-quark expansion, while
the form factors themselves are inaccessible due to endpoint divergences~\cite{Beneke:2000wa}.
These relations are remnants of the symmetry arising from the simultaneous large-energy and heavy-quark limits,
and are broken by radiative corrections and $1/m_b$ and $1/E_M$ power corrections.
Nevertheless, these relations are useful to discuss approximate relations between transversity amplitudes,
and have lead to the design of a basis of decay observables with reduced theoretical
uncertainties~\cite{DescotesGenon:2012zf}. Even for small energies $E_M$ the symmetries imposed by
the heavy-quark expansion alone give rise to strong correlations between the form factors, such that
a similar basis of decay observables can be constructed~\cite{Bobeth:2010wg,Bobeth:2012vn}.
An extension to the full phase space was achieved in Ref.~\cite{Descotes-Genon:2013vna}.

For transitions of a $B$-meson to a pseudo-scalar meson $P$ a common basis of the three nonzero form factors reads
\begin{equation}
 \label{eq:btosll:P-FFs}
\begin{aligned}
  \bra{P(k)} \bar{D} \gamma^\mu b \ket{\bar{B}(p)}
    = & \left[(p + k)^\mu - \frac{M_B^2 - M_P^2}{q^2} q^\mu \right]\, f_+(q^2)
    +   \frac{M_B^2 - M_P^2}{q^2} q^\mu \, f_0(q^2)\,, \\
  \bra{P(k)} \bar{D}\sigma^{\mu\nu} q_\nu b \ket{\bar{B}(p)}
    = & \frac{i}{M_B + M_P} \left[2 q^2p^\mu-(M_B^2 - M_P^2 + q^2)q^\mu\right] \,f_T(q^2)\,.
\end{aligned}
\end{equation}
The hadronic matrix elements of other dimension-three currents vanish by Lorentz symmetry and
parity conservation of the strong interaction. The full basis of local form factors
of $\bar{B}\to \bar{K}$~\cite{Bouchard:2013eph,Bailey:2015dka},
$\bar{B}\to \pi$~\cite{Dalgic:2006dt,Bailey:2008wp,Lattice:2015tia,Flynn:2015mha,Colquhoun:2015mfa,Bailey:2015nbd}
and $\bar{B}_s\to K$~\cite{Bouchard:2014ypa,Flynn:2015mha}
transitions is available at $q^2 \gtrsim 15\,\GeV^2$ from lattice QCD studies.
These form factors are also accessible with light-cone sum rules (LCSRs) for $q^2\lesssim 12\,\GeV^2$.
The full basis of $\bar{B}\to \bar{K}$, $\bar{B}\to \pi$, and $\bar{B}_s\to K$ form factors is available
from LCSRs with final-state distribution amplitudes~\cite{Khodjamirian:2017fxg}. Their uncertainties
are competitive with the extrapolations from lattice QCD results to $q^2 \simeq 0$. The form factors
for $\bar{B}\to \bar{K}$ and $\bar{B}\to \pi$ transitions are also available from LCSRs with $B$-meson
distribution amplitudes~\cite{Gubernari:2018wyi}, albeit with larger parametric and systematic uncertainties.

For transitions of a $B$-meson to a vector-meson $V$ a common basis of the seven nonzero form factors reads
\begin{align}
  \label{eq:btosll:V-FFs}
  \bra{V(k, \eta)} \bar{D} \gamma^\mu b \ket{\bar{B}(p)}
    = & \varepsilon^{\mu\nu\rho\sigma} \eta^*_\nu p_\rho k_\sigma \frac{2 V(q^2)}{M_B + M_V}\,,\\
  \bra{V(k, \eta)} \bar{D} \gamma^\mu \gamma_5 b\ket{\bar{B}(p)}
    = & i \eta^*_\nu \bigg[g^{\mu\nu} (M_B + M_V) A_1(q^2) - \frac{(p + k)^\mu q^\nu}{M_B + M_V}A_2(q^2)
        - \frac{2 M_V q^\mu q^\nu}{q^2} \left(A_0(q^2) - A_3(q^2)\right)\bigg]\,,\\
  \bra{V(k, \eta)} \bar{D} \sigma^{\mu\nu} q_\nu b\ket{\bar{B}(p)}
    = & \varepsilon^{\mu\nu\rho\sigma} \eta^*_\nu p_\rho k_\sigma 2 T_1(q^2)\,,\\
  \bra{V(k, \eta)} \bar{D} \sigma^{\mu\nu} q_\nu \gamma_5 b\ket{\bar{B}(p)}
    = & i \eta^*_\nu \bigg[\left(g^{\mu\nu} (M_B^2 - M_B^2) - (p + k)^\mu q^\nu\right) T_2(q^2)
        + q^\nu \left(q^\mu - \frac{q^2}{M_B^2 - M_V^2} (p + k)^\mu\right) T_3(q^2)\bigg]\,.
\end{align}
with $\eta$ the polarization vector of the vector meson $V$ and
\begin{equation}
    A_3(q^2) \equiv \frac{M_B + M_V}{2 M_V} A_1(q^2) - \frac{M_B - M_V}{2 M_V} A_2(q^2)\,.
\end{equation}
Since vector mesons are resonances
and decay via the strong interaction, the description of their form factors incurs additional
inherent uncertainties. At present it is standard to treat the vector mesons as quasi-stable
states in the narrow-width approximation, see \eg~\cite{Altmannshofer:2008dz}. This holds
for lattice QCD studies as well as sum-rule bases analyses, with one exception. A recent
pilot study of $\bar{B}\to \bar{K} \pi$ form factors, which include the $\bar{K}^*$ resonance,
have found $\mathcal{O}(10\%)$ effects due to the finite width~\cite{Descotes-Genon:2019bud}.
However, their findings indicate universal changes to all form factors, which cancel in
ratios such as angular observables and $\CP$ or isospin asymmetries. Branching ratios are
the most strongly affected observables\footnote{
It should be noted that these effects further increase the tension between measurements and predictions
of the branching ratios.
}.

The full basis of local form factors
of $\bar{B}\to \bar{K}^*$ and $\bar{B}_s\to \phi$
transitions are available at $q^2 \gtrsim 15\,\GeV^2$ from a single lattice QCD study~\cite{Horgan:2013hoa}.
An independent corroboration is presently in progress~\cite{Flynn:2015ynk,Flynn:2016vej,Lizarazo:2016myv}.
The form factors for $\bar{B}\to \rho$ transitions are not available from lattice QCD simulations, and should
rather be treated as $\bar{B}\to \pi\pi$ form factors~\cite{Faller:2013dwa,Kang:2013jaa}.
These form factors are also accessible with light-cone sum rules (LCSRs) for $q^2\lesssim 12\,\GeV^2$.
The full basis of $\bar{B}\to \bar{K}^*$ and $\bar{B}_s\to \phi$ form factors is available
from LCSRs with final-state distribution amplitudes~\cite{Straub:2015ica}. Their uncertainties
are competitive with the extrapolations from lattice QCD results to $q^2 \simeq 0$. The form factors
for $\bar{B}\to \bar{K}^*$~\cite{Gubernari:2018wyi} and $\bar{B}_s\to \phi$~\cite{Gubernari:2020eft}
transitions are also available from LCSRs with $B$-meson distribution amplitudes,
albeit with larger parametric and systematic uncertainties.

\paragraph{Nonlocal hadronic matrix elements}
The nonlocal hadronic matrix elements ${\cal H}_{B\to M}^\mu(q^2)$ can be organized in a similar ways
as the form factors, in terms of \textit{nonlocal form factors} $\mathcal{H}^{B\to M}_i$ and Lorentz structures
$S_i^\mu$. In the SM are three-independent nonlocal form factors per $B\to V$ transitions, and one independent
nonlocal form factor per $B\to P$ transition.
The treatment of nonlocal hadronic matrix elements in the rare semileptonic decays constitutes
presently the largest systematic uncertainty for theory predictions and the inference of the
WET Wilson coefficients; see \eg ref.~\cite{Khodjamirian:2010vf} and ref.~\cite{Hurth:2020rzx}, respectively.
Beyond these parametric uncertainties, hard-to-quantify systematic uncertainties are introduced:
present studies of the nonlocal effects work under the assumption that BSM physics does not
induce four-quark operators beyond the ones listed in eq.~\eqref{eq:wet-btoD-4quark}, or modify the Wilson coefficients of four-quark or radiative operators with $i=1U,2U,3,\dots,6,8'$. The total number of independent
nonlocal form factors needed in the most general basis of WET coefficients has not yet been determined.

To date, the basis for most of the phenomenological analyses of the $\decay{\bar{B}}{\bar{K}^{(*)}\ellell}$ data
is an operator product expansion (OPE) in terms of either local (local OPE) or light-cone operators (LCOPE).
The local OPE can be applied to the nonlocal operators $T_i(x)$ if all components of the distance $x$ are small,
$x^\mu \ll \Lambda_\text{had}^{-1}\,\forall \mu$. This is the case in the limit $q^2\to \infty$, and
can be assumed to hold well already for $q^2 \simeq m_b^2$. This OPE can be carried out either with matching
onto local operators involving a heavy-quark effective theory field~\cite{Grinstein:2004vb} or a QCD-field~\cite{Beylich:2011aq}
for the $b$ quark. Here, we will only discuss the OPE with matching onto QCD fields.
Analytic results for the matching coefficients of the leading-power dimension-three operators are
known to $\mathcal{O}(\alpha_s)$~\cite{Asatrian:2019kbk}. The matrix elements of the dimension-three operators
are exactly the local form factors ${\cal F}^{(\prime)}$ and ${\cal F}^{T(\prime)}$.
The matching coefficients for the dimension-four
and dimension-five operators are known to leading order in $\alpha_s$~\cite{Beylich:2011aq}.
Their matrix elements are similar to local form factors of $\bar{D} \dots b$ currents that contain
strings of covariant derivatives. These matrix elements are presently known within the framework of
QCDF for large energies $E_M$ of the final state hadron~\cite{Beneke:2001at,Beneke:2004dp};
see the comments on QCDF calculations of local form factor above. For the four-quark operators, these
matrix elements free of endpoint divergences at leading power. For the radiative operator $\op{8}$,
however, endpoint divergences emerge, and the QCDF framework breaks down. These contributions are
instead obtained from light-cone sum rule calculations~\cite{Ball:2006eu,Dimou:2012un}.
At large $q^2$ above the open charm threshold, quark-hadron duality implies that integrated measurements
in the accessible phase space should concide with the integrated OPE predictions,
albeit with hard-to-quantify systemtic uncertainties
(dubbed duality violation)~\cite{Grinstein:2004vb,Bobeth:2010wg,Beylich:2011aq,Lyon:2014hpa}.

The LCOPE includes at leading power the local limit, \ie, the results for the dimension-three operators
of the local OPE. Within the LCOPE, the dominant contributions from four-quark operators have been studied.
For the four-quark operators, the higher dimensional operators are suppressed by inverse power of
$q^2 - 4 m_c^2$~\cite{Khodjamirian:2010vf}.
Together with the numerically large results for the the matrix elements of the the next-to-leading power operators,
this result has drawn into question whether the theory predictions are reliable in the
phenomenologically interesting region $q^2\simeq 1\dots 6\GeV^2$~\cite{Jager:2012uw,Ciuchini:2015qxb,Hurth:2016fbr}.
A recent study~\cite{Gubernari:2020eft} revisit the steps undertaken in ref.~\cite{Khodjamirian:2010vf} and comes to the following conclusions:
a) There is only one operator at next-to-leading power in the LCOPE, and the result for its matching coefficient
is confirmed. b) The hadronic matrix elements of this operators had been incompletely calculated, due
to the absence of half of the Lorentz structures in the LCSR setup of ref.~\cite{Khodjamirian:2010vf}.
The complete calculation reduced the hadronic matrix elements of the operator in question by a factor of 10~\cite{Gubernari:2020eft}. Further reduction by another factor of 10 arises from updated numerical inputs.

Nevertheless, it has been advocated~\cite{Khodjamirian:2010vf,Ciuchini:2015qxb,Bobeth:2017vxj} to replace the
present approach of calculating the nonlocal matrix elements in the semileptonic region with the following
approach:
\begin{itemize}
    \item The nonlocal form factors should be calculated at spacelike $q^2$, away from hadronic branch
    cuts in that variable.
    A choice of $q^2 \lesssim -1\GeV^2$ accelerates the convergence of the LCOPE.
    
    \item The gap between spacelike $q^2$ and the phenomenologically interesting region at timelike $q^2$
    should be bridged by some form of analytic continuation.
    This step has been addressed in the literature in a few different ways, which range from more phenomenological~\cite{Lyon:2014hpa,Ciuchini:2015qxb,Arbey:2018ics,Aaij:2016cbx,Brass:2016efg,Blake:2017fyh}
    to more formal~\cite{Khodjamirian:2010vf,Khodjamirian:2012rm,Bobeth:2017vxj}.
    The more formal ones involve dispersion relations~\cite{Khodjamirian:2010vf,Khodjamirian:2012rm}
    or a formal series expansion~\cite{Bobeth:2017vxj}.
\end{itemize}
Very recently, a new parametrization~\cite{Gubernari:2020eft} has been proposed that provides a handle on the truncation error
in the series expansion method. This handle, in form a of a dispersive bound, had not yet been used in
any phenomenological analysis, and studies to that effect are ongoing.

\subsubsection{Decay rates}
\label{sec:btoslldecayrates}
Compared to the very rare fully leptonic $B$ decays, 
semileptonic \btosll\ decays exhibit higher decay rates and are thus generally easier accessible experimentally. 
The semileptonic $b\to s\ell^+\ell^-$ decays $\decay{\Bd}{\Kp\ell^+\ell^-}$ and $\decay{\Bd}{\Kstarz\ell^+\ell^-}$ were first observed by the Belle Collaboration~\cite{Abe:2001dh,Ishikawa:2003cp}. 
Since then, branching fractions of exclusive \btosll\ decays have been studied by the 
BaBar~\cite{Aubert:2008ps}, Belle~\cite{Wei:2009zv}, CDF~\cite{Aaltonen:2011qs}, CMS~\cite{Khachatryan:2015isa} and LHCb collaborations~\cite{Aaij:2014pli,Aaij:2014bsa,Aaij:2014kwa,Aaij:2014lba,Aaij:2015esa,Aaij:2015nea,Aaij:2016kqt,Aaij:2018jhg,Aaij:2015xza}. 
At hadron colliders, the muonic \btosmm\ decays are more easily accessible compared to the \btosee\ modes, 
as muons can be triggered and reconstructed with higher efficiency in the high multiplicity hadronic environment. 
For the example of the LHCb detector,
the first level trigger thresholds in 2012 were set to $\Et(e^\pm)>3\gevcc$ and $\pt(\mu^\pm)>1.78\gevc$ for the electron and muon trigger, respectively, resulting in a lower trigger efficiency for electron modes~\cite{Aaij:2012me,Albrecht:2013fba}. 
This is in contrast to the situation at the $B$-factory experiments that 
exhibit similar efficiencies for \btosmm\ and \btosee\ modes. 
Measurements from the $B$-factory experiments thus typically combine the two lepton flavours for branching fraction measurements to achieve higher signal yields. 

\begin{figure}
\centering
\includegraphics[width=0.49\textwidth]{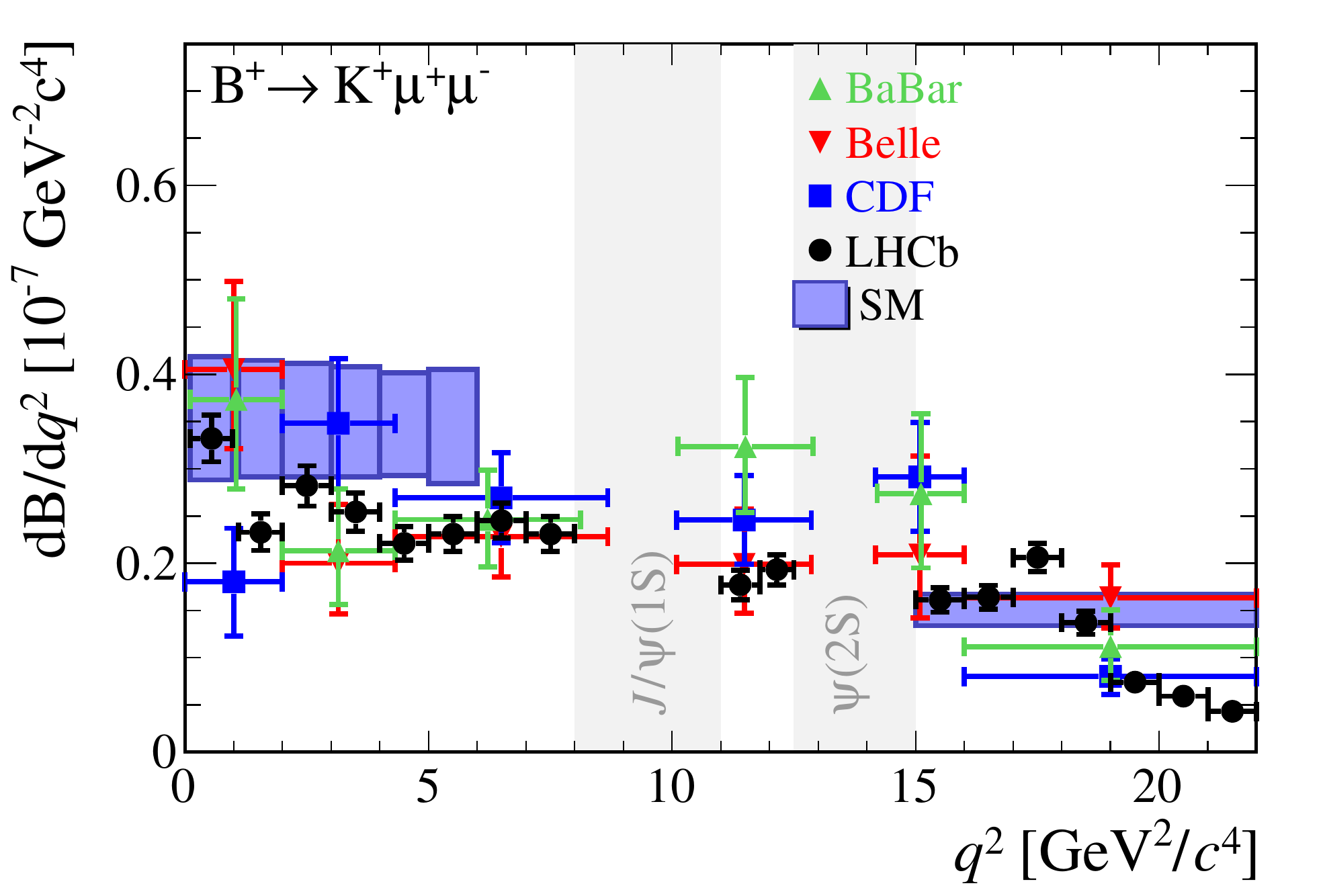}
\includegraphics[width=0.49\textwidth]{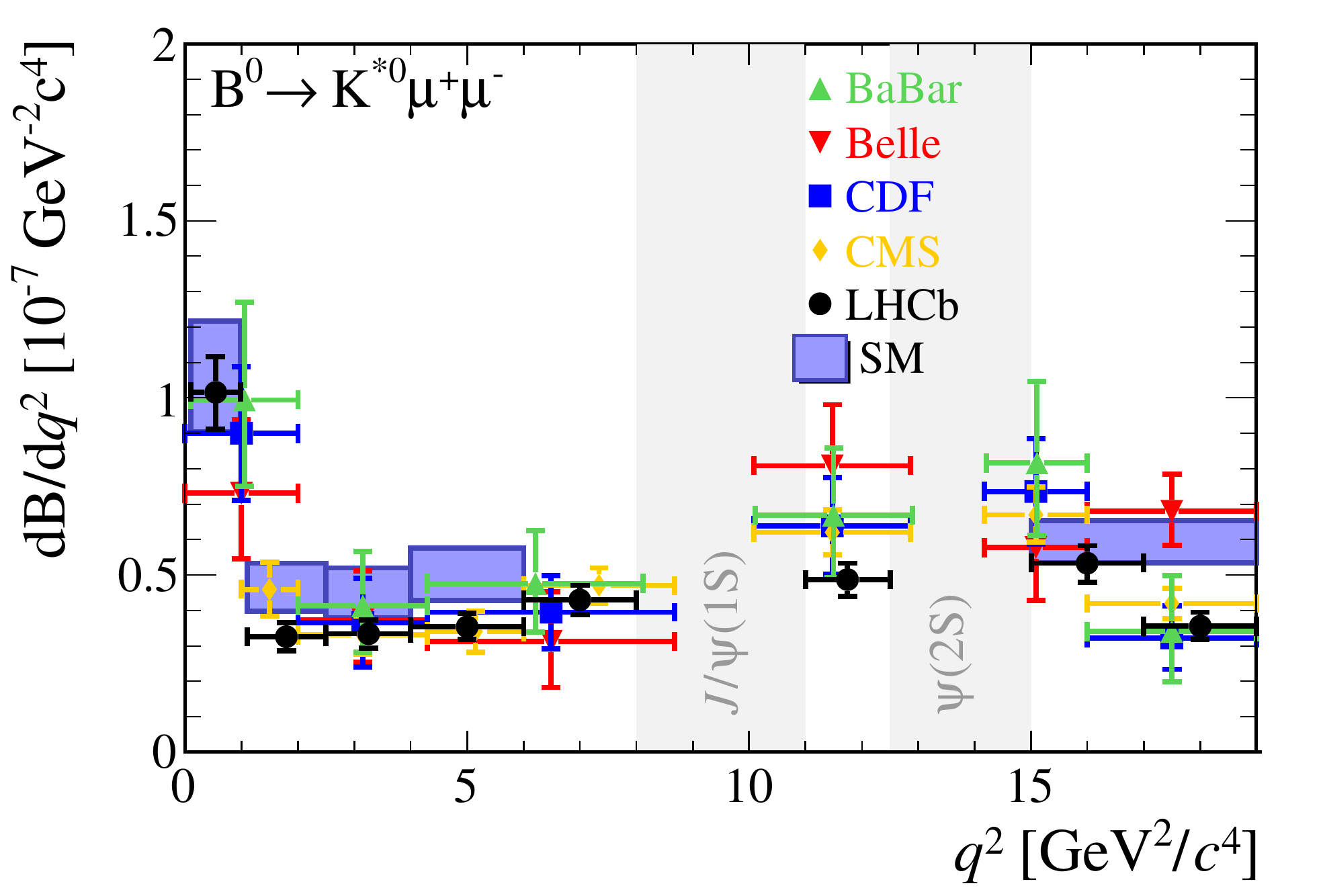}
\includegraphics[width=0.49\textwidth]{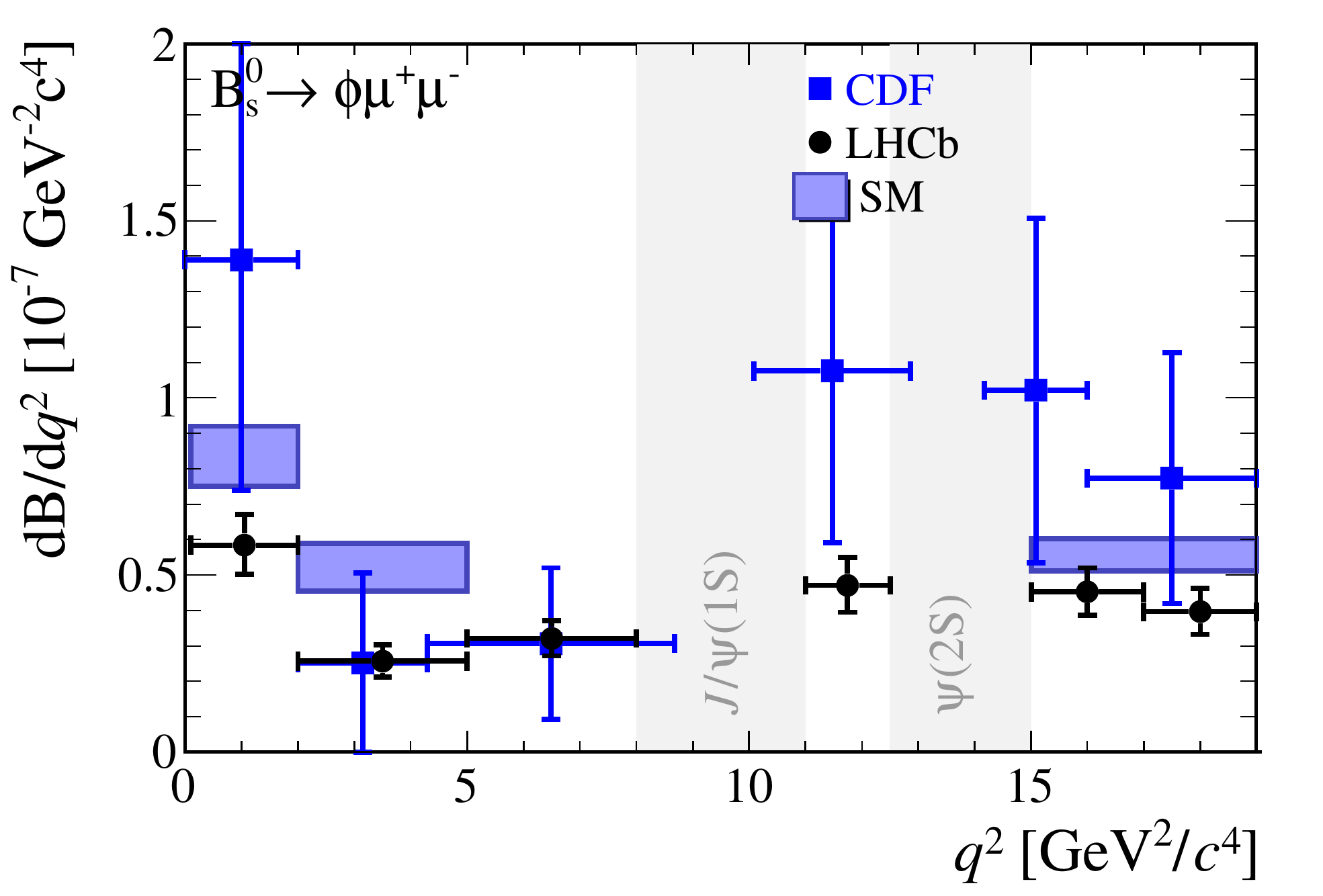}
\includegraphics[width=0.49\textwidth]{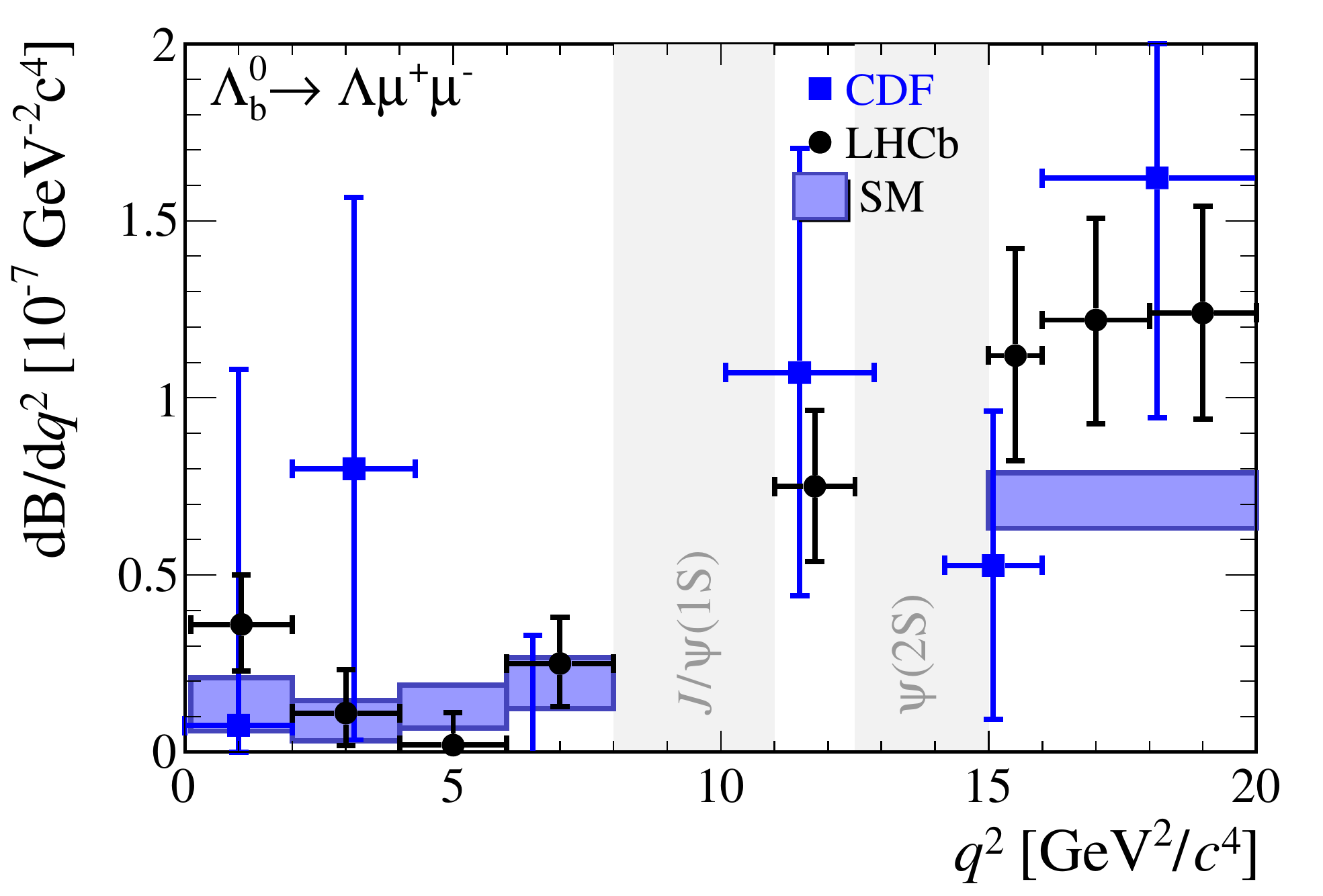}
\caption{Status of the branching fraction measurements for the exclusive decays (top left) $\decay{\Bu}{\Kp\mumu}$, (top right) $\decay{\Bd}{\Kstarz\mumu}$,   (bottom left) $\decay{\Bs}{\phi\mumu}$, and (bottom right) $\decay{\Lb}{\Lambda\mumu}$.
    The experimental data are from the 
    BaBar~\cite{Lees:2012tva}, Belle~\cite{Wei:2009zv}, CDF~\cite{Aaltonen:2011qs}, CMS~\cite{Khachatryan:2015isa} and LHCb~\cite{Aaij:2014pli,Aaij:2015esa,Aaij:2016flj,Aaij:2015xza} collaborations, respectively.
    The measurements by the $B$-factory experiments combine electron and muon lepton flavours and the charged and neutral isospin partners. 
    The SM predictions are taken from Refs.~\cite{Straub:2018kue,Straub:2015ica,Horgan:2013hoa,Horgan:2015vla}.\label{fig:btosmm_bfs}}
\end{figure}

Figure~\ref{fig:btosmm_bfs} summarises the experimental status for the branching fractions of the decays $\decay{\Bu}{\Kp\mumu}$~\cite{Aaij:2014pli}, $\decay{\Bd}{\Kstarz\mumu}$~\cite{Aaij:2014pli}, $\decay{\Bd}{\KS\mumu}$~\cite{Aaij:2014pli}, $\decay{\Bs}{\phi\mumu}$~\cite{Aaltonen:2011qs,Aaij:2015esa}, and $\decay{\Lb}{\Lambda\mumu}$~\cite{Aaltonen:2011qs,Aaij:2015xza}. 
The differential branching fractions ${\rm d}{\cal B}/{\rm d}q^2$ are given in bins of $q^2$, the invariant mass of the dilepton system squared. 
The $q^2$-regions that contain the charmomium resonances $\jpsi$ and $\psitwos$ are excluded,
as the $\decay{B}{X\jpsi(\psitwos)}$ tree-level decays dominate in these regions. 
For the highest statistics rare modes $\decay{\Bu}{\Kp\mumu}$ and $\decay{\Bd}{\Kstarz\mumu}$ measured by LHCb in Refs.~\cite{Aaij:2014pli,Aaij:2016flj},
the $q^2$ region $0.98<q^2<1.1\gevgevcccc$ containing the $\phi$ resonance (${\cal B}(\decay{\phi}{\mumu})=(2.86\pm0.19)\times 10^{-4}$~\cite{pdg2020}) is excluded as well. 
The measurements from the $B$-factory experiments combine the two lepton flavours and furthermore the two isospin partners, \ie\ the rare decays that differ only by the light ($u$ or $d$) spectator quark. 

The SM predictions from Refs.~\cite{Straub:2018kue,Straub:2015ica,Horgan:2013hoa,Horgan:2015vla} are shown as shaded boxes. 
The data lie generally below the SM predictions, in particular at low $q^2$. 
The largest tension is found for the decay $\decay{\Bs}{\phi\mumu}$~\cite{Aaij:2015esa}, corresponding to around $3\,\sigma$ in the $q^2$ range $1<q^2<6\gevgevcccc$.
The situation for the decay $\decay{\Lb}{\Lambda\mumu}$ is less clear, it was however recently pointed out in Ref.~\cite{Blake:2019guk}, that the $\Lb$ hadronisation fraction $f_{\Lb}$ used in Ref.~\cite{Aaij:2015xza} included results from LEP and Tevatron experiments. 
Updating $f_{\Lb}$ to only include the Tevatron average 
lowers the measured branching fraction significantly.  
For the other $\decay{b}{s\ellell}$ decay modes the tensions range from around $1$--$3\,\sigma$. 
These tensions are primarily driven by the precise measurements of the muonic modes by the LHCb collaboration and constitute one of the \textit{flavour anomalies} in the rare decays. 
It should be noted, that the SM prediction of the branching fractions are affected by significant hadronic uncertainties,
both from the non-perturbative form-factor calculations, and, more critically, from the nonlocal contributions discussed in Sec.~\ref{sec:btosll:theory}. 

At hadron colliders, the branching fractions of rare \btosll\ decays are typically measured relative to the corresponding $\decay{B}{X\jpsi(\to\ellell)}$ charmonium decays.  
For the example of the rare decay $\decay{\Bu}{\Kp\mumu}$ this results in the differential branching fraction for a $q^2$ bin of width $q^2_{\rm max}-q^{2}_{\rm min}$ given by
\begin{align}
  \frac{{\rm d}{\cal B}(\decay{\Bu}{\Kp\mumu}) }{{\rm d}q^2} =& \frac{N(\decay{\Bu}{\Kp\mumu})}{N(\decay{\Bu}{\Kp\jpsi})}\frac{\epsilon(\decay{\Bu}{\Kp\jpsi})}{\epsilon(\decay{\Bu}{\Kp\mumu})} \frac{{\cal B}(\decay{\Bu}{\Kp\jpsi}) {\cal B}(\decay{\jpsi}{\mumu})}{q^2_{\rm max}-q^{2}_{\rm min}}, 
\end{align}
where $N(\decay{\Bu}{\Kp\mumu}$ and $N(\decay{\Bu}{\Kp\jpsi}$ are the yields of signal and normalisation mode and $\epsilon(\decay{\Bu}{\Kp\mumu})$ and $\epsilon(\decay{\Bu}{\Kp\jpsi})$ the corresponding efficiencies.  
As the charmonium decays are reconstructed in the same final state as the rare signal modes,
many experimental systematic uncertainties cancel in the efficiency ratio. 
The charmonium decays are furthermore also used to control Monte Carlo simulation, \eg\ to determine trigger efficiencies through data-driven techniques~\cite{Aaij:2012me}.
The branching fractions of the normalisation modes $\decay{\Bu}{\Kp\jpsi}$, $\decay{\Bu}{\Kstarz\jpsi}$,  $\decay{\Bs}{\phi\jpsi}$, and $\decay{\Lb}{\Lambda\jpsi}$ are currently determined at relative precisions of $2.7\%$, $3.9\%$, $4.5\%$ and $7.4\%$, respectively~\cite{pdg2020}, which constitutes the dominant experimental systematic uncertainty for these measurements.  
Some of the measurements of rare \btosmm\ decays (\eg\ the branching fraction of the rare decay $\decay{\Bu}{\Kp\mumu}$) are therefore already at this point systematically limited by the precision on the branching fraction of the corresponding normalisation mode, which is fully correlated between all $q^2$ bins.  
At the $B$-factory experiments the branching fractions are instead normalised via the $B$-counting approach which allows to determine the number of produced $B$-mesons to the sub-percent level~\cite{Bevan:2014iga}. 
Improved measurements of the branching fractions of the normalisation modes at Belle~II seem feasible and would be highly desirable. 

The decay $\decay{\Bd}{\Kstarz\ellell}$ involves the spin $1$ vector meson $\Kstarz(892)$,
which experimentally can be reconstructed in the ${\Kp\pim}$ final state. 
Contributions to this final state can also arise from the so-called S-wave, where the $\Kp\pim$ system is in a spin $0$ configuration, 
resulting from either the decay of spin $0$ resonances or via non-resonant decay. 
The S-wave contribution can be separated from the \Kstarz\ P-wave by exploiting their different angular distributions. 
Using an angular analysis, the fraction of S-wave ($F_S$) has been determined to be ${\cal O}(5\%)$ in the $m_{K\pi}$ region of $100\mevcc$ around the known $\Kstarz$ mass, depending on the $q^2$ region~\cite{Aaij:2016flj}. 

\paragraph{Isospin asymmetries}
The isospin asymmetry 
is the asymmetry of the partial widths $\Gamma$ of the isospin partners that only differ by the light spectator quark. 
For the rare decays $\decay{B^{0}}{K^{(*)0}\ellell}$ and $\decay{B^{+}}{K^{(*)+}\ellell}$ it is defined as
\begin{align}
  {\cal A}_{\rm I} =& \frac{\Gamma(\decay{B^{0}}{K^{(*)0}\ellell})-\Gamma(\decay{B^{+}}{K^{(*)+}\ellell})}{\Gamma(\decay{B^{0}}{K^{(*)0}\ellell})+\Gamma(\decay{B^{+}}{K^{(*)+}\ellell})}\\
  =& \frac{{\cal B}(\decay{B^{0}}{K^{(*)0}\ellell})-(\tau_0/\tau_+){\cal B}(\decay{B^{+}}{K^{(*)+}\ellell})}{{\cal B}(\decay{B^{0}}{K^{(*)0}\ellell})+(\tau_0/\tau_+){\cal B}(\decay{B^{+}}{K^{(*)+}\ellell})},\nonumber
\end{align}
where $\tau_{0,+}$ denote the $\Bd$ and $\Bu$ lifetimes.
As form-factor uncertainties cancel at leading order in the ratio, the isospin asymmetries are precisely predicted to be at the ${\cal O}(1\%)$ level in the SM~\cite{Feldmann:2002iw,Khodjamirian:2012rm,Lyon:2013gba}. 
For the vector-meson decays $\decay{B}{K^{*}\ellell}$ the isospin asymmetry rises towards the photon pole at low $q^2$, and is sizeable for radiative modes as discussed in Sec.~\ref{sec:radiative_isospin}. 

\begin{figure}
  \centering
  \includegraphics[width=0.49\textwidth]{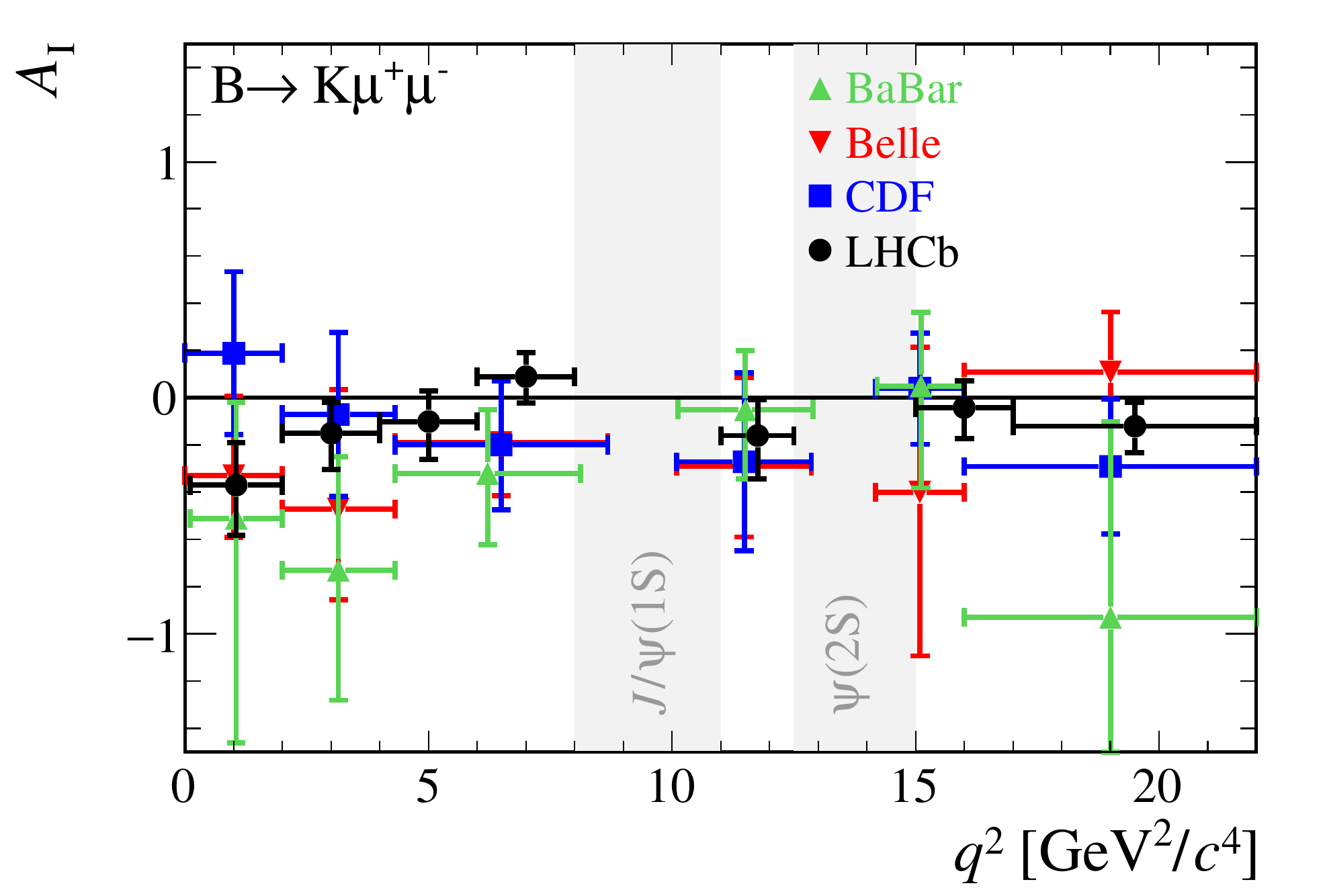}
  \includegraphics[width=0.49\textwidth]{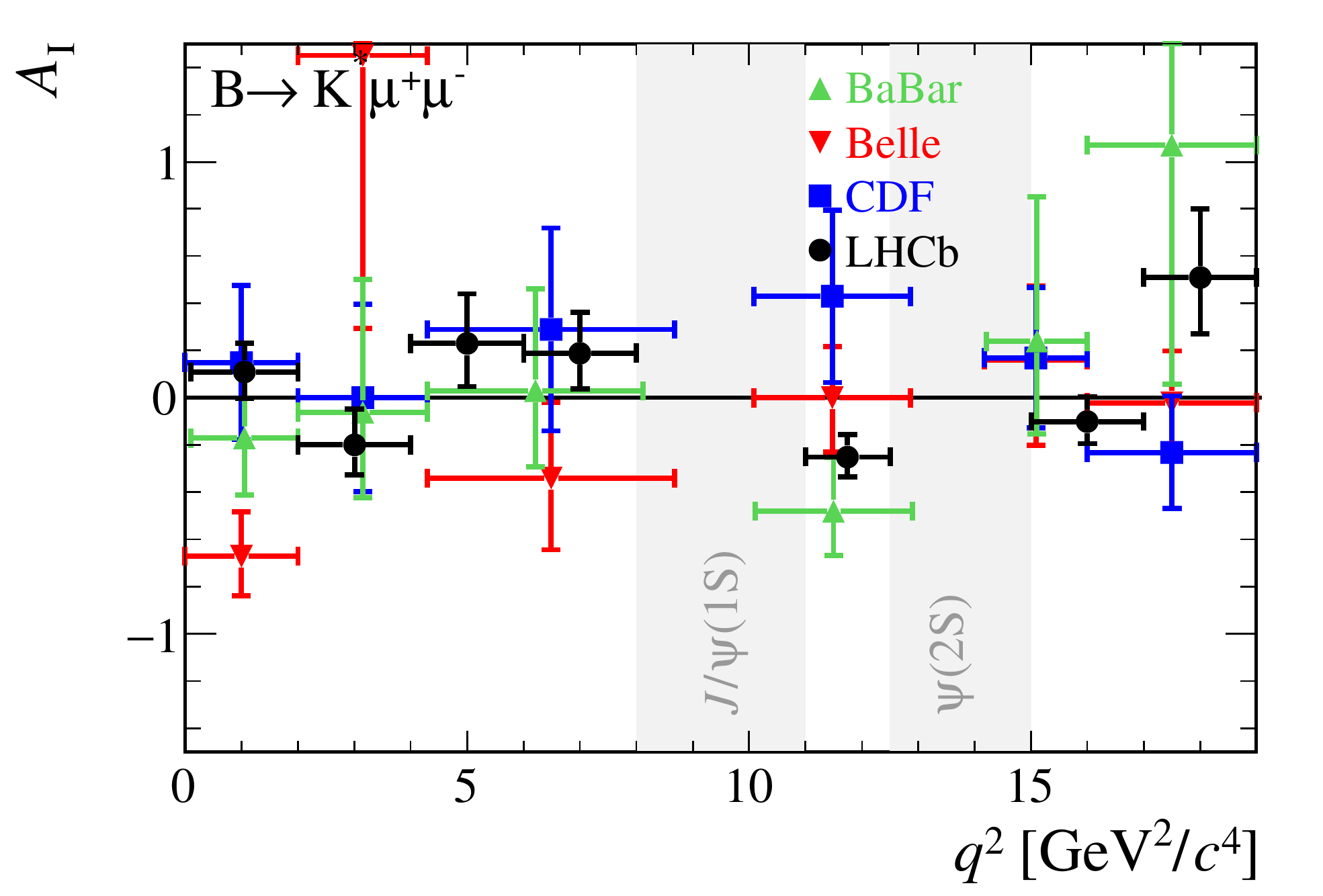}
  \caption{Isospin asymmetries of the decays (left) $\decay{B^{(0,+)}}{K^{(0,+)}\mumu}$ and (right) $\decay{B^{(0,+)}}{K^{*(0,+)}\mumu}$.
    The experimental data are from the 
    BaBar~\cite{Lees:2012tva}, Belle~\cite{Wei:2009zv}, CDF~\cite{Miyake:2012exl},
    and LHCb~\cite{Aaij:2014pli,Aaij:2016flj} collaborations. 
    The measurements by the $B$-factory experiments combine electron and muon lepton flavours. 
    The SM prediction is at the percent level~\cite{Feldmann:2002iw,Khodjamirian:2012rm,Lyon:2013gba}.\label{fig:isospinasymmetries}}
\end{figure}

Figure~\ref{fig:isospinasymmetries} shows the available experimental data from Refs.~\cite{Lees:2012tva,Wei:2009zv,Miyake:2012exl,Aaij:2014pli,Aaij:2016flj}. 
The measurements are found to be in agreement with the SM prediction. 
Though most experimental data for ${\cal A}_{\rm I}(\decay{B}{K\ellell})$ is found to lie below the SM prediction, 
the most precise measurement of ${\cal A}_{\rm I}(\decay{B}{K\mumu})$ by the LHCb collaboration is in good agreement with the SM at $1.5\,\sigma$~\cite{Aaij:2014pli}. 

\paragraph{\CP\ asymmetries}
Similarly to isospin asymmetries, direct \CP\ asymmetries of rare $\decay{b}{s\ellell}$ decays defined as
\begin{align}
{\cal A}_{\CP} =& \frac{\Gamma(\decay{\bar{B}}{\bar{K}^{(*)}\ellell})-\Gamma(\decay{B}{K^{(*)}\ellell})}{\Gamma(\decay{\bar{B}}{\bar{K}^{(*)}\ellell})+\Gamma(\decay{B}{K^{(*)}\ellell})},
\end{align}
can be predicted precisely in the SM due to cancellation of hadronic uncertainties. 
The SM prediction is expected to be at the level of ${\cal O}(10^{-3})$~\cite{Altmannshofer:2008dz}.

\begin{figure}
  \centering
  \includegraphics[width=0.49\textwidth]{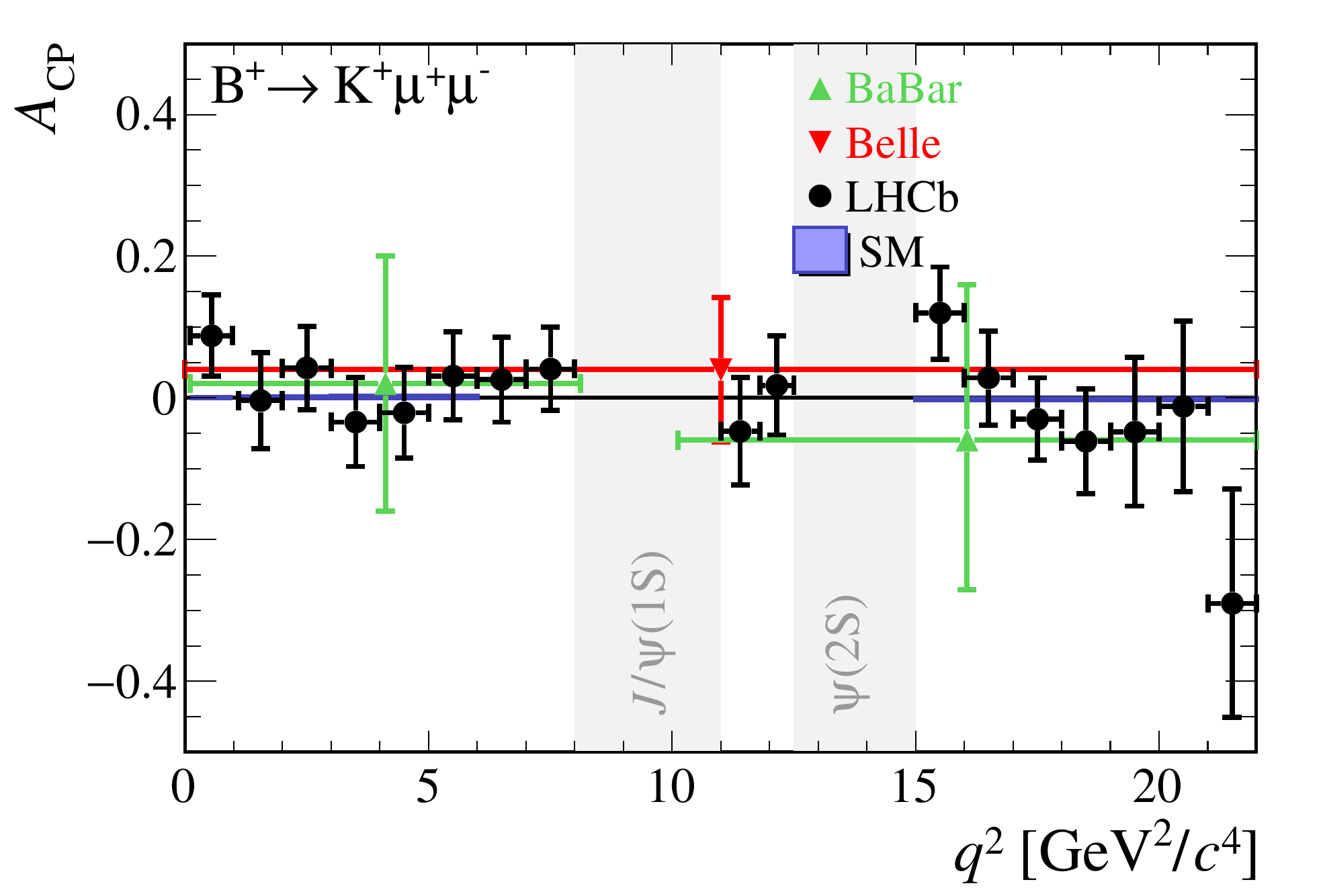}
  \includegraphics[width=0.49\textwidth]{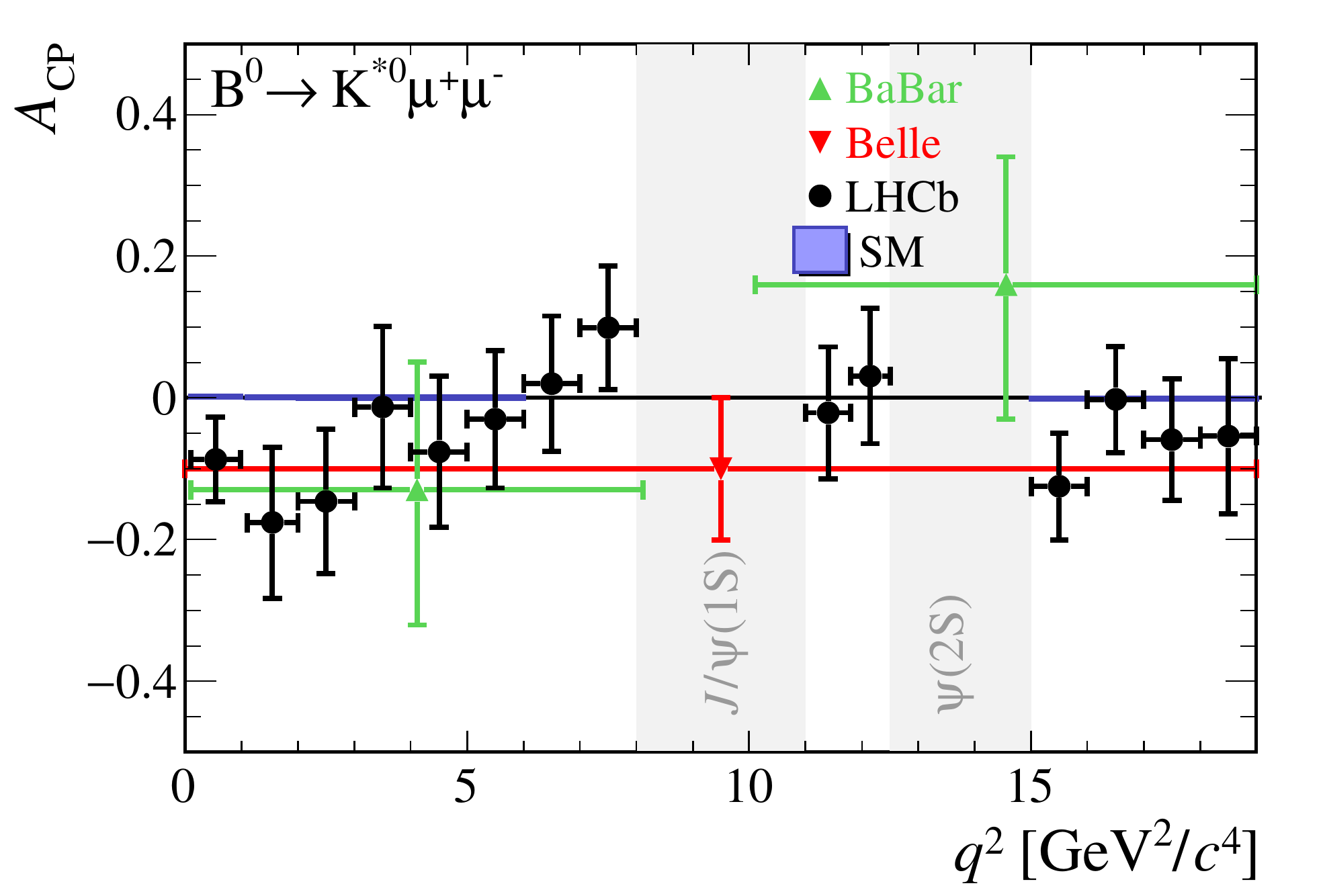}
  \caption{
    \CP-asymmetries of the decays (left) $\decay{B^{+}}{K^{+}\mumu}$ and (right) $\decay{B^{0}}{K^{0}\mumu}$.
    The experimental data are from the 
    BaBar~\cite{Lees:2012tva}, Belle~\cite{Wei:2009zv}, 
    and LHCb~\cite{Aaij:2014bsa} collaborations. 
    The measurements by the $B$-factory experiments combine electron and muon lepton flavours and isospin partners. 
    The SM prediction is at the ${\cal O}(10^{-3})$ level~\cite{Straub:2018kue,Straub:2015ica,Horgan:2013hoa,Horgan:2015vla}.\label{fig:cpasymmetries}}
\end{figure}

Experimentally, LHCb determines ${\cal A}_{\CP}(\decay{B}{K^{(*)}\mumu})$ relative to the corresponding $\decay{B}{K^{(*)}\jpsi}$ charmonium modes (for which ${\cal A}_{\CP}$ is known to be small, \eg\ ${\cal A}_{\CP}(\decay{\Bu}{\Kp\jpsi})=(1.8\pm3.0)\times 10^{-3}$~\cite{pdg2020}) via
\begin{align}
{\cal A}_{\CP}(\decay{B}{K^{(*)}\mumu}) =& {\cal A}_{\CP}^{\rm raw}(\decay{B}{K^{(*)}\mumu}) - {\cal A}_{\CP}^{\rm raw}(\decay{B}{K^{(*)}\jpsi})  ,
\end{align}
as production asymmetries ${\cal A}_{P}$ and detection asymmetries ${\cal A}_D$ affect the modes similarly and thus largely cancel. 
Figure~\ref{fig:cpasymmetries} shows the experimental data from the 
BaBar~\cite{Lees:2012tva}, Belle~\cite{Wei:2009zv}, 
and LHCb~\cite{Aaij:2014bsa} collaborations, which are in good agreement with the SM prediction~\cite{Straub:2018kue,Straub:2015ica,Horgan:2013hoa,Horgan:2015vla}. 
Integrated over the full $q^2$ range, the most precise measurement by the LHCb collaboration results in
\begin{align}
  {\cal A}_{\CP}(\decay{\Bu}{\Kp\mumu}) =& -0.035\pm 0.024_{\rm stat} \pm 0.003_{\rm syst},\\
  {\cal A}_{\CP}(\decay{\Bd}{\Kstarz\mumu}) =& +0.012 \pm 0.017_{\rm stat} \pm 0.001_{\rm syst}\nonumber  
\end{align}
in excellent agreement with the SM~\cite{Aaij:2014bsa}. 

\paragraph{Branching fractions of \btodll\ processes}
Compared to rare $\decay{b}{s\ellell}$ processes, the short-distance $\decay{b}{d\ellell}$ electroweak penguin amplitudes are further CKM suppressed by $|V_{\rm td}/V_{\rm ts}|\approx 0.209$~\cite{Charles:2004jd},
resulting in branching fractions of ${\cal O}(10^{-8})$ in the SM. 
The $\decay{b}{d\ellell}$ decay $\decay{\Bu}{\pip\mumu}$ has been first observed by the LHCb collaboration~\cite{Aaij:2015nea}. 
In addition, also first evidence for the decays $\decay{\Bd}{\pip\pim\mumu}$ (in the $\pip\pim$ mass region containing the $\rho^0(770)$ vector meson) and $\decay{\Bs}{\Kstarzb\mumu}$ has been found~\cite{Aaij:2014lba,Aaij:2018jhg}. 
Furthermore, the baryonic $\decay{b}{d\ellell}$ decay $\decay{\Lb}{p\pi^-\mumu}$ was observed by the LHCb collaboration~\cite{Aaij:2017ewm}. 
The corresponding branching fractions are measured to be~\cite{Aaij:2015nea,Aaij:2014lba,Aaij:2018jhg,Aaij:2017ewm} 
\begin{align}
  {\cal B}(\decay{\Bu}{\pip\mumu}) =& (1.83\pm 0.24_{\rm stat} \pm 0.05_{\rm syst})\times 10^{-8},\\
  {\cal B}(\decay{\Bd}{\pip\pim\mumu}) =& (2.11 \pm 0.51_{\rm stat} \pm 0.22_{\rm syst})\times 10^{-8}, \nonumber\\
  {\cal B}(\decay{\Bs}{\Kstarzb\mumu}) =& (2.9 \pm 1.0_{\rm stat} \pm 0.4_{\rm syst})\times 10^{-8}, ~~~\text{and}\nonumber\\
  {\cal B}(\decay{\Lb}{p\pi^-\mumu}) =& (6.9\pm 1.9_{\rm stat}\pm {1.1}_{\rm syst}{{}^{+1.3}_{-1.0}}_{\rm norm})\times 10^{-8}.\nonumber
\end{align}
The measurements are in good agreement with SM predictions~\cite{Ali:2013zfa,Hambrock:2015wka,Bailey:2015nbd,Beneke:2004dp,Wu:2006rd,Faustov:2013pca,Kindra:2018ayz},
with the $\decay{\Bu}{\pip\mumu}$ data slightly below the predictions in Refs.~\cite{Hambrock:2015wka,Bailey:2015nbd},
consistent with the observed tensions in $\decay{b}{s\ellell}$ decays.
Note that the first $q^2$ bin in the decay $\decay{\Bu}{\pip\mumu}$ includes possible contributions from the light resonances $\rho^0$ and $\omega$, \ie\ from $\decay{\Bu}{\rho^0(\omega)\pip}$ decays~\cite{Aaij:2015nea}. 

The direct \CP\ asymmetry ${\cal A}_{\CP}$ in $\decay{b}{d\ellell}$ decays is expected to be larger than in $\decay{b}{s\ellell}$ decays, 
as non-local contributions proportional to $V_{ub}V_{ud}^*$ and $V_{\rm cb}V_{\rm cd}^*$ have the same Cabbibo suppression as the short-distance amplitude and different (CKM and strong) phases~\cite{Hambrock:2015wka}. 
The LHCb collaboration finds for the decay $\decay{\Bu}{\pip\mumu}$~\cite{Aaij:2015nea}
\begin{align}
{\cal A}_{\CP}(\decay{\Bu}{\pip\mumu}) =& -0.11\pm 0.12_{\rm stat} \pm 0.01_{\rm syst}
\end{align}
in good agreement with the SM prediction~\cite{Hambrock:2015wka}. 

\subsubsection{Angular analyses}
\label{sec:angularanalyses}
Besides searches for NP through branching fraction measurements,
semileptonic $\decay{b}{s\ellell}$ decays furthermore allow to probe the SM through   
analyses of the angular distributions of the final state particles.
The theory expressions within the full basis of dimension-six WET operators
for the entire set of angular observables in $\bar{B}\to \lbrace P,V\rbrace$ transitions
are discussed in Ref.~\cite{Bobeth:2012vn}.
Both the angular distributions of rare 
decays with pseudoscalar mesons in the final state (denoted by $\decay{B}{P\ellell}$),
as well as decays with final state vector mesons (denoted by $\decay{B}{V\ellell}$) have been studied experimentally~\cite{Aaij:2014tfa,Aubert:2006vb,Wei:2009zv,Aaltonen:2011ja,Sirunyan:2018jll,Aaboud:2018krd,Lees:2015ymt,Wehle:2016yoi,Khachatryan:2015isa,Sirunyan:2017dhj,Sirunyan:2020hlk,Aaij:2020nrf,Aaij:2015oid,Aaij:2015esa}. 

\paragraph{Angular analyses of $\decay{B}{P\ellell}$ decays}
Rare $\decay{B}{P\ellell}$ decays are described by a single decay angle denoted by $\thetal$, which is defined as the angle between direction of the $\ell^+$ and the direction of the pseudoscalar final state meson in the dilepton rest-frame. 
The differential decay rate of the decay $\decay{B}{K\ellell}$, depending on $\ctl$ and $q^2$, is given by
\begin{align}
  \frac{{\rm d}^2\Gamma(\decay{B}{K\ellell})}{{\rm d}q^2\,{\rm dcos}\thetal} =& \frac{3}{4}(1-F_{\rm H})(1-\cos^2\thetal) + \frac{1}{2}F_{\rm H} + A_{\rm FB}\ctl,
\end{align}
where $F_{\rm H}$ denotes the so-called flat term and $A_{\rm FB}$ the forward-backward asymmetry~\cite{Bobeth:2007dw}. 
Both $F_{\rm H}$ and $A_{\rm FB}$ are negligibly small in the SM, but can be enhanced through 
new (pseudo)scalar or tensor contributions. 

The angular distributions of the decay $\decay{B}{K\ellell}$ have been analysed by BaBar, Belle, CDF, CMS and LHCb~\cite{Aaij:2014tfa,Aubert:2006vb,Wei:2009zv,Aaltonen:2011ja,Sirunyan:2018jll}. 
The results are in good agreement with each other and the SM predictions. 
Due to the closeness of the SM point of $A_{\rm FB}$ and $F_{\rm H}$ to physical parameter boundaries, 
care must be taken when evaluating the experimental parameter uncertainties, \eg\ by employing coverage correction techniques~\cite{Feldman:1997qc}. 

\begin{figure}
  \centering
\includegraphics[width=0.49\textwidth]{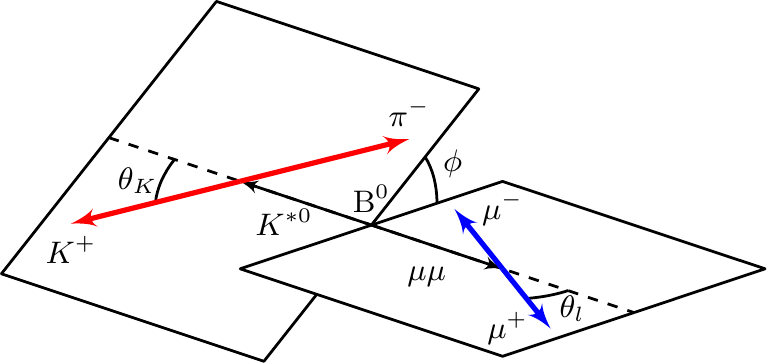}
\caption{Angular conventions for the rare $\decay{B}{V\ellell}$ decay $\decay{\Bd}{\Kstarz(\to \Kp\pim)\ellell}$.
  Note that the angles \ctl\ and \ctk\ are shown in the rest frames of the $\mumu$ and $\Kstarz$ systems,
  where the momenta of $\mup$ and $\mun$ ($\Kp$ and $\pim$) are back-to-back.\label{fig:decayangles}}
\end{figure}

\paragraph{Angular analyses of $\decay{B}{V\ellell}$ decays}
Rare $\decay{B}{V\ellell}$ decays exhibit a more complex angular structure compared to $\decay{B}{P\ellell}$ decays. 
For the decay $\decay{\Bd}{\Kstarz(\to \Kp\pim)\ellell}$ the final state is described by the three decay angles $\thetal$, $\phi$, and $\thetak$ in addition to $q^2$ and, unless the width of the $\Kstarz$ is neglected, $m_{K\pi}$.
The three decay angles are sketched in Fig.~\ref{fig:decayangles}. 
The angle $\thetal$ is defined as the angle between the direction of the $\mup$ ($\mun$) and the direction opposite the $\Bd$ ($\Bdb$) in the rest frame of the dilepton system. 
The angle $\thetak$ is the angle between the direction of the $\Kp$ ($\Km$) and the direction opposite the $\Bd$ ($\Bdb$) in the rest frame of the $\Kstarz$ ($\Kstarzb$) system. 
The angle $\phi$ is the angle between the plane defined by the diumon pair and the plane defined by the kaon and the pion in the $\Bd$ ($\Bdb$) rest frame.
It should be noted, that this is the experimental convention for the decay angles as detailed in the Appendix of Ref.~\cite{Aaij:2013iag},
and that different angular conventions exist in the theory literature (\eg Refs.~\cite{Altmannshofer:2008dz,Bobeth:2008ij}),
for a detailed discussion see the Appendix of Ref.~\cite{Gratrex:2015hna}. 

The four-differential angular decay rate for the decays $\decay{\Bdb}{\Kstarzb\ellell}$ and $\decay{\Bd}{\Kstarz\ellell}$ is given by~\cite{Altmannshofer:2008dz}
\begin{align}
\frac{{\rm d}^4\Gamma(\decay{\Bdb}{\Kstarzb\ellell})}{{\rm dcos}\thetal\,{\rm d}\phi\,{\rm dcos}\thetak\,{\rm d}q^2} =& \frac{9}{32\pi}\sum_i I_i(q^2) f_i(\ctl,\phi,\ctk)\label{eq:diffdecayrate}\\
\frac{{\rm d}^4\Gamma(\decay{\Bd}{\Kstarz\ellell})}{{\rm dcos}\thetal\,{\rm d}\phi\,{\rm dcos}\thetak\,{\rm d}q^2} =& \frac{9}{32\pi}\sum_i \bar{I}_i(q^2) f_i(\ctl,\phi,\ctk), \nonumber
\end{align}
where the $\parenbar{I}_i(q^2)$ are bilinear combinations of the six 
transversity amplitudes $A_{0,\parallel,\perp}^{L,R}$, where the supersripts $L$ and $R$ refer to the chirality of the leptonic current.
When accounting for non-negligible lepton masses and possible scalar and pseudoscalar contributions two additional decay amplitudes $A_t$ and $A_{\rm scalar}$ need to be introduced in addition~\cite{Altmannshofer:2008dz}.
Table~\ref{tab:Iis} explicitly gives the dependence on the transversity amplitudes and the definition for the corresponding angular terms $f_i(\ctl,\phi,\ctk)$. 
\begin{table}
{ \renewcommand*{\arraystretch}{1.7}
\begin{tabular}{cll}\hline
$i$ & $I_i(q^2)$ & $f_i(\ctl,\phi,\ctk)$ \\ \hline\hline
  $1s$ &  $\frac{2+\beta_\ell^2}{4} (|A_\perp^L|^2 + |A_\parallel^L|^2 + |A_\perp^R|^2 + |A_\parallel^R|^2) + \frac{4m_\mu^2}{\qsq} \Re{ \big[A_\perp^{L}A_\perp^{R*}+A_\parallel^{L}A_\parallel^{R*}\big]}$     &  $\sin^2\theta_K$ \\
$1c$ &  $  |A_{0}^{L}|^2 +|A_{0}^{R}|^2  + \frac{4m_\mu^2}{\qsq}\big(|A_t|^2 + 2 \Re{\big[A_0^L A_0^{R*}\big]}\big) +\beta_\ell^2 |A_\text{scalar}|^2 $  & $\cos^2\theta_K$\\
$2s$ &  $\frac{\beta_\ell^2}{4} \bigl(|A_\perp^L|^2 + |A_\parallel^L|^2 + |A_\perp^R|^2 + |A_\parallel^R|^2\bigr)$ & $\sin^2\theta_K\cos2\theta_\ell$\\
$2c$ &  $ -\beta_\ell^2 \bigl( |A_{0}^{L}|^2 + |A_{0}^{R}|^2 \bigr)$ & $\cos^2\theta_K\cos2\theta_\ell$\\
$3$ &  $\frac{1}{2} \beta_\ell^2 \bigl( |A_{\perp}^{L}|^2 -|A_{\parallel}^{L}|^2 + |A_{\perp}^{R}|^2 - |A_{\parallel}^{R}|^2  \bigr)$ & $\sin^2\theta_K\sin^2\theta_\ell\cos2\phi$\\
$4$ &  $\frac{1}{\sqrt{2}}\beta_\ell^2 \Re {\bigl[ A_{0}^{L}A_{\parallel}^{L*} + A_{0}^{R}A_{\parallel}^{R*} \bigr] }$ & $\sin2\theta_K\sin2\theta_\ell\cos\phi$\\
$5$ &  $\sqrt{2} \beta_\ell \Bigl( \Re{\bigl[A_{0}^{L}A_{\perp}^{L*} - A_{0}^{R}A_{\perp}^{R*}\bigr]}  -\frac{m_{\mu}}{\sqrt{q^2}} \Re{ \bigl[ A_{\parallel}^{L}A_\text{scalar}^{*}+A_{\parallel}^{R}A_\text{scalar}^{*}  \bigr]} \Bigr)$ & $\sin2\theta_K\sin\theta_\ell\cos\phi$\\
$6s$ & $ 2 \beta_\ell \Re{\bigl[  A_{\parallel}^{L}A_{\perp}^{L*} - A_{\parallel}^{R}A_{\perp}^{R*}  \bigr] } $ & $\sin^2\theta_K\cos\theta_\ell$ \\
$6c$ & $ 4\beta_\ell \frac{m_\mu}{\sqrt{\qsq}} \Re{ \bigl[A_0^L A_{\text{scalar}}^* + A_0^R A_{\text{scalar}}^*\bigr]}$ & $\cos^2\theta_K\cos\theta_\ell$\\
$7$ &  $\sqrt{2} \beta_\ell \Bigl( \Im{\bigl[  A_{0}^{L}A_{\parallel}^{L*} - A_{0}^{R}A_{\parallel}^{R*}  \bigr]} +\frac{m_{\mu}}{\sqrt{q^2}} \Im{ \bigl[ A_{\perp}^{L}A_\text{scalar}^{*}+A_{\perp}^{R}A_\text{scalar}^{*}  \bigr] } \Bigr)$ & $\sin2\theta_K\sin\theta_\ell\sin\phi$\\
$8$ &  $\frac{1}{\sqrt{2}} \beta_\ell^2 \Re{\bigl[  A_{0}^{L}A_{\perp}^{L*} + A_{0}^{R}A_{\perp}^{R*} \bigr] }$ & $\sin2\theta_K\sin2\theta_\ell\sin\phi$\\
$9$ &  $ \beta_\ell^2 \Im{\bigl[  A_{\parallel}^{L}A_{\perp}^{L*} + A_{\parallel}^{R}A_{\perp}^{R*}  \bigr] }$ & $\sin^2\theta_K\sin^2\theta_\ell\sin2\phi$\\
\hline
  $10$ & $\frac{1}{2} \Bigl(|A_{\rm S}^{L}|^2 + |A_{\rm S}^{R}|^2  + \frac{4m_\mu^2}{q^2} \bigl( |A_{t}|^2+2\Re{\bigl[A_{\rm S}^{L}A_{\rm S}^{R*}\bigr]}  \bigr) \Bigr)$ & $1$ \\
$11$  &  $\sqrt{3} \Bigl(\Re{\bigl[A_{\rm S}^{L}A_0^{L*} + A_{\rm S}^{R}A_0^{R*} + \frac{4m_\mu^2}{q^2}(A_{\rm S}^{L}A_{0}^{R*} + A_{\text{scalar},t}A_{t}^{*})\bigr]}+ \Re{\bigl[\frac{4m_\mu^2}{q^2}A_{0}^{L}A_{\rm S}^{R*}\bigr]} \Bigr)$  & $\ctk$  \\
$12$ & $- \frac{1}{2}\beta_\ell^2 \bigl(|A_{\rm S}^{L}|^2 + |A_{\rm S}^{R}|^2\bigr)$ & $\cos2\theta_\ell$\\
$13$  &  $-\sqrt{3}  \beta_\ell^2 \Re{\bigl[  A_{\rm S}^{L}A_0^{L*} + A_{\rm S}^{R}A_0^{R*}\bigr] } $ & $\ctk\cos2\theta_\ell$ \\
$14$ & $\sqrt{\frac{3}{2}}  \beta_\ell^2 \Re{ \bigl[ A_{\rm S}^{L}A_\parallel^{L*} + A_{\rm S}^{R}A_\parallel^{R*}\bigr]} $ & $\sin\theta_K \sin2\theta_\ell\cos\phi$\\
$15$ & $2\sqrt{\frac{3}{2}}  \beta_\ell \Re{\bigl[A_{\rm S}^{L}A_\perp^{L*} - A_{\rm S}^{R}A_\perp^{R*}\bigr] }$ & $\sin\theta_K\sin\theta_\ell\cos\phi$\\
$16$ & $2\sqrt{\frac{3}{2}}  \beta_\ell \Im{\bigl[A_{\rm S}^{L}A_\parallel^{L*} - A_{\rm S}^{R}A_\parallel^{R*}\bigr]}$ & $\sin\theta_K\sin\theta_\ell\sin\phi$\\
$17$ & $\sqrt{\frac{3}{2}}  \beta_\ell^2 \Im{\bigl[A_{\rm S}^{L}A_\perp^{L*} + A_{\rm S}^{R}A_\perp^{R*}\bigr]} $ & $\sin\theta_K\sin2\theta_\ell\sin\phi$  \\
\hline\end{tabular}}
\caption{
Combinations of transversity amplitudes $I_i(q^2)$ and angular terms $f_i(\ctl,\phi,\ctk)$. 
The factor $\beta_\ell$ is given by $\beta_\ell=\sqrt{1-4m_\ell^2/q^2}$. 
The coefficients $\bar{I}_i(q^2)$ are given by exchanging $A\to \bar{A}$, \ie\ by complex conjugation of all weak phases in the decay amplitudes. 
\label{tab:Iis}}
\end{table}

Using the angular coefficients $\parenbar{I}_i$ in Eq.~\ref{eq:diffdecayrate} it is possible to define the \CP-averaged angular observables $S_i$ and the \CP-asymmetries $A_i$ according to~\cite{Altmannshofer:2008dz}
\begin{align}
  S_i =& \frac{I_i + \bar{I}_i}{{\rm d}(\Gamma+\bar{\Gamma})/{\rm d}q^2},\label{eq:pwaveobservables}\\
  A_i =& \frac{I_i - \bar{I}_i}{{\rm d}(\Gamma+\bar{\Gamma})/{\rm d}q^2}.\nonumber
\end{align}
The normalisation factor 
is given by
\begin{align}
  \frac{{\rm d}(\Gamma+\bar{\Gamma})}{{\rm d}q^2} =& \tfrac{3}{4}(2I_1^s + I_1^c) - \tfrac{1}{4}(2I_2^s + I_2^c) + \tfrac{3}{4}(2\bar{I}_1^s + \bar{I}_1^c) - \tfrac{1}{4}(2\bar{I}_2^s + \bar{I}_2^c), \label{eq:pwavenorm}
\end{align}
which implies
\begin{align}
  \tfrac{3}{4}(2S_1^s+S_1^c)-\tfrac{1}{4}(2S_2^s+S_2^c)=&1.
\end{align}
Some of the angular observables are better known under more descriptive names,
namely the forward-backward asymmetry $A_{\rm FB}$ and the longitudinal polarisation fraction of the $\Kstarz$, $F_{\rm L}$, 
defined as
\begin{align}
  A_{\rm FB} =& \tfrac{3}{8}(2S_6^s + S_6^c) \overset{m_\ell=0}{=} \tfrac{3}{4}S_6^s ~~~\text{and}\\
  F_{\rm L} =& -S_2^c \overset{m_\ell=0}{=} S_1^c,\nonumber
\end{align}
where the second equalities are only valid when neglecting lepton masses. 
It is possible to construct ratios of observables in which the hadronic form-factor uncertainties cancel at leading order.
Examples for these optimised observables at low $q^2$ are the $P_i^{(\prime)}$ observables~\cite{Kruger:2005ep,Matias:2012xw,DescotesGenon:2012zf,Descotes-Genon:2013vna} given by
\begin{align}
  P_1 =& \frac{S_3}{2S_2^s} = A_{\rm T}^{(2)},\\
P_2 =& \frac{S_6^s}{8S_2^s},\nonumber\\
P_3 =& -\frac{S_9}{4S_2^s},\nonumber\\
P_{4,5,6,8}^\prime =& \frac{S_{4,5,7,8}}{2\sqrt{-S_2^sS_2^c}}.\nonumber
\end{align}
There is furthermore the $H_{\rm T}^{(i)}$ family of observables that are optimised for large $q^2$~\cite{Bobeth:2010wg,Bobeth:2012vn}. 
It should be noted that the experimental information in the different bases 
are equivalent in the asymptotic regime of large statistics, 
and it is possible to convert from one complete basis to another, if experimental correlations between the observables in a specific basis are provided. 

As discussed in Sec.~\ref{sec:btoslldecayrates} the final state $\Kp\pim\ellell$ 
also includes contributions from the S-wave, where the $\Kp\pim$ system is in a spin $0$ configuration. 
This irreducible background to the signal decay $\decay{\Bd}{\Kstarz\ellell}$ needs to be accounted for. 
The S-wave introduces two additional amplitudes $A_s^{L,R}$, resulting in eight additional angular coefficients
given in Tab.~\ref{tab:Iis} to account for the S-wave and the S-wave/P-wave interference~\cite{Hofer:2015kka}.
Experimentally, the determination of the S-wave fraction $F_S$ is of particular importance,
as the presence of the S-wave scales all P-wave observables by the factor $(1-F_S)$,
as the P-wave observables in Eq.~\ref{eq:pwaveobservables} are normalised with respect to the P-wave rate (Eq.~\ref{eq:pwavenorm}). 
The $m_{K\pi}$ distribution can be exploited to constrain the S-wave contribution. 

\begin{figure}
  \centering
\includegraphics[width=0.49\textwidth]{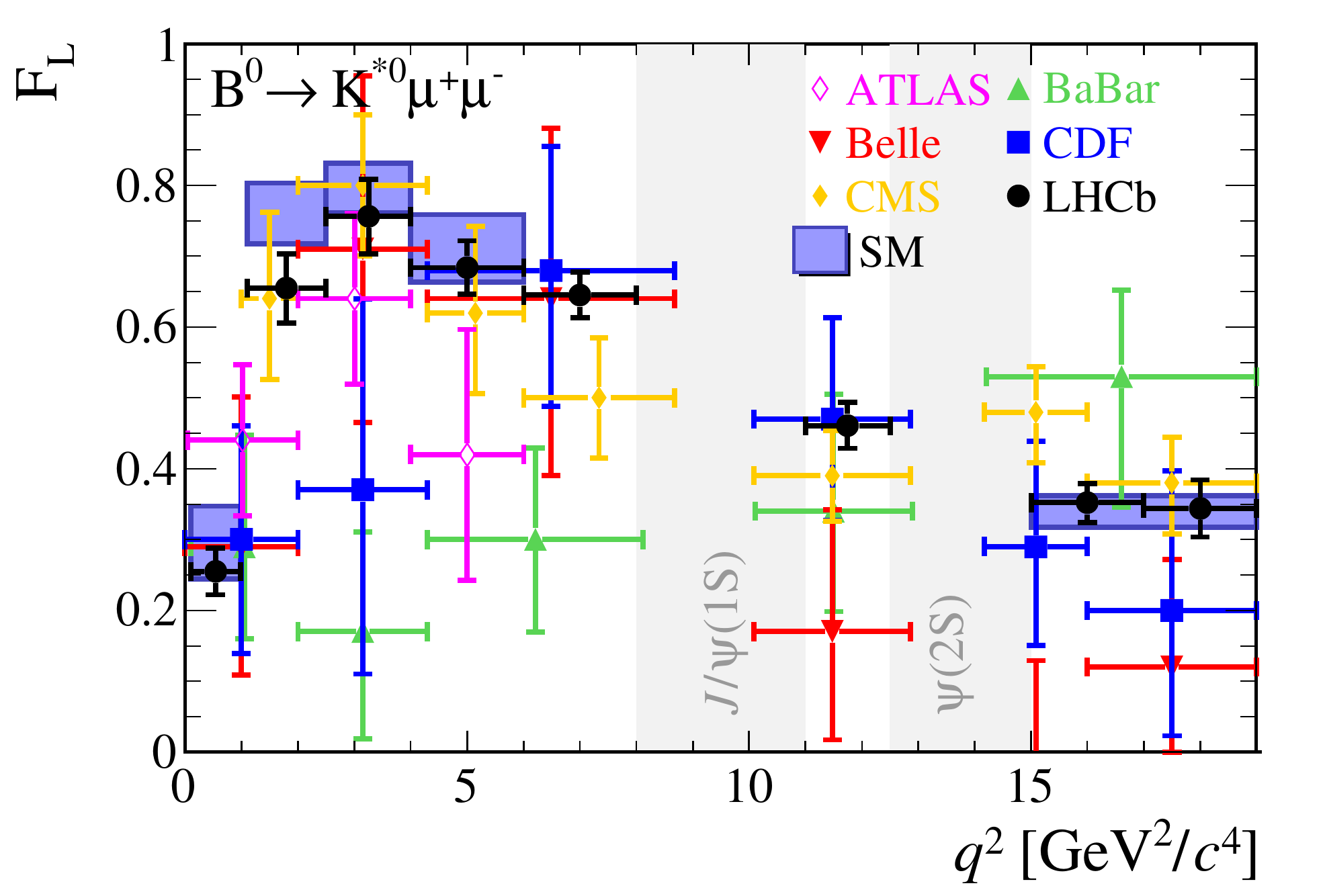}
\includegraphics[width=0.49\textwidth]{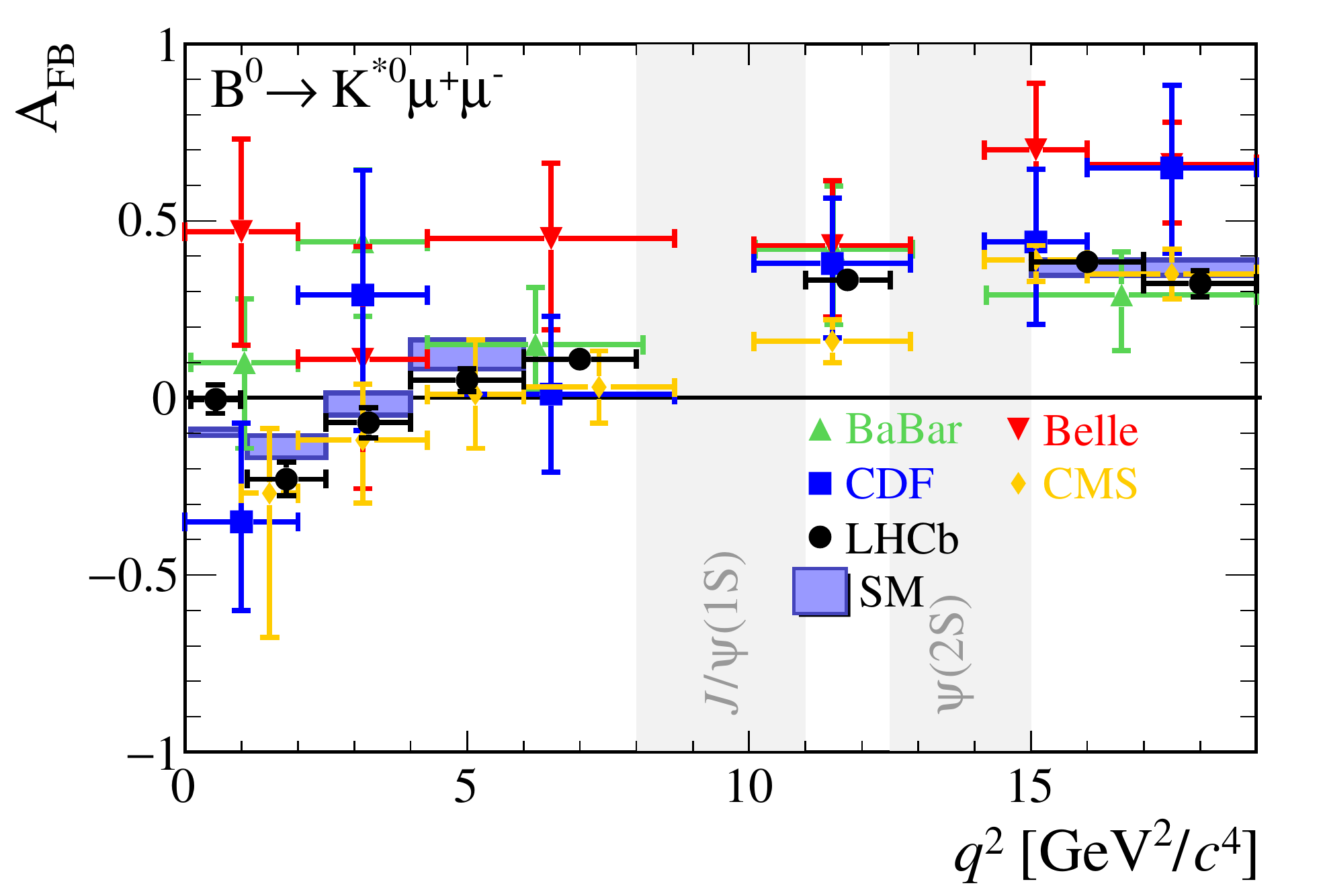}
  \caption{Angular observables (left) $F_{\rm L}$ and (right) $A_{\rm FB}$ as measured by the ATLAS~\cite{Aaboud:2018krd}, BaBar~\cite{Lees:2015ymt}, Belle~\cite{Wei:2009zv}, CDF~\cite{Aaltonen:2011ja}, CMS~\cite{Khachatryan:2015isa} and LHCb~\cite{Aaij:2020nrf} collaborations. The results from the $B$-factory experiments combine lepton flavours and isospin partners. The CDF results combine isospin partners. Overlaid is the SM prediction from Refs.~\cite{Straub:2018kue,Straub:2015ica,Horgan:2013hoa,Horgan:2015vla}.\label{fig:flafb}}
\end{figure}

\begin{figure}
  \centering
\includegraphics[width=0.49\textwidth]{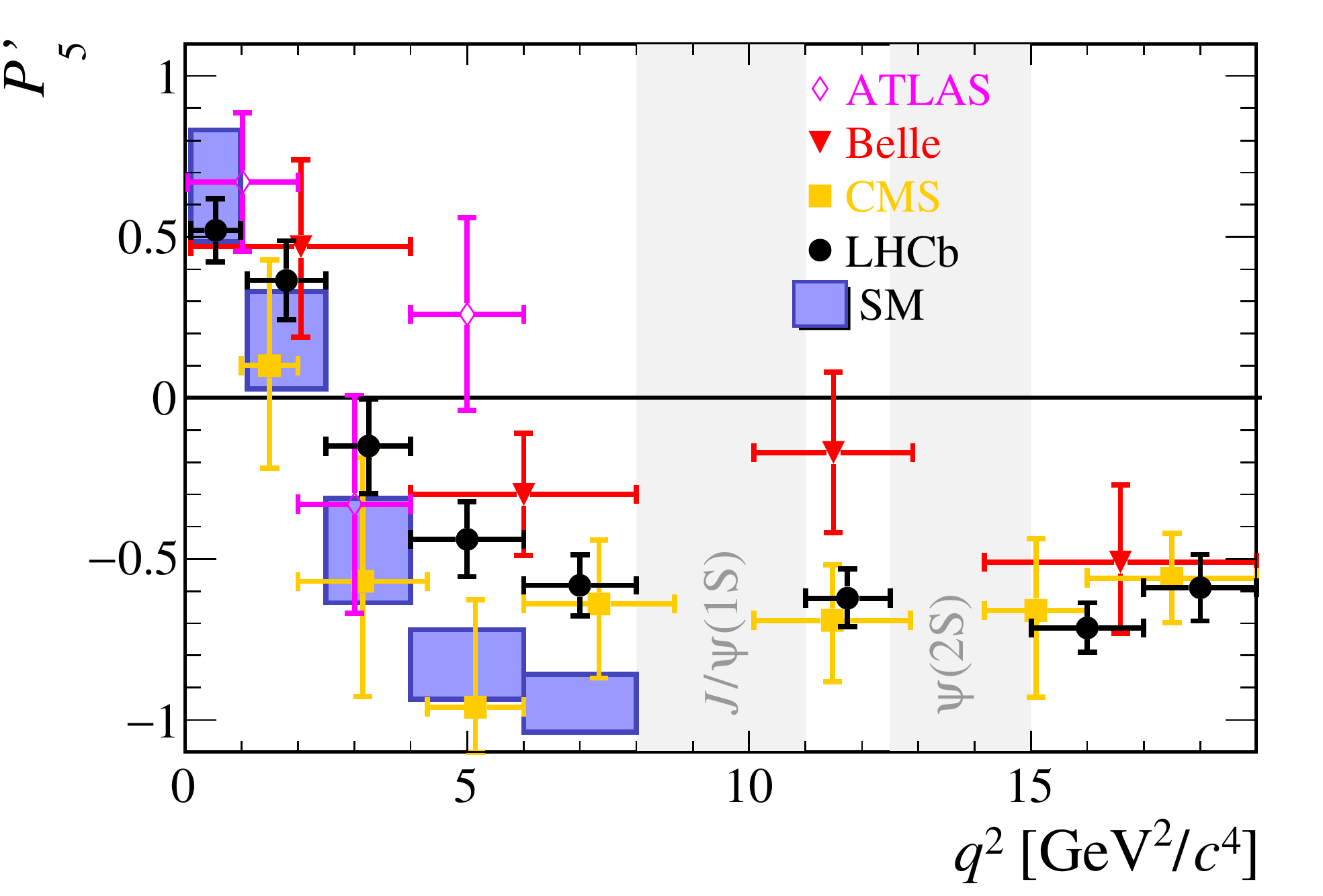}
  \caption{Angular observable $P_5^\prime$ as measured by the ATLAS~\cite{Aaboud:2018krd}, Belle~\cite{Wehle:2016yoi}, CMS~\cite{Sirunyan:2017dhj} and LHCb~\cite{Aaij:2020nrf} collaborations. Overlaid is the SM prediction from Refs.~\cite{Descotes-Genon:2014uoa,Khodjamirian:2010vf}. The result from Belle combines lepton flavours and isospin partners.\label{fig:s5p5p}}
\end{figure}

Angular analyses of the decay $\decay{B}{\Kstar\ellell}$ have been performed by the 
ATLAS~\cite{Aaboud:2018krd}, BaBar~\cite{Aubert:2006vb,Lees:2015ymt}, Belle~\cite{Wei:2009zv,Wehle:2016yoi}, CDF~\cite{Aaltonen:2011ja}, CMS~\cite{Khachatryan:2015isa,Sirunyan:2017dhj,Sirunyan:2020hlk}, and LHCb~\cite{Aaij:2015oid,Aaij:2020nrf} collaborations. 
The observables $F_{\rm L}$ and $A_{\rm FB}$ are shown in Fig.~\ref{fig:flafb}. 
Overall good agreement of the experimental data~\cite{Aaboud:2018krd,Lees:2015ymt,Wei:2009zv,Aaltonen:2011ja,Khachatryan:2015isa,Aaij:2020nrf} with the SM predictions~\cite{Straub:2018kue,Straub:2015ica,Horgan:2013hoa,Horgan:2015vla} is observed. 
For $F_{\rm L}$, some tension of the BaBar measurement~\cite{Lees:2015ymt} with the SM and the other experimental data can be seen at low $q^2$.
For $A_{\rm FB}$, the most precise measurements~\cite{Khachatryan:2015isa,Aaij:2020nrf} are found in good agreement with the SM, with the central values below the SM prediction at low $q^2$. 
Figure~\ref{fig:s5p5p} shows the angular observable~$P_5^\prime$. 
The most precise measurement by the LHCb collaboration~\cite{Aaij:2020nrf} shows tensions with the SM prediction at low $q^2$,
in the $q^2$ regions $4<q^2<6\gevgevcccc$ and $6<q^2<8\gevgevcccc$ these correspond to local significances of $2.5$ and $2.9\,\sigma$, respectively. 
An analysis of the full set of angular observables measured in Ref.~\cite{Aaij:2020nrf} indicates 
a global tension with the SM prediction corresponding to $3.3\,\sigma$\footnote{During the review of this manuscript an angular analysis of the decay $\decay{\Bu}{\Kstarp\mumu}$ by the LHCb collaboration was made public~\cite{Aaij:2020ruw}, which reports a consistent tension with the SM prediction at the level of $3.1\,\sigma$.}.  
Also the results from ATLAS~\cite{Aaboud:2018krd} and Belle~\cite{Wehle:2016yoi} indicate some tension 
in the low-$q^2$ region,
whereas the CMS result~\cite{Sirunyan:2017dhj} is in better agreement with the SM.  
The tensions with SM predictions in the angular observables of the decay $\decay{\Bd}{\Kstarz\ellell}$ constitutes one of the \textit{flavour anomalies}.

The formalism discussed in this section also applies to other $\decay{B}{V\ellell}$ decays like $\decay{\Bd}{\rhoz(\to\pip\pim)\ellell}$, $\decay{\Bs}{\Kstarzb(\to\Km\pip)\ellell}$, and $\decay{\Bs}{\phi(\to\Kp\Km)\ellell}$. 
So far, no angular analysis of the suppressed $\decay{b}{d\ellell}$ decays $\decay{\Bd}{\rhoz\ellell}$ and $\decay{\Bs}{\Kstarzb\ellell}$ have been performed. 
However, the LHCb collaboration has performed an angular analysis of the rare $\decay{b}{s\ellell}$ mode $\decay{\Bs}{\phi\mumu}$~\cite{Aaij:2015esa}.
As the final state of the decay $\decay{\Bs}{\phi(\to\Kp\Km)\mumu}$ is not flavour specific\footnote{Note that this is also the case for the decay $\decay{\Bd}{\rhoz(\to\pip\pim)\ellell}$}, 
only a subset of the angular observables discussed in this section is accessible. 
In particular, the angular observable $P_5^\prime$ can unfortunately not be measured with an untagged analysis. 
Determination of the decay flavour would require an identification of the initial \Bs\ production flavour and a decay-time dependent analysis (see Ref.~\cite{Descotes-Genon:2015hea}) which is experimentally challenging. 
The observables $F_{\rm L}$, $S_{3,4,7}$ and the \CP-asymmetries $A_{5,6,8,9}$ measured by the untagged LHCb analysis~\cite{Aaij:2015esa} are found to be in good agreement with SM predictions~\cite{Straub:2018kue,Straub:2015ica,Horgan:2013hoa,Horgan:2015vla}. 

\subsubsection{Experimental prospects}
An active program to study branching fractions and angular distributions of semileptonic rare decays is currently underway at the LHC. 
This effort is joined by the Belle~II experiment which has started data taking data with the fully instrumented detector in March 2019. 

The LHCb collaboration continues to explore the combined Run 1 and 2 data sample, and updated analyses for several modes (\eg\ $\decay{\Bd}{\Kstarz\mumu}$, $\decay{\Bs}{\phi\mumu}$) using the full data sample are expected. 
Also new approaches to the analysis of $b\to s\ellell$ decays are being discussed.
Following a first $q^2$-unbinned analysis of the decay $\decay{\Bu}{\Kp\mumu}$~\cite{Aaij:2016cbx}, 
similar $q^2$-unbinned approaches to the more complicated vector mode $\decay{\Bd}{\Kstarz\mumu}$ are being explored~\cite{Hurth:2017sqw,Blake:2017fyh,Chrzaszcz:2018yza,Mauri:2018vbg}. 
In the longer term future, the LHCb upgrades~\cite{Bediaga:2018lhg} will provide large samples of rare $b$-hadron decays\footnote{The yields for the decay $\decay{\Bd}{\Kstarz\mumu}$ are projected to be of ${\cal O}(400\,{\rm k})$ for the LHCb Upgrade~II~\cite{Bediaga:2018lhg}} that will allow to study $b\to s\ellell$ and $b\to d\ellell$ transitions with unprecedented precision. 
While measurements of branching fractions would be limited by the knowledge of the branching fractions of the normalisation modes $\decay{B}{(K,K^*,\ldots)\jpsi}$, 
the precision of measurements of ${\cal A}_{\CP}$ and ${\cal A}_I$ is expected to be at the percent level with the LHCb Upgrade~II data sample~\cite{Bediaga:2018lhg}. 
The uncertainty for angular observables of the decay $\decay{\Bd}{\Kstarz\mumu}$ is expected to reach $\lesssim 1\%$ for the $q^2$ binning currently in use, 
for exclusive $\decay{b}{d\mumu}$ decays angular analyses with a precision similar to the current precision for $\decay{b}{s\mumu}$ modes are expected to be feasible~\cite{Bediaga:2018lhg}. 
The ATLAS (CMS) experiments expect the $P_5^\prime$ uncertainties with the $3000\invfb$ data sample to reduce by up to a factor of 9 (15), compared to their Run~1 results~\cite{Cerri:2018ypt}. 

The Belle~II collaboration expects sensitivities to the $\decay{\Bd}{\Kstarz\ellell}$ angular observables of a few percent with the full $50\invab$ data sample~\cite{Kou:2018nap}, 
uncertainties of similar size also are expected for exclusive $\decay{b}{s\ellell}$ branching fractions. 
The Belle~II data sample would furthermore allow to significantly improve the current knowledge on the branching fractions of the normalisation modes $\decay{B}{(K,K^*,\ldots)\jpsi}$. 
The very similar performance for $\decay{b}{s\ee}$ and $\decay{b}{s\mumu}$ decays at Belle~II will be particularly beneficial for the lepton flavour universality tests discussed in Sec.~\ref{sec:lfu} below. 

\subsection{Lepton Flavour Universality tests in rare $B$ decays}
\label{sec:lfu}
\subsubsection{Lepton flavour universality tests $R_h$}
Lepton universality constitutes a central property of the SM and is well established in decays of mesons~\cite{Ablikim:2013pqa,Lazzeroni:2012cx,Aguilar-Arevalo:2015cdf}, leptons~\cite{pdg2020,Pich:2013lsa} and gauge bosons~\cite{ALEPH:2005ab}. 
Any differences in decay rates originate purely from lepton mass effects in the SM. 
The ratios 
\begin{align}
R_{h} =& \frac{\Gamma(\decay{B}{h\mumu})}{\Gamma(\decay{B}{h\ee})}
\label{eq:rh}
\end{align}
constitute precise probes of the SM, as hadronic uncertainties arising from form-factors and charm-loop contributions largely cancel in the ratio. 
In the intermediate $q^2$ region $1<q^2<6\gevgevcccc$ the ratios $R_{K,K^*}$ are therefore precisely predicted to be $R_{K,K^*}=1.00\pm 0.01$ in the SM~\cite{Hiller:2003js,Bobeth:2007dw},
with QED corrections not exceeding the ${\cal O}(1\%)$ level~\cite{Bordone:2016gaq}. 
Below $\approx 1\gevgevcccc$ lepton mass effects become significant, reducing the $R_h$ ratio below $1$ in the SM. 

\begin{figure}
    \centering
    \includegraphics[width=0.4\textwidth]{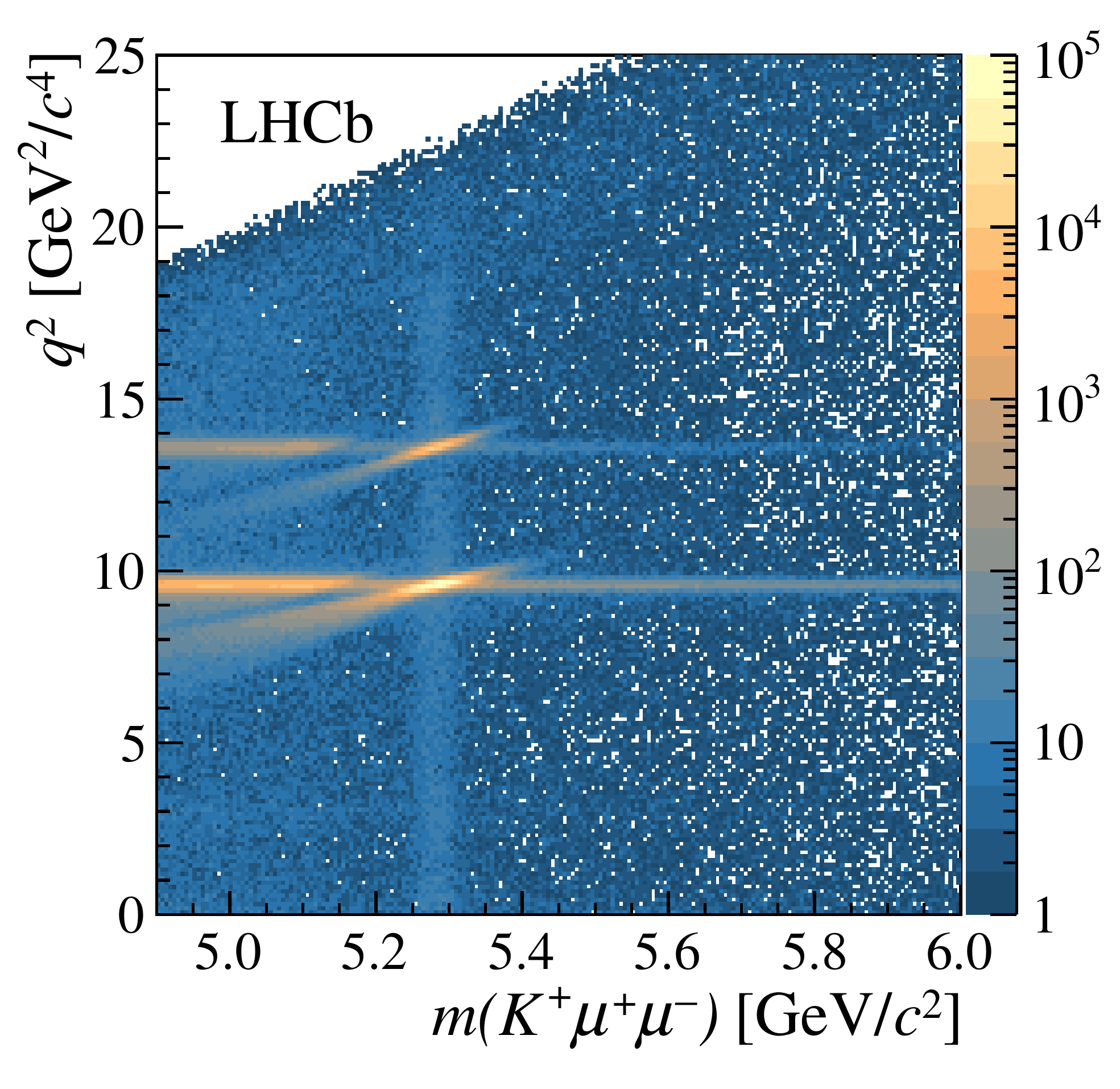}\hspace*{0.025\textwidth}
    \includegraphics[width=0.4\textwidth]{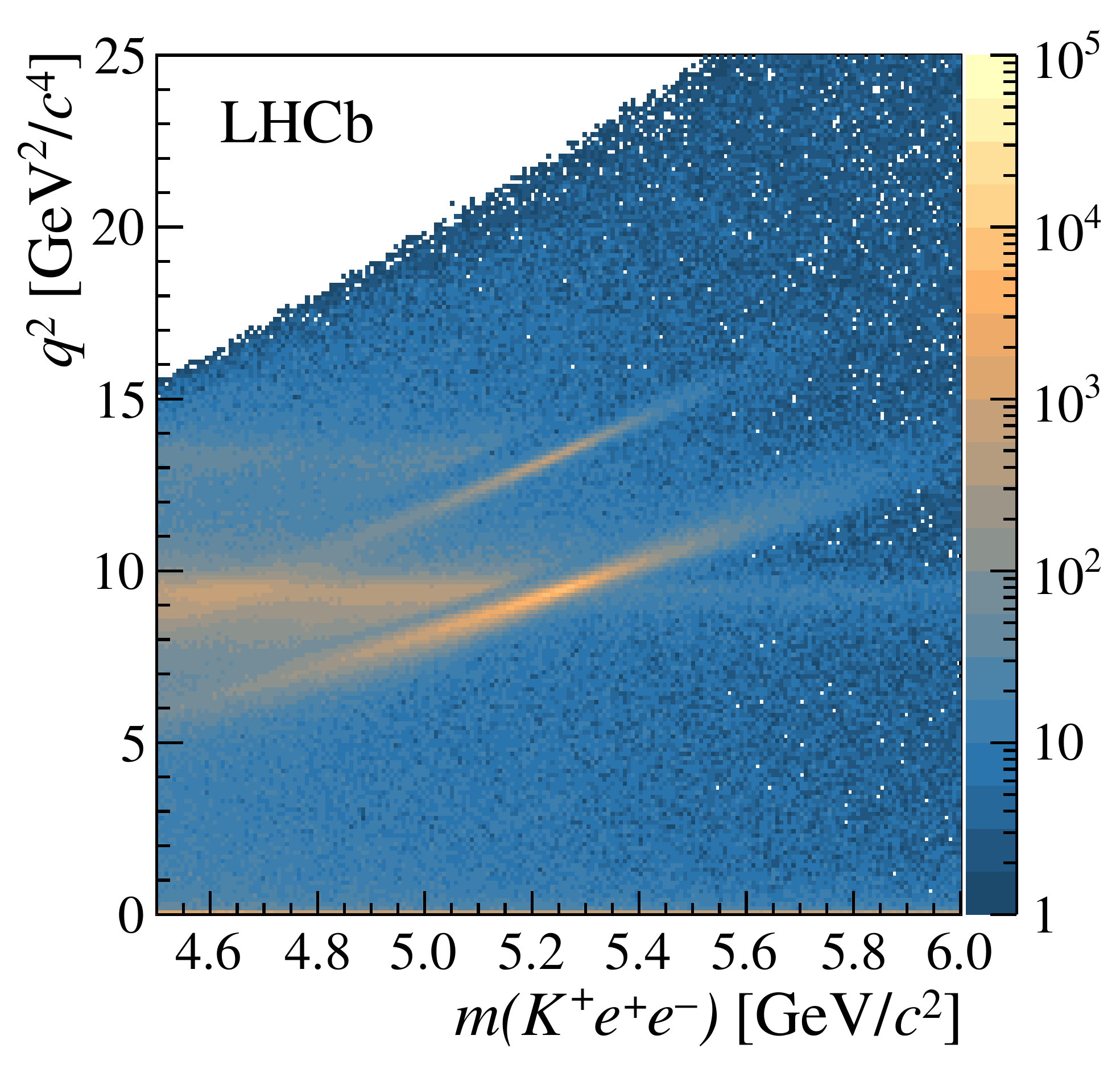}
    \caption{Reconstructed mass $m(\Kp\ellell)$ \vs\ $q^2$ for (left) the $\Kp\mumu$ and (right) the $\Kp e^+e^-$ final state. 
    The control and normalisation modes $\decay{\Bu}{\Kp\jpsi(\psitwos)}$ are clearly visible, as is the degraded resolution for the electron mode due to Bremsstrahlung. 
    The multivariate classifier to suppress combinatorial background events is not applied in this figure. 
    Figure reproduced from Ref.~\cite{Aaij:2019wad}\label{fig:rkeemumu}}
\end{figure}

As mentioned in Sec.~\ref{sec:btoslldecayrates} the triggering and reconstruction of the di-electron final state is more experimentally challenging at hadron colliders due to lower trigger 
efficiencies. 
A further experimental challenge is the emission of Bremsstrahlung photons by the electrons. 
While a dedicated Bremsstrahlung recovery procedure is employed by LHCb to add photons compatible with Bremsstrahlung emission to the reconstructed electron momenta, this still  significantly deteriorates the resolution of the reconstructed $B$ mass. This can be clearly seen in Fig.~\ref{fig:rkeemumu} 
for the $\Kp\mumu$ and $\Kp\ee$ final states, 
where the difference in resolution between the dimuon and dielectron mode is apparent.
The $R_h$ measurements at LHCb are therefore limited by the electron mode, 
whereas the yields and resolutions for muon and electron modes at the $B$ factory experiments are similar. 
At LHCb, the $R_{K,K^*}$ ratios are measured via the double ratio
\begin{align}
\label{eq:doubleRatio}
  R_{K,K^*} =& \frac{{\cal B}(\decay{B}{K^{(*)}\mumu})}{{\cal B}(\decay{B}{K^{(*)}\jpsi(\to\mumu)})}\times\frac{{\cal B}(\decay{B}{K^{(*)}\jpsi(\to\ee)})}{{\cal B}(\decay{B}{K^{(*)}\ee})}\\
  =& \frac{N_{\decay{B}{K^{(*)}\mumu}}}{N_{\decay{B}{K^{(*)}\jpsi(\to\mumu)}}}\times 
\frac{\epsilon_{\decay{B}{K^{(*)}\jpsi(\to\mumu)}}}{\epsilon_{\decay{B}{K^{(*)}\mumu}}} \times
\frac{N_{\decay{B}{K^{(*)}\jpsi(\to\ee)}}}{N_{\decay{B}{K^{(*)}\ee}}}\times
\frac{\epsilon_{\decay{B}{K^{(*)}\ee}}}{\epsilon_{\decay{B}{K^{(*)}\jpsi(\to\ee)}}}
\nonumber
\end{align}
where $N_{\decay{B}{K^{(*)}\ellell}}$ and $N_{\decay{B}{K^{(*)}\jpsi(\to\ellell)}}$ denote the yields of rare and normalisation modes and $\epsilon_{\decay{B}{K^{(*)}\ellell}}$ and $\epsilon_{\decay{B}{K^{(*)}\jpsi(\to\ellell)}}$ the corresponding efficiencies. 
This approach is experimentally advantageous as many systematic effects cancel in the efficiency ratios. 
Important further experimental cross-checks are provided by the ratios $r_{\jpsi}=\Gamma(\decay{B}{K^{(*)}\jpsi(\to\mumu)})/\Gamma(\decay{B}{K^{(*)}\jpsi(\to\ee)})$ and $R_{\psitwos}=\Gamma(\decay{B}{K^{(*)}\psitwos(\to\mumu)})/\Gamma(\decay{B}{K^{(*)}\psitwos(\to\ee)})$, both of which are known to be unity to high precision. 

\begin{figure}
  \centering
\includegraphics[width=0.49\textwidth]{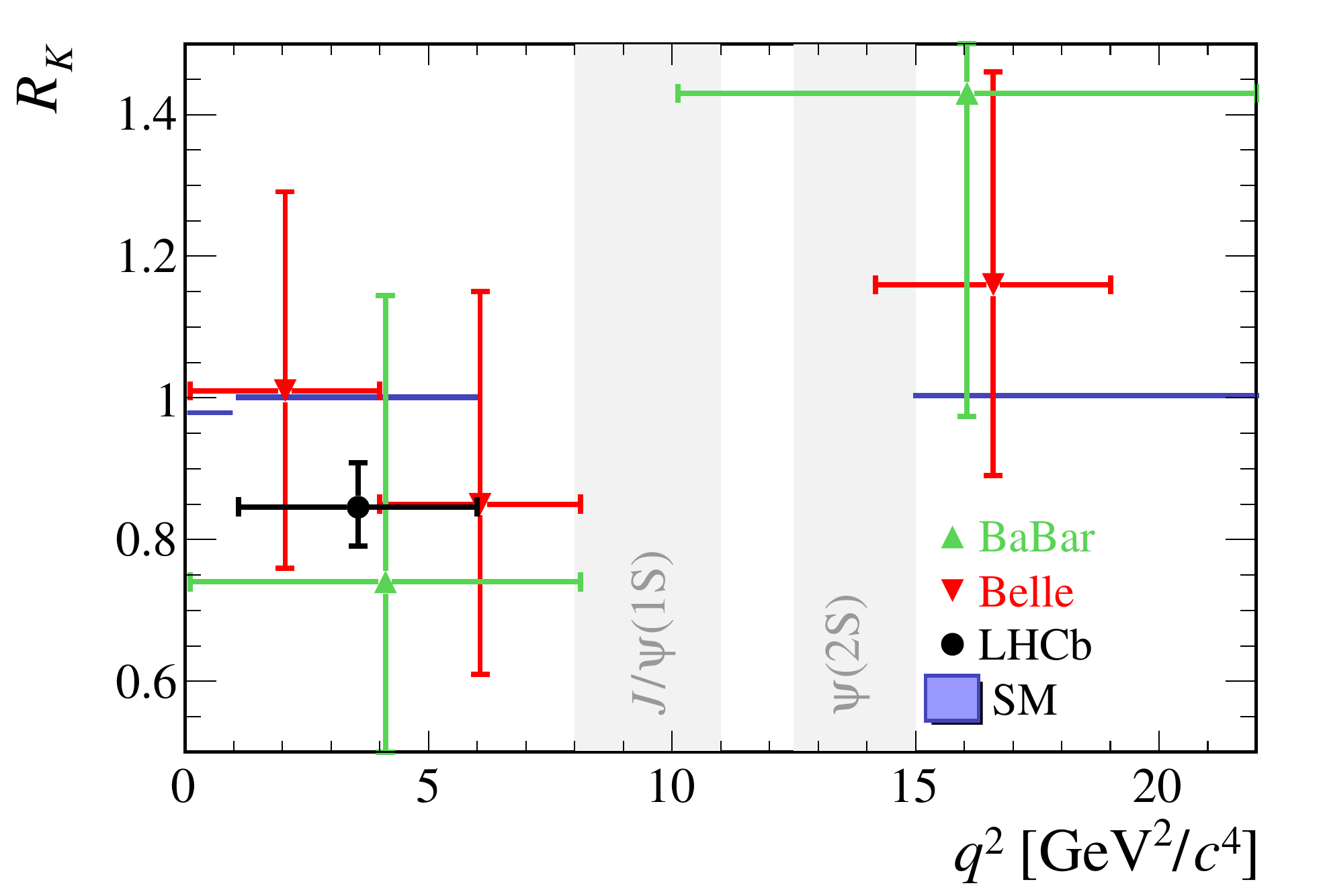}
\includegraphics[width=0.49\textwidth]{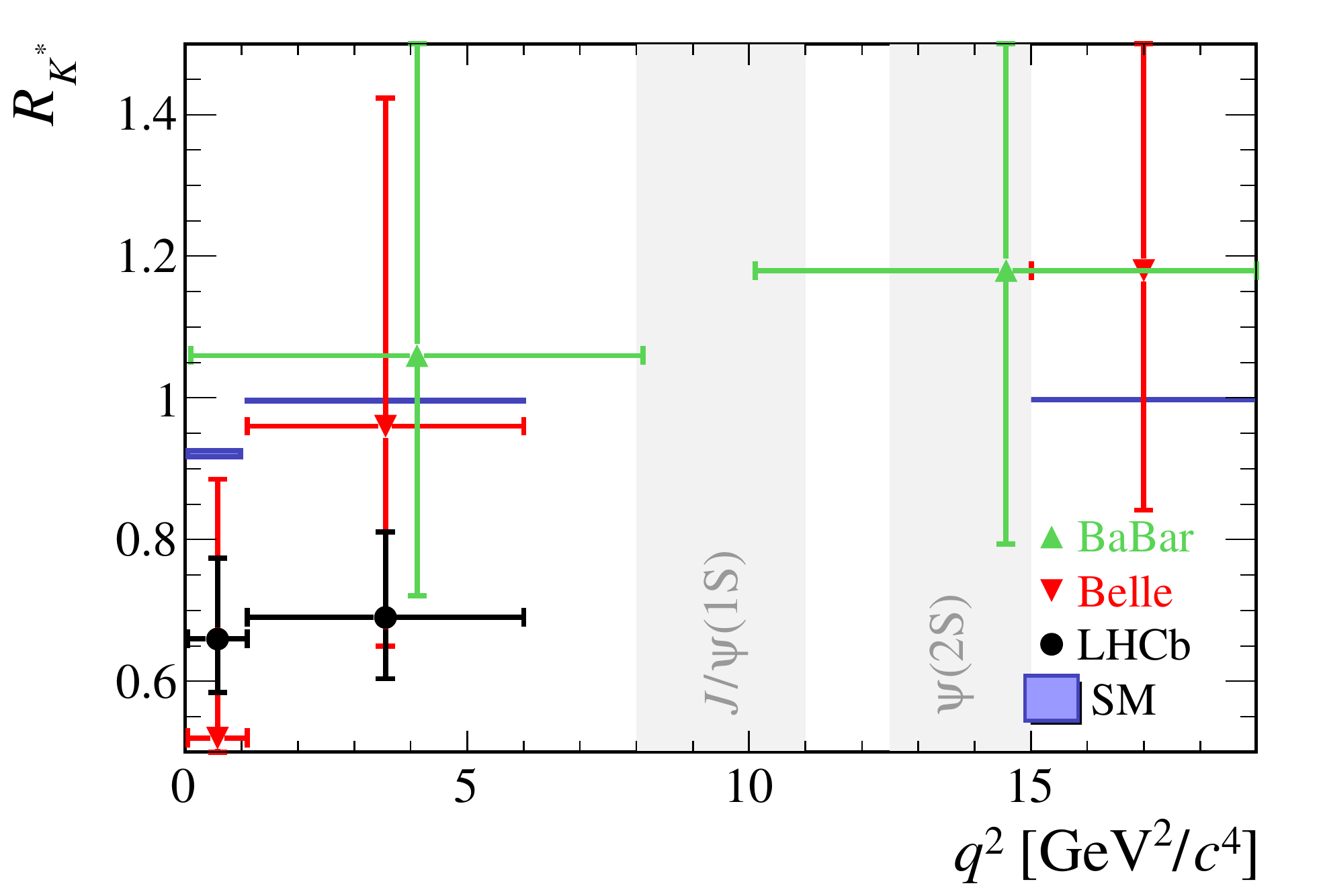}
\caption{
  Lepton universality tests (left) $R_K$ and (right) $R_{K^*}$ measured by the BaBar~\cite{Lees:2012tva}, Belle~\cite{Abdesselam:2019lab,Abdesselam:2019wac}, and LHCb~\cite{Aaij:2017vbb,Aaij:2019wad} collaborations. Overlaid is the SM prediction from Refs.~\cite{Straub:2018kue,Straub:2015ica,Horgan:2013hoa,Horgan:2015vla}\label{fig:rkrkst}}
\end{figure}

Figure~\ref{fig:rkrkst} gives the experimental data on $R_K$ and $R_{K^*}$ by the BaBar, Belle and LHCb collaborations~\cite{Lees:2012tva,Abdesselam:2019lab,Abdesselam:2019wac,Aaij:2017vbb,Aaij:2019wad}. 
The most precise measurements, provided by the LHCb collaboration~\cite{Aaij:2017vbb,Aaij:2019wad}, are given by
\begin{align}
  R_K(1.1<q^2<6.0\gevgevcccc) =& 0.846^{+0.060}_{-0.054}\text{(stat)}{}^{+0.016}_{-0.014}\text{(syst)}\\
  R_{K^*}(0.045<q^2<1.1\gevgevcccc) =& 0.66^{+0.11}_{-0.07}\text{(stat)}\pm 0.03\text{(syst)}\nonumber\\
  R_{K^*}(1.1<q^2<6.0\gevgevcccc) =& 0.69^{+0.11}_{-0.07}\text{(stat)}\pm 0.05\text{(syst)},\nonumber  
\end{align}
and are in tensions with the SM predictions at $2.5$, $2.1$, and $2.4\,\sigma$, respectively\footnote{It should be noted that the low-$q^2$ bin $0.045<q^2<1.1\gevgevcccc$ for $R_{K^*}$ contains a significant contribution from the photon pole which is known to be lepton flavour universal.}. 
The measurements are 
strongly statistically limited, 
in particular there is no systematic limitation from the precision of the SM prediction,
which is largely free of hadronic uncertainties. 
The tensions in the lepton universality tests $R_{K,K^{*}}$ constitute a further \textit{flavour anomaly} in the rare decays\footnote{During the review of this manuscript an updated measurement of $R_K$ by the LHCb collaboration was made public~\cite{Aaij:2021vac}, which reports a tension with the SM prediction at the level of $3.1\,\sigma$.}. 

\paragraph{Baryonic tests for lepton flavour universality}
Lepton flavour universality can also be tested in baryonic $\decay{b}{s \ell^+ \ell^-}$ decays. So far only the channel $\decay{\Lambda^0_b}{p K^- \ellell}$ has been measured, using roughly half of the combined Run~1 and~2 data set collected by the LHCb collaboration~\cite{Aaij:2019bzx}. This decay is experimentally advantageous over the decay $\decay{\Lb}{\Lambda \ellell}$ as the $p K^-$ system forms a common decay vertex with the lepton pair, whereas the $\Lambda$ has a significant decay length, making the reconstruction and selection of the decay more challenging. 

The observable measured here is $R_{pK}$, defined analogously to Eq.~\ref{eq:rh}.  
It is measured by using a double ratio with the control modes  $\decay{\Lambda^0_b}{p K^- \jpsi (\to \mumu, \ee )}$, in full analogy to Eq.~\ref{eq:doubleRatio}. The measurement is performed in the $q^2$ range $0.1<q^2<6.0\gevgevcccc$ and the $p K^-$ mass range $m(p K^-) < 2.6\gevcc$. The lepton flavour universality ratio is measured to be
\begin{equation}
     R_{pK}(0.1<q^2<6.0\gevgevcccc) = 0.86^{+0.14}_{-0.11}{}_\text{(stat)}\pm 0.05_\text{(syst)},
\end{equation}
in agreement with unity within one standard deviation. This result is also in agreement with the deviations observed in lepton-universality tests with B mesons discussed above. More data is needed to confirm or exclude the presence of New Physics contributions in these decays.
There is currently no theory framework available to compute $R_{pk}$ in the $q^2$ bin used by LHCb.

\subsubsection{Angular lepton flavour universality tests}
\begin{figure}
  \centering
\includegraphics[width=0.49\textwidth]{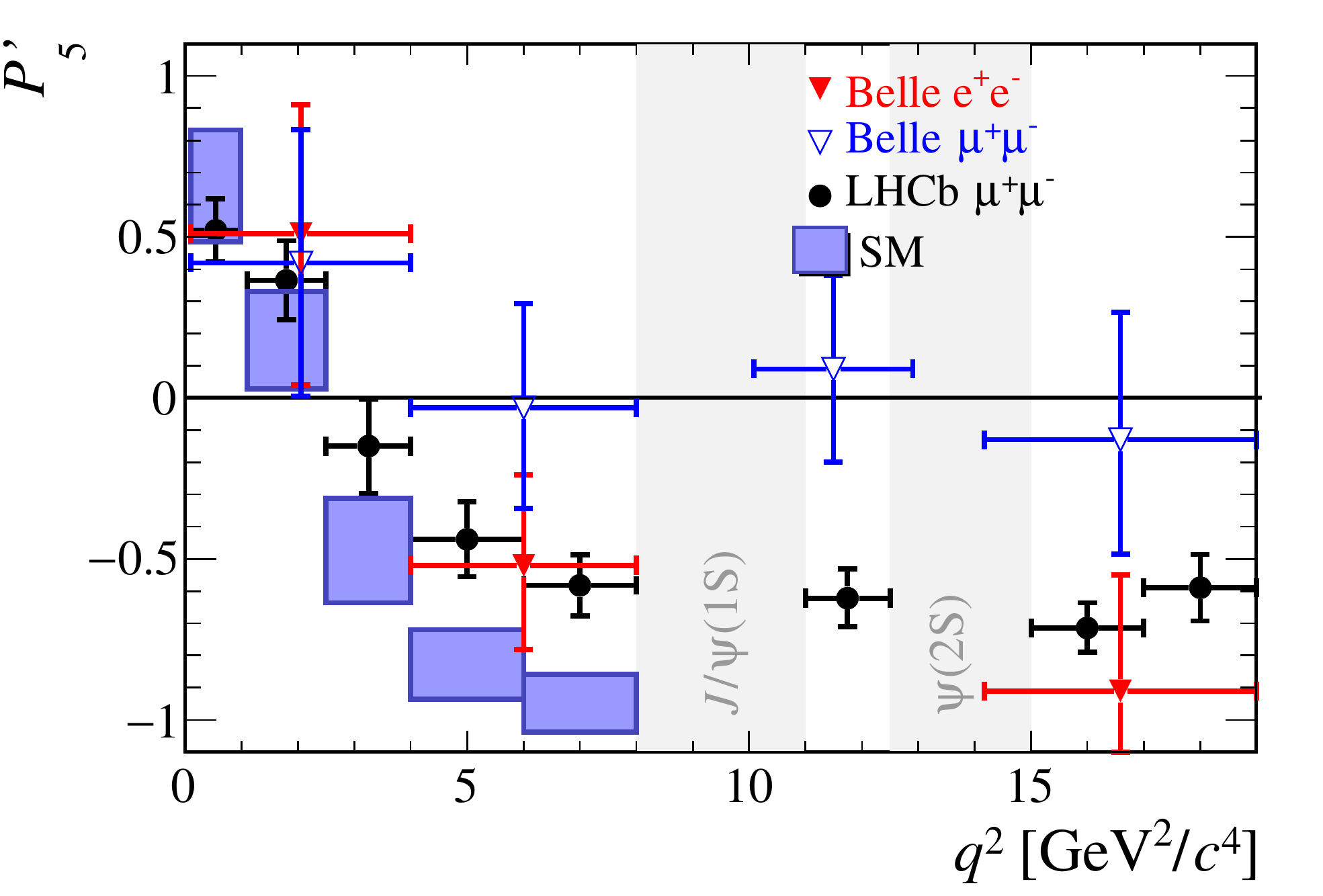}
  \caption{Angular observable $P_5^{\prime e}$ and $P_5^{\prime\mu}$ by the Belle collaboration~\cite{Wehle:2016yoi}, together with the LHCb measurement of $P_5^{\prime\mu}$~\cite{Aaij:2020nrf}. Overlaid is the SM prediction from Refs.~\cite{Descotes-Genon:2014uoa,Khodjamirian:2010vf}.\label{fig:p5plfu}}
\end{figure}

Similar to ratios of decay rates, also differences of angular observables of $b\to s\mumu$ and $b\to s\ee$ decays constitute precise tests of lepton flavour universality~\cite{Capdevila:2016ivx,Serra:2016ivr}. 
In the $q^2$ region of interest for the flavour anomalies, the Belle collaboration has published a measurement of both the di-electron and the di-muon mode and the angular lepton universality tests $Q_4=P_4^{\prime\mu}-P_4^{\prime e}$ and $Q_5=P_5^{\prime\mu}-P_5^{\prime e}$~\cite{Wehle:2016yoi}. 
Figure~\ref{fig:p5plfu} shows $P_5^{\prime\mu}$ and $P_5^{\prime e}$ in comparison with the $P_5^{\prime\mu}$ measurement by the LHCb collaboration. 
The observable $P_5^{\prime\mu}$ in the $q^2$ range $4<q^2<8\gevgevcccc$ is found to be in tension with the SM prediction~\cite{Capdevila:2016ivx,Descotes-Genon:2014uoa} at $2.6\,\sigma$. 
The observable $P_5^{\prime e}$ in the same $q^2$ region is in better agreement with the SM prediction at $1.3\,\sigma$. 
The difference $Q_5$ in the $q^2$ region $4<q^2<8\gevgevcccc$ is found to be in agreement with the SM at $0.8\,\sigma$. 

\subsubsection{Experimental prospects}
The lepton flavour universality tests in rare $\decay{b}{s\ellell}$ decays are currently statistically limited, with the experimental uncertainty much larger than the uncertainty on the precise SM prediction. 
The large data samples that will be available in the LHCb upgrades and at Belle~II are thus particularly valuable. 
At Belle~II, the very similar performance for reconstruction of the electron and muon modes are very beneficial. 
The Belle~II collaboration projects a sensitivity of $3.6\%$ ($3.2\%$) for $R_K$ ($R_{K^*}$) in the $q^2$ range  $1<q^2<6\gevgevcccc$ for the 
full $50\invab$ data sample~\cite{Kou:2018nap}. 
The precision for angular observables in exclusive $\decay{b}{s\ee}$ and $\decay{b}{s\mumu}$ modes will be similar at Belle~II, for the lepton flavour universality test $Q_5$ in the $q^2$ range $4<q^2<6\gevgevcccc$ Belle~II expects an uncertainty of $4\%$ with the full data sample~\cite{Kou:2018nap}. 

The LHCb collaboration projects uncertainties of $\sim 2\%$ for $R_{K}$ and $R_{K^*}$ with the $50\invfb$ data sample of the LHCb upgrade and $\lesssim 1\%$ for the full Upgrade~II data sample~\cite{Bediaga:2018lhg}. 
While the electron modes will still constitute the limiting factor in the LHCb upgrades, the move to a full software trigger will significantly improve the efficiency for $\decay{b}{s\ee}$ modes~\cite{CERN-LHCC-2014-016}. 

\subsection{Lepton flavour violating $B$ decays}
\label{sec:lfv}

Lepton flavour violation (LFV) for neutral lepton has been established through the observation of neutrino oscillations. 
However, for charged leptons LFV is negligible in the Standard Model --- see the discussion in Sec.~\ref{sec:effectivetheory} ---
and any observation of such a process would constitute a clear sign of physics beyond the Standard Model.
In the light of the recent flavour anomalies, as discussed in the previous chapters, several extensions to the SM have been proposed that link the violation of lepton universality to LFV, predicting in particular significantly enhanced decay rates in \btosemu processes~\cite{Glashow:2014iga,Crivellin:2015era,Hiller:2016kry}. The decay rates predicted could be just below the current experimental limits. 
It is interesting to study the complementary between LFV in $B$-meson decays and charged lepton decays like $\decay{\ell^-}{\ell^{\prime -} \gamma} $ and $\decay{\ell^-}{3 \ell^{\prime -} } $, which will be a way to differentiate different New Physics models in case non-vanishing decay rates are found. 

\begin{figure}
  \centering
\includegraphics[width=0.49\textwidth]{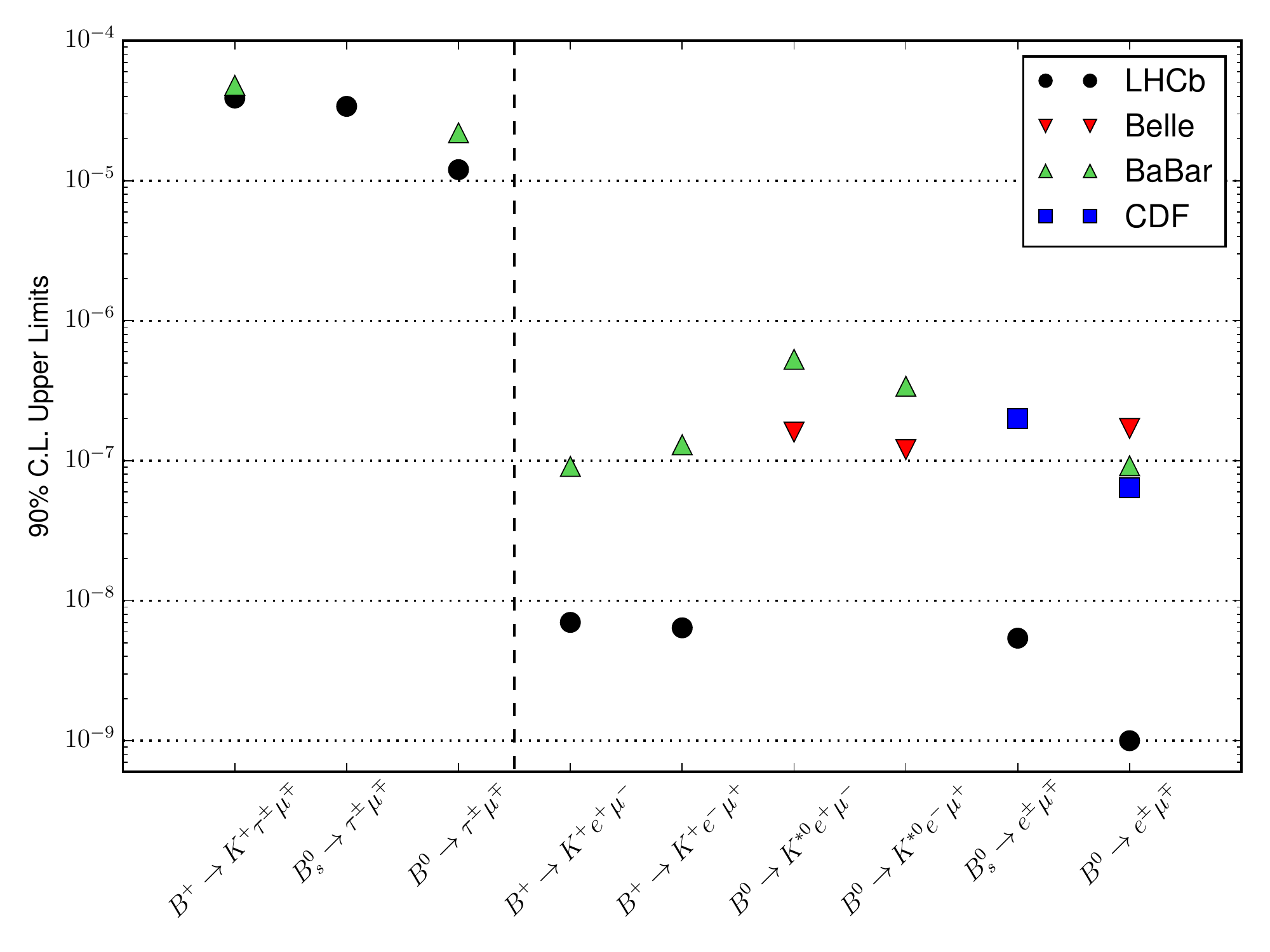}
\caption{\label{fig:lfv} Selection of searches for Lepton flavour violation in $B$-decays, limits extracted from Refs.~\cite{B2emu_Belle,B2K*emu_Belle,B2K/K*emu_BaBar,B2Ktaumu_BaBar,B2taumu_BaBar,B2emu_BaBar,B2emu_CDF}.}
\end{figure}

Experimentally, \btosemu processes have been tested in the channels \decay{\Bu}{K^{+}\emu}$, \decay{\Bd}{K^{*0}\emu}$ and $\decay{B}{\emu}$. The older limits are set by the CDF, Babar and Belle experiments at the order of $10^{-7}$ while the most recent limits published by the LHCb experiment start to explore the range of $10^{-8}$--$10^{-9}$~\cite{B2emu_Belle,B2K*emu_Belle,B2K/K*emu_BaBar,B2Ktaumu_BaBar,B2taumu_BaBar,B2emu_BaBar,B2emu_CDF}. The results are summarized in Fig.~\ref{fig:lfv}. LHCb and its upgrades are expected to dominate the field of lepton-flavour violating $B$-meson decays in $\mu^\pm e^\mp$ channels while the situation is less clear in the $\tau^\pm \mu^\mp$ and $\tau^\pm e^\mp$ channels. In purely leptonic decays, the Belle~2 experiment is expected to dominate the field of $\tau^-$ lepton decays with interesting contributions from LHCb possible in the best accessible decay modes like $\decay{\tau^-}{\mumu \mu^-}$.

\subsection{Rare $b\to s\tau^+\tau^-$ decays}
\label{sec:btostautau}
Semileptonic $\decay{b}{s\tau^+\tau^-}$ decays are theoretically interesting as they test electroweak penguin processes involving third generation leptons. Experimentally, they are still largely uncharted territory, with bounds measured some orders of magnitude away from the SM expectations. The most accessible modes will be $\decay{\Bu}{K^+ \tau^+\tau^-}$, $\decay{\Bd}{K^0 \tau^+\tau^-}$ and $\decay{\Bd}{K^{*0} \tau^+\tau^-}$, all with predicted branching fractions in of ${\cal O}(10^{-7})$~\cite{Straub:2018kue,Du:2015tda}.
The Belle collaboration has published a search for the decay $\decay{\Bu}{\Kp \tau^+\tau^-}$, resulting in an upper limit of
\begin{align}
    {\cal B}(\decay{\Bu}{\Kp\tau^+\tau^-})  < & 2.25\times 10^{-3}
\end{align}
at $90\%$ CL~\cite{TheBaBar:2016xwe}. 
The final Belle~2 data set is expected to reach a few times $10^{-5}$~\cite{Kou:2018nap}, still two orders of magnitude away from the SM expectation. LHCb and its upgrades may be able to contribute to these searches.

\section{Tree-level $b\to c\ell\nu$ decays}
\label{sec:treelevel}
\subsection{Theory}
\label{sec:treetheory}
The transition amplitudes for exclusive semileptonic $b\to c\ell^-\bar\nu$ decays
factorize --- to leading order in the electromagnetic interaction --- into
leptonic current and hadronic matrix elements of $b\to c$ currents.
Universal radiative corrections have been computed at the level of the effective
field theory~\cite{Sirlin:1980nh}. As for the
$b\to s$ transitions, the hadronic matrix elements are parametrized in terms
of scalar-valued functions of the squared momentum transfer $q^2$, the so-called
form factors. In the SM, the relevant currents are local $\bar{c}\gamma^\mu b$ and $\bar{c}\gamma^\mu\gamma_5 b$ currents.
For phenomenological analyses beyond the SM, form factors for the full basis of local $b\to c$ current are needed.
Below, we will first discuss results for the hadronic form factors from QCD-based methods only.
The first such method is lattice QCD, which is a first-principle method that simulates hadrons
in a finite volume on a discrete spacetime lattice. The second method, light-cone sum rules (LCSRs)
with $B$-meson distribution amplitudes, uses semi-local quark-hadron duality to compute
form factors based on universal non-local $B$-meson-to-vacuum matrix elements.\\

In this review, we focus on exclusive $b\to c\ell^-\bar\nu$ decay observables that provide SM tests,
in particular the lepton-flavour universality ratios
\begin{equation}
    R_{H_c} \equiv \frac{\mathcal{B}(\bar{H_b}\to H_c \tau^-\bar\nu)}{\mathcal{B}(\bar{H_b}\to H_c \mu^-\bar\nu)}\,.
\end{equation}
Here $X_c$ denotes a single hadron with $C=+1$, and $H_b$ is either a $\bar{B}$ meson or the $\Lambda_b$ baryon.
Predictions for the total branching fractions require accurate knowledge of the hadronic form
factors in the entire semileptonic phase space $m_\ell^2 \leq q^2 \leq (M_{H_b} - M_{H_c})^2$.
We therefore also discuss parametrizations that extrapolate the results of either of the aforementioned
QCD-based methods or that interpolate between them.\\

\paragraph{Form factors in $\bar{B}\to D$ transitions}
The hadronic form factors for $\bar{B}\to D$ transitions in the SM have been obtained
from two independent lattice QCD studies~\cite{Na:2015kha,Lattice:2015rga}.
They provide precise predictions for the two form factors $f_+$ and $f_0$ at large $q^2$ values,
with uncertainties as low as $\sim 1\%$.
The full set of three form factors has been computed using light-cone sum rules (LCSRs) with $B$-meson
distribution amplitudes for four different $q^2$ points with $q^2 \leq 0$, \ie,
outside the semileptonic phase space~\cite{Gubernari:2018wyi}. This work supersedes the results
of the first LCSR calculation of some of these form factors~\cite{Faller:2008tr}.\\

\paragraph{Form factors in $\bar{B}\to D^*$ transitions}
The hadronic form factor $h_{A_1}$ for $\bar{B}\to D^*$ transitions has been obtained
from two independent lattice QCD studies~\cite{Bailey:2014tva,Harrison:2017fmw},
but only in the kinematic endpoint $q^2_{\text{max}} = (M_B - M_{D^*})^2$. An average of the two results
is available from the HFLAV collaboration~\cite{Amhis:2019ckw}.
Similar to $\bar{B}\to D$ transitions, the full set of seven form factors has been computed
using light-cone sum rules (LCSRs) with $B$-meson
distribution amplitudes for four different $q^2$ points with $q^2 \leq 0$, \ie,
outside the semileptonic phase space~\cite{Gubernari:2018wyi}. This work supersedes the results
of the first LCSR calculation of some of these form factors~\cite{Faller:2008tr}.\\

\paragraph{Form factors in $\Lambda_b\to\Lambda_c$ transitions}
The vector and axial vector currents give rise to six independent hadronic form factors in $\Lambda_b\to\Lambda_c$
transitions. These are available from a single lattice QCD study~\cite{Detmold:2015aaa}, with excellent control
of the uncertainties in the entire semileptonic phase space. Lattice QCD results for the four form factors of the
tensor currents were published in Ref.~\cite{Datta:2017aue}, based on the same simulations that were used in Ref.~\cite{Detmold:2015aaa}.\\

\paragraph{Form factors in $B_c\to J/\psi$ transitions}
The four hadronic form factors arising in matrix element of the vector and axial vector currents are
available from a single lattice QCD study~\cite{Harrison:2020gvo} in the entire semileptonic phase space.
The uncertainties are under good control, with relative uncertainties $\sim 10\%$.\\

\paragraph{Parametrisation and extrapolation to the entire semileptonic phase space}
We focus on two approaches to the parametrisation of the hadronic form factors.
\begin{itemize}
    \item The Boyd/Grinstein/Lebed (BGL) parametrisation~\cite{Boyd:1997kz} uses analyticity and unitarity to
    derive dispersive bounds on the parameters of the form factors. These bounds restrict the
    parameter space a-priori to a bounded set, the open hypercube $(-1, +1)^N$, where $N$ is the
    total number of form factor paramters. This is commonly known as the weak bound.
    The bound becomes progressively stronger when more exclusive semileptonic $b\to c$
    transitions are considered simultaneously. The bound in such a global analysis is commonly known
    as the strong unitarity bound~\cite{Bigi:2017jbd}.\\
    
    A common simplification of the BGL parametrization is the model by Bourrely/Caprini/Lellouch (BCL)~\cite{Bourrely:2008za},
    which replaces the BGL outer functions and Blaschke factors with simpler objects, at
    the expense of requiring a truncated expansion in the computation of the dispersive bound.
    The $\Lambda_b\to\Lambda_c$ form factors are studied exclusively in adhoc simplified parametrisations
    similar to the BCL one.\\
    
    \item The heavy-quark expansion makes use of the fact that both the $b$ and the $c$ quark are heavy,
    compared to the intrinsic hadronic scale $\Lambda_\text{had} \simeq \text{a few hundred}\MeV$.
    Expanding the exclusive form factors in the inverse heavy quark masses reduces all form factors
    in the four transitions $\bar{B}^{(*)} \to D^{(*)}$ to a hand full of independent
    \textit{Isgur-Wise functions}; see Ref.~\cite{Neubert:1993mb} for an extensive review and
    refs.~\cite{Bernlochner:2017jka,Bordone:2019vic,Bordone:2019guc} for recent developments. The use
    of the lattice QCD and LCSR results made possible a simultaneous fit of all Isgur-Wise
    functions at order $1/m_c^2$~\cite{Bordone:2019vic}, which was subsequently extended beyond the $SU(3)_F$ symmetry limit~\cite{Bordone:2019guc}.\\
    
    At the same order in the heavy-quark expansion, the $\Lambda_b\to\Lambda_c$ form factors feature
    fewer independent Isgur-Wise functions than the $\bar{B}^{(*)} \to D^{(*)}$ form factors~\cite{Falk:1992ws}.
    A phenomenological study of the lattice QCD results~\cite{Bernlochner:2018kxh} yields the surprising
    insight that the form factors in vector and axial currents on the one hand and form factors in tensor currents
    cannot be simultaneously described withing the heavy-quark expansion framework at order $1/m_c^2$.
\end{itemize}
The tensions observed in the LFU ratios since 2012 have triggered renewed interest in both types of parametrisation
of the exclusive form factors%
~\cite{%
Bigi:2016mdz,%
Bernlochner:2017jka,Bigi:2017jbd,Bigi:2017njr,Jaiswal:2017rve,%
Bernlochner:2018kxh,Bernlochner:2018bfn,%
Bordone:2019vic,Bordone:2019guc%
}.

\paragraph{Predictions for the LFU ratios}
Using the HQE predictions for the form factors in $\bar{B}_{(s)}\to D_{(s)}^{(*)}$ transitions
and accounting for $SU(3)_F$-breaking in the leading and next-to-leading Isgur-Wise functions,
one obtains for the LFU ratios~\cite{Bordone:2019guc}
\begin{equation}
\label{eq:RDDstDsDsst}
\begin{aligned}
    R_{D}     = & 0.2989 \pm 0.0032\,, &
    R_{D_s}   = & 0.2970 \pm 0.0034\,, \\
    R_{D^*}   = & 0.2472 \pm 0.0050\,, &
    R_{D_s^*} = & 0.2450 \pm 0.0082\,. \\
\end{aligned}
\end{equation}
Similarly, the HQE prediction for $R_{\Lambda_c}$, based on purely theoretical inputs for the form factors from lattice QCD
simulations, reads~\cite{Bernlochner:2018kxh,Bernlochner:2018bfn}
\begin{align}
\label{eq:RLc}
    R_{\Lambda_c} =& 0.331 \pm 0.010\,.
\end{align}
The SM prediction for $R_{J/\psi}$ reads~\cite{Harrison:2020nrv}
\begin{align}
    \label{eq:RJpsi}
    R_{J/\psi}    =& 0.2582 \pm 0.0038\,.
\end{align}

\subsection{Experimental results}
\subsubsection{Exclusive decays $\decay{\bar{B}}{D^{(*)}\ell^-\bar{\nu}_\ell}$}
The exclusive $b\to c\ell^-\bar{\nu}_\ell$ decays $\decay{\bar{B}}{D^{(*)}\ell^-\bar{\nu}_\ell}$, where $\ell=e,\mu$, are copiously produced and have been precisely measured by BaBar, Belle, CLEO and the LEP experiments. 
The $B$-factory experiments are particularly well suited for the study of semileptonic $b\to c\ell^-\bar{\nu}_\ell$ decays due to the known kinematics of the initial ($\Upsilon(4S)$) state and their large angular coverage.  
Furthermore, the $B$-factory experiments can reconstruct the second (non-signal, ``recoil'') $B$-meson produced from the $\Upsilon(4S)$ decay , 
either in hadronic decays (\textit{hadronic tag}) or semileptonic decays (\textit{semileponic tag}). The \textit{$B$-tagging} improves the resolution of the neutrino kinematics at the cost of signal efficiency. 
Typical tagging efficiencies for the hadronic (semileptonic) tag are around $0.2$--$0.4\%$ ($0.3$--$0.6\%$)~\cite{Bevan:2014iga}. 

No significant contribution from NP is expected for 
the decays $\decay{\bar{B}}{D^{(*)}\ell^-\bar{\nu}_\ell}$ ($\ell=e,\mu$) 
and the modes are instead used for the determination of the CKM matrix element $V_{\rm cb}$ and to obtain information on  the form factor parameters. 
For a detailed discussion on the 
status of the determination of $V_{\rm cb}$ from exclusive $b\to c\ell^-\bar{\nu}_\ell$ decays see Ref.~\cite{Amhis:2019ckw}. 

The BaBar, Belle and CLEO experiments have also performed 
separate measurements of the branching fractions of electron and muon modes. 
The resulting branching fraction ratios are in good agreement with lepton universality, lepton non-universality between the electron and muons modes is constrained to the level of a few percent~\cite{Aubert:2008yv,Glattauer:2015teq,Waheed:2018djm,Abdesselam:2017kjf,Adam:2002uw}.

\subsubsection{Lepton universality tests $R_D$ and $R_{D^*}$}
Due to the high mass of the $\tau$ lepton of $m_\tau\approx 16.8 m_\mu\approx 3480 m_e$, contributions from NP could be enhanced in $b\to c\tau^-\bar{\nu}_\tau$ transitions. 
Measurements of 
$\decay{\bar{B}}{D^{(*)}\tau^-\bar{\nu}_\tau}$ decays, which are less well constrained than the corresponding electron and muon modes are thus of high interest. 
The lepton flavour universality tests
\begin{align}
R_{D^{(*)}} =& \frac{{\cal B}(\decay{\bar{B}}{D^{(*)}\tau^-\bar{\nu}_\tau})}{{\cal B}(\decay{\bar{B}}{D^{(*)}\ell^-\bar{\nu}_\ell})}
\end{align}
with $\ell=e,\mu$ can be predicted to high precision in the SM as the form factor uncertainties cancel to a large degree~\cite{Bigi:2017jbd}. 
The SM predictions  
given in Eq.~\ref{eq:RDDstDsDsst}
lie significantly below unity due to the large effect of the $\tau$ mass. 

Experimentally, the $\decay{\bar{B}}{D^{(*)}\tau^-\bar{\nu}_\tau}$ decay can be reconstructed in different $\tau$ decay modes, summarised in Tab.~\ref{tab:taudecays}:
\begin{itemize}
\item The leptonic $\tau$ decays $\decay{\tau^-}{\mu^-\bar{\nu}_\mu\nu_\tau}$ and $\decay{\tau^-}{e^-\bar{\nu}_e\nu_\tau}$. 
  The $R_D^{(*)}$ ratios thus can be calculated with the same light lepton in the final state of the $\decay{\bar{B}}{D^{(*)}\tau^-\bar{\nu}_\tau}$ signal mode and the $\decay{\bar{B}}{D^{(*)}\ell^-\bar{\nu}_\ell}$ normalisation mode,
  which reduces systematic effects from the lepton reconstruction and identification. 
\item Hadronic $\tau^-$ decay modes like $\decay{\tau^-}{\pim(\piz)\nu_\tau}$ (referred to as \textit{one-prong}) and $\decay{\tau^-}{\pim\pip\pim(\piz)\nu_\tau}$ (\textit{three-prong}).
  The hadronic $\decay{\tau^-}{\pim\pip\pim(\piz)\nu_\tau}$ decays allow the reconstruction of the $\tau^-$ decay vertex,
  which constrains the topology of the signal decay and furthermore allows for improved background rejection. 
\end{itemize}
\begin{table}
    \centering
    \begin{tabular}{lr}\hline
       $\tau^-$ decay  & ${\cal B}~[\%]$ \\\hline\hline
        $\decay{\tau^-}{\mu^-\bar{\nu}_\mu\nu_\tau}$ & $17.39\pm 0.04$\\
        $\decay{\tau^-}{e^-\bar{\nu}_e\nu_\tau}$ & $17.82\pm 0.04$\\\hline
        $\decay{\tau^-}{\pim\nu_\tau}$ & $10.82\pm0.05$\\
        $\decay{\tau^-}{\pim\piz\nu_\tau}$ & $25.49\pm0.09$\\
        $\decay{\tau^-}{\pim\pip\pim\nu_\tau}$ & $9.02\pm0.05$\\
        $\decay{\tau^-}{\pim\pip\pim\piz\nu_\tau}$ & $4.49\pm0.05$\\
    \hline\end{tabular}
    \caption{Relevant decay modes of the $\tau^-$ lepton, branching fractions are taken from Ref.~\cite{pdg2020}. The multi-hadron final states do not include contributions from \KS\ decays.\label{tab:taudecays}}
\end{table}

\paragraph{Measurements of $R_{D^{(*)}}$ at the $B$-factory experiments}
At the $B$-factory experiments the normalisation modes combine the two light lepton flavours. 
The lepton universality tests $R_{D^{(*)}}$ at the $B$-factory experiments have been performed 
using both hadronic~\cite{Lees:2012xj,Lees:2013uzd,Huschle:2015rga,Hirose:2016wfn,Hirose:2017dxl} and semileptonic~\cite{Belle:2019rba} $B$-tagging. 
The $\tau$ lepton has been reconstructed in the leptonic~\cite{Lees:2012xj,Lees:2013uzd,Huschle:2015rga,Belle:2019rba} and hadronic~\cite{Hirose:2016wfn,Hirose:2017dxl} one-prong modes. 

The most precise measurement to date has been performed by the Belle collaboration using semileptonic $B$-tagging and leptonic $\tau$ decays~\cite{Belle:2019rba}. 
$R_D$ and $R_{D^*}$ are determined in a two-dimensional fit to a multivariate classifier, trained to separate $\decay{\bar{B}}{D^{(*)}\tau^-\bar{\nu}_\tau}$ signal from $\decay{\bar{B}}{D^{(*)}\ell^-\bar{\nu}_\ell}$ normalisation events, and 
and $E_{\rm ECL}$, the energy in the electromagnetic calorimeter not associated with reconstructed particles,
which is used to separate signal/normalisation from background events. 
The result of~\cite{Belle:2019rba}
\begin{align}
  R_D =& 0.307\pm 0.037\pm 0.016\\
  R_{D^*} =& 0.283\pm 0.018\pm 0.014\nonumber
\end{align}
is in agreement with the most recent SM prediction in Eq.~\eqref{eq:RDDstDsDsst} at $0.2$ and $1.5\,\sigma$, respectively. 

\begin{figure}
  \centering
  \includegraphics[height=0.49\textwidth]{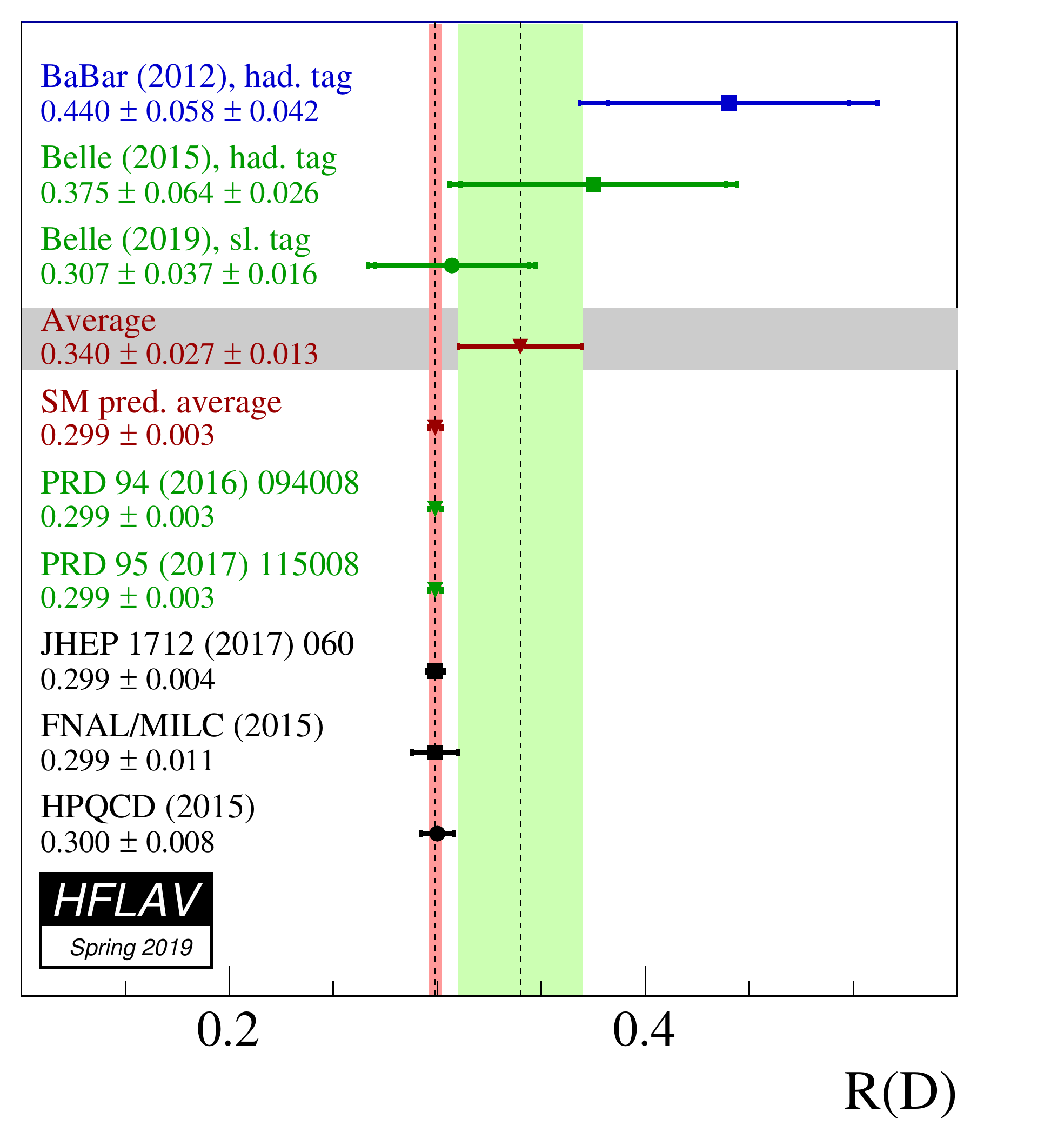}
  \includegraphics[height=0.49\textwidth]{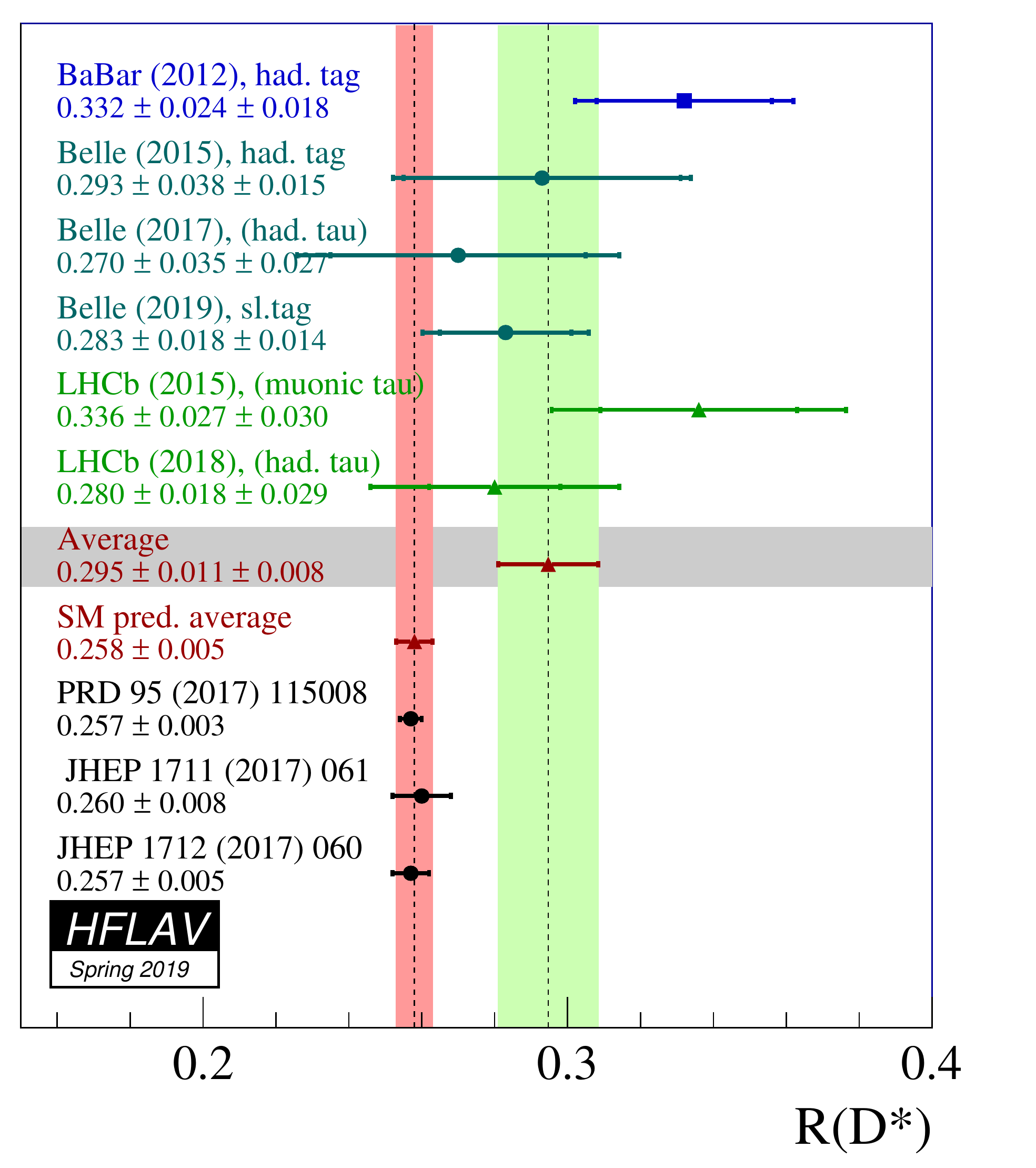}
  \caption{Measurements of the lepton universality tests (left) $R_D$ and (right) $R_{D^*}$ by the BaBar~\cite{Lees:2012xj,Lees:2013uzd}, Belle~\cite{Huschle:2015rga,Hirose:2017dxl,Belle:2019rba} and LHCb~\cite{Aaij:2015yra,Aaij:2017uff,Aaij:2017deq} collaborations. 
    The SM predictions are taken from Refs.~\cite{Lattice:2015rga,Na:2015kha,Bigi:2016mdz,Bernlochner:2017jka,Bigi:2017jbd,Jaiswal:2017rve}. 
    Figure reproduced from Ref.~\cite{Amhis:2019ckw}.\label{fig:rdrdst}}
\end{figure}

An overview of the experimental results on $R_{D^{(*)}}$ 
is given in Fig.~\ref{fig:rdrdst}. 
The measured central values for both $R_D$ and $R_{D^*}$ generally lie above the SM predictions with the largest tension (corresponding to $2.0\,\sigma$ for $R_D$ and $2.7\,\sigma$ for $R_{D^*}$) from the measurement of the BaBar collaboration~\cite{Lees:2012xj,Lees:2013uzd}. 

\paragraph{Measurements of $R_{D^{(*)}}$ at the LHC}
The LHCb collaboration has performed two measurements of $R_{D^*}$, one using muonic $\decay{\tau^-}{\mu^-\bar{\nu}_\mu\nu_\tau}$ decays~\cite{Aaij:2015yra}, 
and one using $\decay{\tau^-}{\pim\pip\pim(\piz)\nu_\tau}$ three-prong decays~\cite{Aaij:2017uff,Aaij:2017deq}. 

Due to the neutrinos in the final state of the signal decay $\decay{\Bdb}{D^{*+}\tau^-(\to\mu^-\bar{\nu}_\mu\nu_\tau)\bar{\nu}_\tau}$ 
the momentum of the $B$-meson can not be reconstructed analytically at the LHC. 
Instead, the $B$-momentum can be approximated with a resolution of around $18\%$ by using information on the $B$ decay vertex and thus the $B$ flight direction. 
A three-dimensional fit 
in the missing mass squared ($m_{\rm miss}^2$), the muon energy in the $B$ rest frame ($E_\mu^*$), and the four -momentum transfer ($q^2$)
is performed to determine the contributions from the $\decay{\Bdb}{D^{*+}\tau^-\bar{\nu}_\tau}$ signal- and $\decay{\Bdb}{D^{*+}\mu^-\bar{\nu}_\mu}$ normalisation decays. 
The resulting value of $R_{D^*}=0.336\pm 0.027\pm 0.030$ is found to be in agreement with the SM prediction at $2.1\,\sigma$~\cite{Aaij:2015yra}. 

The determination of $R_{D^*}$ with hadronic $\decay{\tau^-}{\pim\pip\pim(\piz)\nu_\tau}$ decays  
uses the decay $\decay{\Bd}{D^{*-}3\pi}$ as normalisation mode which exhibits the same final state as the signal, resulting in reduced systematic uncertainties.
A three-dimensional fit to the $\tau$ lifetime, $q^2$, and the output of a multivariate classifier is performed to determine the yields of signal and normalisation mode. 
The resulting value of $R_{D^*}=0.291\pm 0.019\pm 0.026\pm 0.013$ is in agreement with the SM prediction at $1\,\sigma$~\cite{Aaij:2017uff,Aaij:2017deq}. 

\begin{figure}
  \centering
  \includegraphics[width=0.59\textwidth]{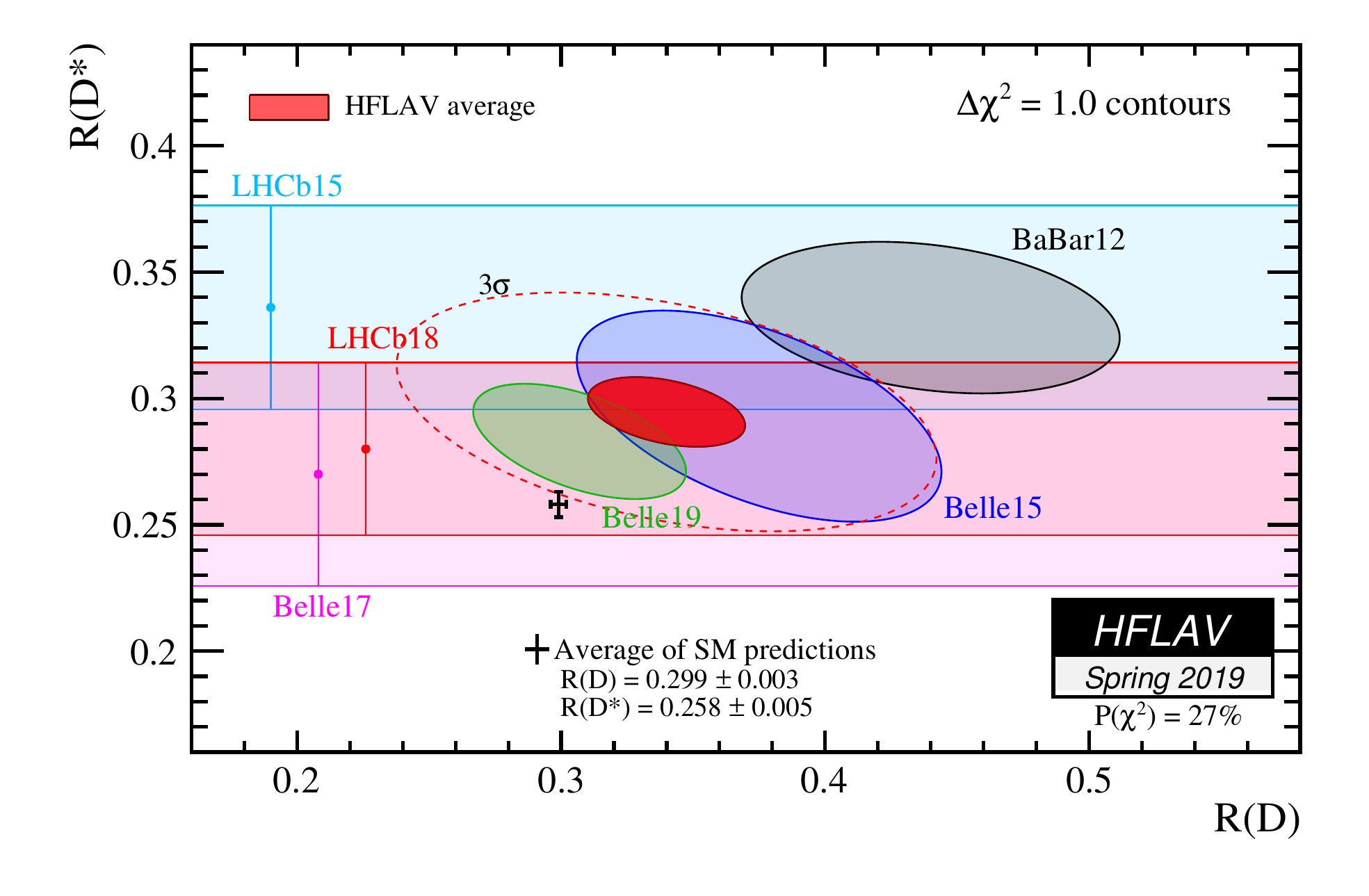}
  \caption{
    Two-dimensional fit of the experimental data on $R_D$ and $R_{D^*}$ from 
    the BaBar~\cite{Lees:2012xj,Lees:2013uzd}, Belle~\cite{Huschle:2015rga,Hirose:2017dxl,Belle:2019rba} and LHCb~\cite{Aaij:2015yra,Aaij:2017uff,Aaij:2017deq} collaborations by the heavy flavour averaging group~\cite{Amhis:2019ckw}. 
    The SM predictions are taken from Refs.~\cite{Lattice:2015rga,Na:2015kha,Bigi:2016mdz,Bernlochner:2017jka,Bigi:2017jbd,Jaiswal:2017rve},
    and do not yet include the update predictions in Eq.~\eqref{eq:RDDstDsDsst}.
    Figure reproduced from Ref.~\cite{Amhis:2019ckw}.\label{fig:rdrdstchi2}}
\end{figure}

\paragraph{Combination of $R_D$ and $R_{D^*}$ data}
Figure~\ref{fig:rdrdstchi2} shows 
the current experimental data on $R_{D^{(*)}}$ from 
the BaBar~\cite{Lees:2012xj,Lees:2013uzd}, Belle~\cite{Huschle:2015rga,Hirose:2017dxl,Belle:2019rba} and LHCb~\cite{Aaij:2015yra,Aaij:2017uff,Aaij:2017deq} collaborations in two dimensions.
The experimental average from a fit of the experimental input by the heavy flavour averaging group (HFLAV) results in a tension with SM predictions at $3.1\,\sigma$~\cite{Amhis:2019ckw}.
A recent update of the SM prediction~\cite{Bordone:2019guc,Bordone:2019vic}
has increased this tension to just above $4\sigma$. The updated prediction will
be used in an updated HFLAV average~\cite{PrivCommunicationMRotondo}.
This tension constitutes a \textit{flavour anomaly} in $\decay{b}{c\tau^-\bar{\nu}_\tau}$ tree-level decays. 

\subsubsection{Lepton universality test $R_{\jpsi}$}
At the LHC, $b$-hadrons of all species are produced, including \Bc\ mesons.
The LHCb collaboration has performed a measurement of the ratio
\begin{align}
  R_{J/\psi} =& \frac{{\cal B}(\decay{\Bc}{\jpsi\tau^-\bar{\nu}_\tau})}{{\cal B}(\decay{\Bc}{\jpsi\mu^-\bar{\nu}_\mu})}
\end{align}
where the $\tau$ lepton is reconstructed in the muonic decay mode $\decay{\tau^-}{\mu^-\bar{\nu}_\mu\nu_\tau}$. 
The recent lattice QCD study of the SM form factors has reduced the relative uncertainty of the $R_{J/\psi}$ SM prediction
to a level below that of $R_{D^*}$~\cite{Harrison:2020gvo,Harrison:2020nrv}.
The LHCb experiment finds
first evidence for the decay ${\cal B}(\decay{\Bc}{\jpsi\tau^-\bar{\nu}_\tau}$ and measures
$R_{\jpsi}=0.71\pm 0.17\pm 0.18$~\cite{Aaij:2017tyk}. This result is $1.9\,\sigma$ above the updated SM prediction given in Eq.~\eqref{eq:RJpsi}.

\subsubsection{Experimental prospects}
Precision measurements of $\decay{\bar{B}}{\bar{D}^{(*)}\tau^-\bar{\nu}_\tau}$
decays have a high priority at Belle~II\footnote{It should be noted that the Belle~II collaboration recently presented preliminary results on the branching fractions of exclusive $\decay{\bar{B}}{\bar{D}^{*}\ell^-\bar{\nu}_\ell}$ ($\ell=e,\mu$) decays with $34.6\invfb$ of early data~\cite{Abudinen:2020ddg,Abudinen:2020acd}}. %
Besides performing precise measurements of $R_{D^{(*)}}$, the Belle~II collaboration will
be able to perform measurements differential in $q^2$ and angular analyses. 
Angular analyses will allow access to $P_\tau(D^{(*)})$, the polarisation of the $\tau$ lepton, and $P_{D^*}$, the polarisation of the $D^*$ meson, quantities which are interesting probes for NP but which so far have only been determined with limited accuracy~\cite{Hirose:2017dxl,Abdesselam:2019wbt}. 
The Belle~II collaboration expects uncertainties of $\sigma(R_{D})=(\pm 2.0_{\rm stat}\pm 2.5_{\rm syst})\%$, $\sigma(R_{D^*})=(\pm 1.0_{\rm stat}\pm 2.0_{\rm syst})\%$, and $\sigma(P_\tau(D^*))=\pm 0.06_{\rm stat}\pm0.04_{\rm syst}$ with the full $50\invab$ data sample~\cite{Kou:2018nap}.
Belle~II will furthermore explore experimentally challenging inclusive measurements of $\bar{B}\to X_c\tau^-\bar{\nu}_{\tau}$ decays~\cite{Kou:2018nap}. These observables can be predicted in the SM using methods complementary to the approaches used for exclusive decays as discussed in Sec.~\ref{sec:treetheory}, see \eg\ Refs.~\cite{Ligeti:2014kia,Mannel:2017jfk}. 

LHCb will exploit the fact that all $b$-hadron species are produced at the LHC and is expected to probe $\decay{b}{c\tau^-\bar{\nu}_\tau}$ transitions by performing measurements of $R_D$, $R_{D^*}$, $R_{D_s}$, $R_{\jpsi}$, and $R_{\Lambda_c}$~\cite{Bediaga:2018lhg}. 
LHCb projects an ultimate sensitivity to $R_{D^*}$ with the full Upgrade~II data sample of $\sigma(R_{D^*})=0.003$ for both the muonic and the three-prong hadronic $\tau$ decay, assuming that systematic uncertainties scale the same as the statistical uncertainty~\cite{Bediaga:2018lhg}. 
Angular analyses of $\decay{B}{D^*\tau\nu}$ decays are challenging but are expected to be feasible with the large data sample of the LHCb upgrades.

\section{Summary, Interpretation and Outlook}
\label{sec:conclusions}

The \textit{flavour anomalies} are comprised of a series of tensions 
between measurements and SM predictions for a variatey of $b$-hadron decay processes: 
Measurements of branching fractions and angular distributions of $\decay{b}{s\mumu}$ FCNC decays, which are dominated by precise LHCb results, show a consistent pattern of tensions with the SM expectations. The tensions seen in the lepton-flavour universality 
tests $R_K$, $R_K^*$, and $R_{pK}$ 
are compatible with the deviations seen in the \decay{b}{s\mumu} modes.
While individual measurements show tensions with significances between $2$--$3\,\sigma$, the combination of the measurements provides an intriguing and consistent picture. 

Another set of measurements performed by the $B$-factory experiments Babar and Belle, and by LHCb shows
hints of lepton-flavour universality breaking in tree-level $b\to c\ell^- \bar{\nu}_\ell$ transitions.
Specifically, tensions arise in the lepton flavour universality ratios $R_{D}$, $R_{D^*}$ and $R_{J/\psi}$  between tauonic and muonic decays.
The experimental average of the individual $R_{D}$ and $R_{D^*}$ measurements shows a tension with the SM prediction of more than $3\,\sigma$ significance.

\begin{figure}
    \centering
    \includegraphics[width=.4\textwidth]{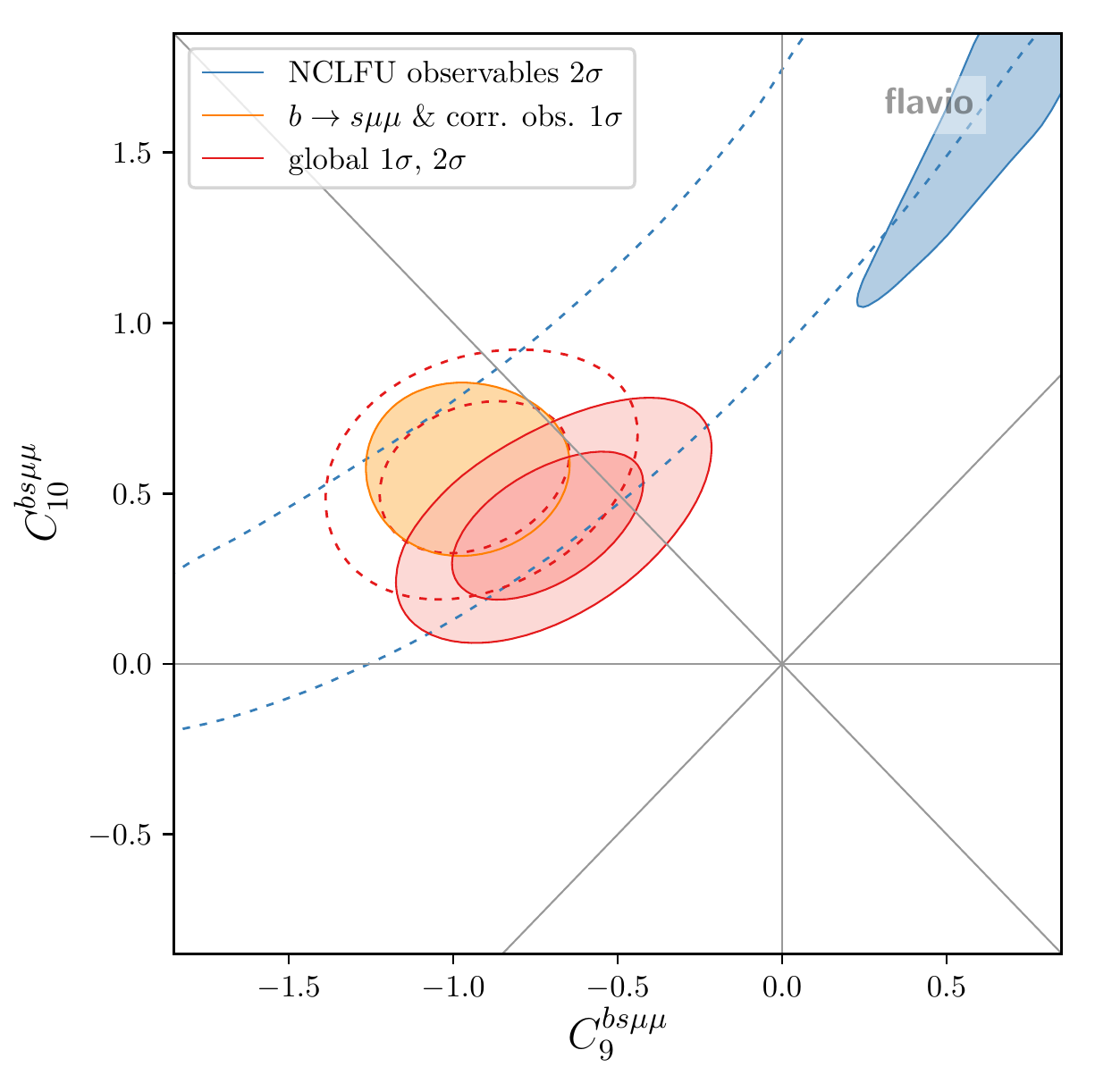}
    \raisebox{.05\textwidth}{
    \includegraphics[width=.5\textwidth,trim=0 0 260 0,clip]{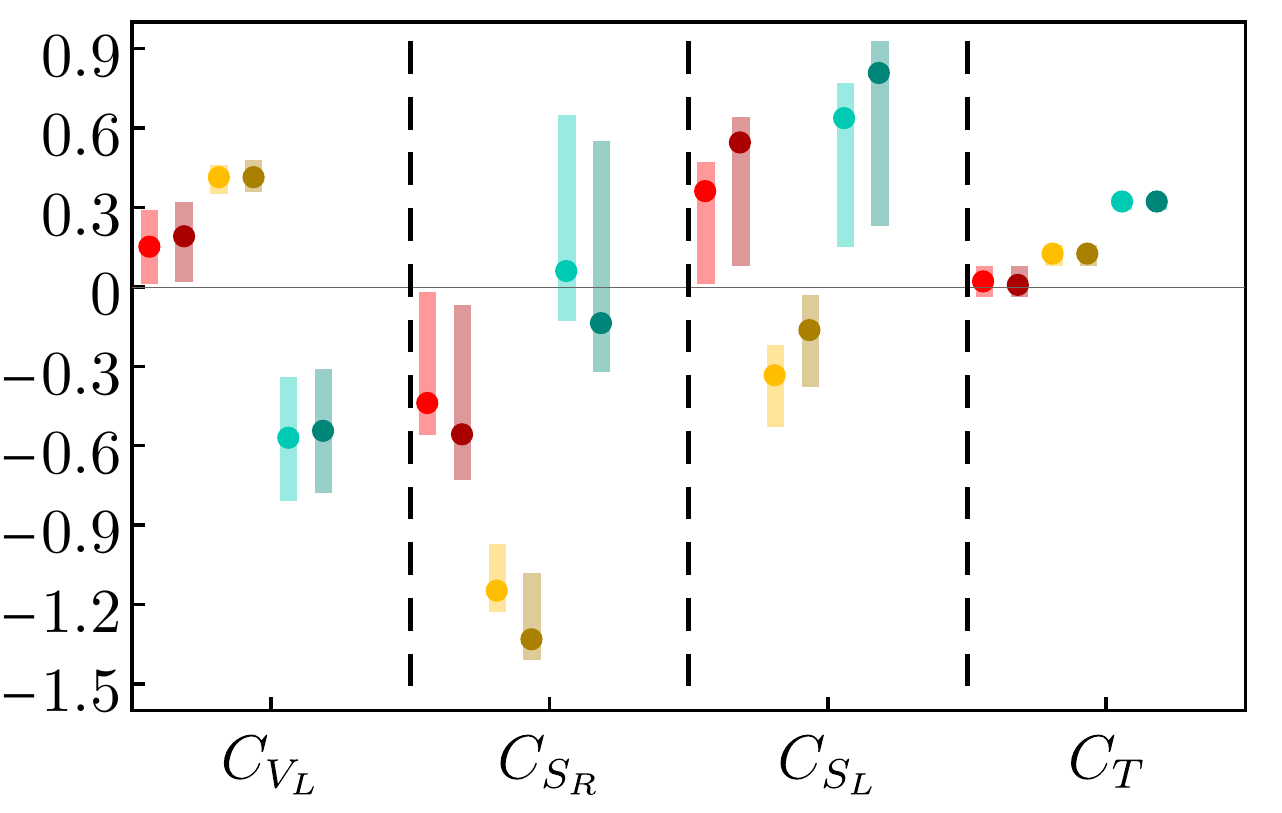}
    }
    \caption{
    \label{fig:wet-fits}
    The metrology of the WET Wilson coefficient determinations from $b\to s\mumu$ processes (left)
    and $b\to c\tau^-\bar{\nu}_\tau$ processes (right) at hand of two prototypical studies~\cite{Aebischer:2019mlg,Murgui:2019czp}.
    The left plot shows the two-dimensional constraint on the BSM contributions to the 
    two numerically leading WET Wilson coefficients $C_9$ and $C_{10}$, exhibiting
    a $\sim 6\sigma$ tensions with the SM point in the origin.
    The right plot shows summaries of the the one-dimensional profile likelihoods for the BSM contributions to the
    WET Wilson coefficients under the assumption that the coefficient $C_{VR}$ is
    lepton-flavour universal, as required by matching on the SM Effective Field Theory~\cite{Cata:2015lta}.
    The colors encode the different minima in the fit, which cannot be disentangled by present data.
    We refer to the respective works for further details.
    }
\end{figure}

For the neutral-current decays, the tensions can be explained by a
modification of the $b\to s$ operators that include a vectorial lepton current ($\op{9}$).
Its Wilson coefficient requires a shift by around $-25\%$ of its SM value~\cite{%
Alguero:2019ptt,%
Ciuchini:2019usw,%
Aebischer:2019mlg,%
Arbey:2019duh%
}; see Fig.~\ref{fig:wet-fits} (left). Global analyses of all available
$b\to s\lbrace \gamma, \ellell\rbrace $ data also indicate the need for shifts in the couplings
of operators with either the axial lepton current ($\op{10}$),
or hadronic currents with non-SM-like chirality ($\op{9',10'}$), or both.
These findings hinge crucially on our current understanding of the hadronic matrix elements
of both local and nonlocal nature, \ie, local and nonlocal form factors.
As a consequence, the interpretation of the $b\to s\mumu$
anomalies as a genuine sign of BSM physics is presently theory limited. Ongoing efforts by
lattice QCD groups to increase the precision of the local form factors (see \eg\  Refs.~\cite{Flynn:2015ynk,Flynn:2016vej,Lizarazo:2016myv})
and current developments in the field of the nonlocal hadronic matrix elements (see \eg\  Ref.~\cite{Gubernari:2020eft})
give hope that future phenomenological analyses can overcome these limitations.

The most precise measurements of $\decay{b}{s\mumu}$ are currently provided by the LHC experiments. 
Most of the analyses, however, do not yet use the full LHC Run~1 and~2 data sample. 
Therefore, a wealth of new experimental results can be expected in the near future, even before the start of LHC Run 3 in 2022. 
It will be crucial to observe if the significance of the anomalies in branching fractions and angular distributions will increase with the added data and will evolve to individually significant measurements, or if more data reduces the tensions seen. 
Another essential ingredient to establish the \textit{flavour anomalies} would be to see significant, consistent measurements from different experiments. 
With the large datasets collected by ATLAS, CMS and LHCb, 
the prospects to resolve or establish these anomalies in the near future are excellent. 

Apart from the LHC, the Belle~II experiment~\cite{Abe:2010gxa} will provide improved measurements of inclusive decays and furthermore play an important role 
by independently scrutinising the \textit{flavour anomaly} measurements published at the  LHC. In $b\to s\ellell$ decays, a sufficiently large data-set for new insights is expected to be collected by 2022--2023~\cite{talkPhilFPCP}.  

Among the charged-current decays, the tensions cannot be clearly explained
as a modification of a single WET coefficient~\cite{Murgui:2019czp}; see Fig.\ref{fig:wet-fits} (right).
In contrast to the neutral-current decays, the theoretical predictions are under
excellent control, largely thanks to the the heavy-quark expansion of the local form factors
and stringent dispersive bounds on the form factor parameters.
Thus, the interpretation of the charged-current semitauonic anomalies as genuine signs
of BSM physics is presently limited by the experimental precision.

The experimental picture in $b \to c \ell^- \bar{\nu}$ currently shows Belle and LHCb data with similar precision. Both LHCb and Belle~II will put a strong focus on clarifying the picture, and will utilise their orthogonal advantages: at the LHC, all $b$-hadron species are produced and can hence be used for precision tests of $b \to c \ell^- \bar{\nu}$ transitions. In contrast, the cleaner environment at Belle~II allows access to tests of the polarization of the $\tau^-$ lepton and $D^{*0}$ meson. Measurements in $b\to c\tau^- \bar{\nu}_\tau$ decays could be the first significant contributions of the Belle~II experiment in the quest to understand the \textit{flavour anomalies}~\cite{talkPhilFPCP}.

The available data and their phenomenological analyses within the WET have highlighted
significant and consistent tensions with the SM expectations. A path to future interpretations
of these tensions is clear: the WET constraints are an indispensable ingredient for BSM
model-building studies. They should be included in global BSM studies within, \eg, the
SM Effective Field Theory (SMEFT) or the Higgs Effective Field Theory. In particular
for SMEFT-based studies, first steps in this direction have been taken with the
SM global likelihood (\texttt{smelli}) software~\cite{Aebischer:2018iyb}.

In light of the anomalies, the present plans for theoretical and experimental improvements,
and the prospects for large future data-sets by the LHC and $B$-factory experiments,
flavour physics in general and $B$ physics in particular is and will remain an exciting
and rewarding field of research.

\section*{Acknowledgements}
C.\,L.\ and D.\ v.\ D.\ gratefully acknowledge support by the Emmy Noether programme of the Deutsche Forschungsgemeinschaft (DFG), under grants LA 3937/1-1 and DY 130/1-1. D.\ v.\ D.\ further acknowledges support by the DFG Collaborative Research Center
110 ``Symmetries and theEmergence of Structure in QCD''.
J.\,A.\ acknowledges support from the Heisenberg programme of the Deutsche Forschungsgemeinschaft (DFG), GZ: AL~1639/5-1 and funding from the European Research Council (ERC) under the European Union's Horizon 2020 research and innovation programme under grant agreement No.~714536: PRECISION.

\clearpage




\bibliographystyle{elsarticle-num} 

\bibliography{main}





\end{document}